\documentclass[times,sort&compress,3p]{elsarticle}
\journal{Journal of Multivariate Analysis}
\usepackage[labelfont=bf]{caption}

\usepackage{multirow,amsmath,amsfonts,amssymb,amsthm,booktabs,color,epsfig,graphicx,hyperref,url,bm}
\usepackage{bbold}
\theoremstyle{plain}
\newtheorem{theorem}{Theorem}

\newtheorem{corollary}{Corollary}

\theoremstyle{definition}

\def\NN{\mathbb{N}}

\def\QQ{\bm{Q}}
\def\GG{\bm{G}}

\def\ss{\bm{s}}
\def\SS{\bm{S}}
\def\uu{\bm{u}}
\def\UU{\bm{U}}
\def\EE{{\rm E}}
\def\VV{{\rm var}}
\def\hh{\bm{h}}
\def\YY{\bm{Y}}
\def\yy{\bm{y}}
\def\bh{\bm{h}}
\def\bvartheta{\bm{\vartheta}}
\def\btheta{\bm{\theta}}

\def\cor{{\rm corr}}

\def\RR{\mathbb{R}}

\def\R{{\rm I\! R}}

\begin{document}

\begin{frontmatter}

\title{A flexible Clayton-like spatial copula with application to bounded support  data}

\author[A1]{Moreno Bevilacqua\corref{mycorrespondingauthor}}
\author[A2]{Eloy Alvarado}
\author[A3]{Christian Caama\~no-Carrillo}

\address[A1]{Facultad de Ingenier\'ia y Ciencias, Universidad Adolfo Ib\'a\~nez, Vi\~na del Mar, Chile\\
Dipartimento di Scienze Ambientali, Informatica e Statistica, Ca’ Foscari University of Venice, Venice, Italy}
\address[A2]{Departamento de Industrias, Universidad T\'ecnica Federico Santa Mar\'ia, Santiago, Chile}
\address[A3]{Departamento de Estad\'istica, Universidad del B\'io-B\'io, Concepci\'on, Chile}

\cortext[mycorrespondingauthor]{Corresponding author. Email address: moreno.bevilacqua@uai.cl\url{}}

\begin{abstract}
The Gaussian copula is a powerful tool that has been widely used to model spatial and/or temporal correlated data with arbitrary marginal distributions. However, this kind of model can potentially be too restrictive since it expresses a  reflection symmetric dependence. In this paper, we propose a new spatial copula model that makes it possible to obtain random fields with arbitrary marginal distributions with a type of dependence that can be reflection symmetric or  not.

Particularly, we propose a new random field with uniform marginal distributions that can be viewed as a spatial generalization of the classical Clayton copula model. It is obtained through a power transformation of a specific instance of a beta random field which in turn is obtained using a transformation of two independent Gamma random fields.

For the proposed random field, we study the second-order properties and  we provide analytic expressions for the bivariate distribution and its correlation.  Finally, in the reflection symmetric case, we study the associated geometrical properties.
As an application of the proposed model we focus on spatial modeling of  data with bounded support. Specifically, we focus on spatial regression models with marginal distribution of the  beta type. In a  simulation study, we investigate the use of the weighted pairwise composite likelihood method for the estimation of this model. Finally, the effectiveness of our methodology is illustrated by analyzing point-referenced vegetation index data using  the Gaussian copula  as benchmark.
Our developments have been implemented in an open-source package for the \textsf{R} statistical environment.
\end{abstract}


\end{frontmatter}

\section{Introduction\label{sec:1}}

Many applications of statistics across a wide range of disciplines such as climatology, environmental sciences and engineering, to name a few, show an increasing interest in the statistical analysis of geo-referenced spatial data. In order to model the inherent uncertainty of the data, Gaussian random fields play a fundamental role; see \cite{Stein:1999,Banerjee-Carlin-Gelfand:2004,Cressie:Wikle:2011}, for instance. Indeed, the Gaussian  random fields can be completely specified in  terms of mean and correlation function and flexible correlation models such as the Matérn \citep{Stein:1999} or the  Generalized Wendland correlation models \citep{bb2019} can be specified to model the geometrical properties of the random field.

Unfortunately, the Gaussian assumption is rarely met in practice. Indeed, in many geostatistical applications, including climatology, oceanography, the environment and the study of natural resources, the Gaussian framework is unrealistic because the observed data have different  features such as asymmetry, heavy tails, positivity and bounded support.

The spatial generalized mixed models  proposed by \citet{Diggle:Tawn:Moyeed:1998} has been widely used to model non-Gaussian spatial data \citep{MoralesNavarrete2021}. Under this framework, non-Gaussian models  are specified using a specific  link function and  a latent Gaussian random field through a conditional independence assumption. However,  the  conditional independence assumption underlying models of this kind leads to random fields with  non-standard marginal distributions and  with a ``forced'' nugget effect that implies no mean square continuity \citep{GELFAND201686,MoralesNavarrete2021}.  This can be a potential problem when modeling data exhibiting continuity and/or differentiability.

More flexible models can be obtained through a suitable transformation of a Gaussian random field or independent copies thereof. This is an appealing approach because the correlation of the transformed random field depends on the correlation of the Gaussian random field and very often the geometrical properties of the non-Gaussian random field are inherited from the Gaussian random field. Notable examples  can be found in  \cite{DeOliveira:2006} for log-Gaussian   random fields, \cite{Palacios:Steel:2006} for Gaussian-log-Gaussian random fields, \cite{Zhang:El-Shaarawi:2010} for Skew-Gaussian random fields, \cite{Xua:Genton:2017} for Tukey $g$-$h$  random fields, \cite{bevilacqua2021non} for $t$ random fields, \cite{bevilacqua2020modeling} for Weibull random fields, \cite{Blasi2022} for sinh-arcsinh random fields, and \citep{MoralesNavarrete2021} for Poisson random fields just to mention a few.

A general powerful modeling tool to obtain  non-Gaussian random fields  with arbitrary marginal distributions can be obtained under the copula framework \citep{Joe:2014}. Unfortunately, the adaptation of the copula framework to the spatial continuous setting  is not a trivial task.
Hereafter, we consider random fields defined on   $A$ a subset of $\R^d$  even if the proposed methodology can be easily extended to
more complicated spaces such as the continuous space-time framework \citep{Gneiting:2002} or the spherical space \citep{gneiting2013,porcubev} or  a linear network \citep{andmo}.
 A random field $\{S(\ss), \ss \in A\}$  with arbitrary marginal cdf $F_{S}$ can be obtained through the transformation
\begin{equation*}\label{copula1}
S(\ss)=F_{S}^{-1}\{U(\ss)\},
\end{equation*}
where $U$ is a random field with uniform marginals and  $F_{S}^{-1}$
is the generalized inverse distribution function. This approach makes it possible to model the marginal distributions and the dependence structure separately.  In particular the dependence is driven by the dependence of the random field $U$.

Inference and prediction of the non-Gaussian random field $S$ can be a challenging task in particular when considering only one realization
from $S$ which is the typical case for spatial data.
It is important to highlight that when  considering independent realizations from $S$, the inference generally simplifies and the set of copula models that can be used is  larger; see for instance \cite{Gentcop} and the references therein.

However, in this paper we assume that only one  realization of $S$ is available which is the typical case for spatial data.
In this case, a popular choice is $U(\ss)=\Phi\{ Z(\ss) \}$ where $\Phi$ is the cdf of the Gaussian distribution and $\{Z(\ss), \ss \in A\}$ is a standard Gaussian random field. This is  the so-called Gaussian copula random field that has been widely applied in the analysis of spatial and/or temporal correlated data; see \citep{bardossy2006copula,Kazianka:Pilz:2010,Masarotto:Varin:2012,graler2014modelling,guolo2014beta,DeOliveira:2006},  just to mention a few. The main reason for success of the Gaussian copula is that the dependence structure is indexed by a correlation function which is helpful in many applications such as spatial statistics, time series or longitudinal data.

However, Gaussian copula random fields suffer from some limitations due to the kind of dependence structure they can model. In particular, the resulting type of dependence is reflection symmetric, which basically means that high values exhibit a spatial dependence similar to low ones. These restrictions are often violated when analyzing Gaussian or non Gaussian real data. For instance  reflection asymmetry has been observed in rainfall or wind speed data \citep{bardossy2006copula,SUROSO2018685,bevilacqua2020modeling}. To overcome these restrictions, a random field $U$ with uniform marginals and more complex dependence structures than $\Phi\{Z(\ss)\}$ is needed.

A   first flexible copula model for spatial data has been  proposed by   \cite{bardossy2006copula} and deeply studied in
\cite{QUESSY201640} and
 \cite{Quessy2019}. The so-called Chi-squared copula has been obtained  by considering the cdf associated with a squared Gaussian random field and the associated uniform random field takes the form $U(\ss)=|1-2\Phi\{Z(\ss)\}|$.  However this kind of model offers only a specific type of bivariate dependence.
As a more flexible alternative, a vine copula approach can be performed \citep{ERHARDT201574,graler2014modelling}, but the computational time explodes as the dimension increases because the number of possible configurations becomes very high.

In this paper, we propose a new flexible random field with uniform marginals. The main features of this model is that, unlike the Gaussian copula random field, the type of dependence can be reflection symmetric or not. In addition, as in the Gaussian copula or Chi-squared copula  case, the dependence structure of the proposed random field is indexed by a correlation function with an additional parameter that characterizes the type of reflection (a)symmetry.

 Specifically, we first define a general class of random fields  with uniform marginal distributions that we call Archimedean  random field because it is  obtained   generalizing  the stochastic representation  associated to the  Archimedean copula \citep{Genest1986,Joe:2014}. Then we focus on a specific instance of the Archimedean  class  that we call Clayton random field because it can be viewed as a spatial generalization of the Clayton copula.
It turns out that the Clayton  random field can be written as a power transformation  of a specific random field with beta marginals which is  obtained  through a transformation  of Gamma random fields obtained as sum of squared Gaussian random fields sharing a common underlying correlation function.

For the Clayton random field, we study the second-order properties and  we provide analytic expressions for the bivariate pdf
and cdf (the associated  bivariate copula density and copula function, respectively) and the correlation function. It turns out that the bivariate pdf, the bivariate cdf  and the correlation function depend on some special functions, namely the Gaussian hypergeometric function, the Appell function of the fourth type, and the Kampé de Fériet function \citep{Gradshteyn:Ryzhik:2007}. Finally, in a special case, we study the associated geometrical properties. Particularly, we show that nice properties such as  mean-square continuity and degrees of mean-square differentiability can be inherited from the underlying Gaussian random field.

As an application of the Clayton  random field, we consider a spatial random field  
 with beta marginal  distribution with a  parametrization that allows to specify a spatial  mean  regression model   \citep{ferrari2004beta}.
Estimating the (reparametrized) beta random field or more generally  random fields with arbitrary marginal distributions obtained from the  Clayton random field  is not an easy task.
In fact, it must be said that the likelihood function associated with the proposed Clayton random field involves an analytically intractable form. As a consequence, likelihood-based  estimation methods  are unfeasible. Exploiting the results on the bivariate distribution of the Clayton  random field, we investigate the use of the weighted pairwise composite likelihood \citep{Lindsay:1988,Varin:2008} as method of estimation. In particular, in our simulation study, we study the performance of the weighted pairwise composite likelihood when estimating the parameters of the (reparametrized) beta random field. The methods proposed in this paper are implemented in the \textsf{R} package \texttt{GeoModels} \citep{Bevilacqua:2018aa} and the \textsf{R} code for reproducing the work is available as an online supplement.

The remainder of the article is organized
as follows. In Section \ref{sec:2}, we first study  an auxiliary random field with beta marginals which is a key tool when defining the Clayton random field. In Section \ref{sec:3}, we
first define  the Archimedean random field  and then we study a specific case, the Clayton random field, and in particular we provide the associated  bivariate pdf and cdf and the correlation function.
In addition we study geometrical properties of the Clayton random field in the special reflection symmetric case.
 In Section \ref{sec:4}, we present a simulation study
in order to investigate the performance of the  weighted pairwise composite likelihood  method when estimating the mean regression parameters and the dependence parameters of the (reparametrized) beta random field obtained from the Clayton random field. In Section \ref{sec:5}, we illustrate  the application
of the proposed methodology
 by analyzing spatially point-referenced vegetation indexes data using a random field with beta marginal distribution. In particular we compare the proposed model using the  Gaussian copula model as a benchmark. All the proofs are deferred to the Appendix.

\section{A random field with beta marginals}\label{sec:2}

Henceforth, given a weakly stationary random field  $\{Q(\ss), \ss \in A \}$ with  $E\{Q(\ss)\}=\mu(\ss)$ and $\VV\left\{Q(\ss)\right\}=\sigma^2$, we denote by $\rho_Q(\hh)=\cor\{Q(\ss_i),Q(\ss_j)\}$ its correlation function, where $\hh=\ss_i-\ss_j \in A$ is the lag separation vector. For any set of  distinct points $(\ss_1,\ldots,\ss_n)^\top$, $n\in \mathbb{N}$, we denote by $\QQ_{ij}=(Q(\ss_i),Q(\ss_j))^\top$, $i \neq j$, the bivariate random vector and by $\QQ=(Q(\ss_1),\ldots, Q(\ss_n))^\top$ the multivariate random vector. Moreover, we denote by $f_{Q}$ and  $F_{Q}$ the marginal probability density function (pdf) and cumulative distribution function (cdf) of  $Q(\ss)$, respectively, with  $f_{\QQ_{ij}}$  $F_{\QQ_{ij}}$ the pdf (probability density function) and cdf (cumulative distribution function)  of $\QQ_{ij}$ and with $f_{\QQ}$ the pdf of $\QQ$
with $F_{\QQ}$ the cdf of $\QQ$
. Additionally, we denote by $\Phi$ and $\phi$ the cdf and pdf of the standard Gaussian random variable, respectively. Finally, if a random field has a $X$ marginal distribution, then we call it $X$ random field.

Let $\{Z(\ss), \ss \in A\}$, a standard Gaussian random field  with correlation function  $\rho_Z(\hh)$ and with some abuse of notation we set $\rho(\hh)=\rho_Z(\hh)$. Henceforth, we call  $Z$ and $\rho(\hh)$ as  the underlying Gaussian random field and correlation function respectively.

To build our random field with beta marginals we  first consider a Gamma random field $\{G_{\psi}(\ss), \ss \in A\}$ with marginal distribution $G_{\psi}(\ss) \sim  \mathcal{G} (\psi/2,1)$, defined as
\begin{equation*}\label{pairchi}
G_{\psi}(\ss)=\sum_{i=1}^\psi Z_i(\ss)^2/2,
\end{equation*}
where $Z_i$, $i\in\{1,\ldots,\psi\}$  are mutually independent copies of $Z$ with  $\rho_{G_{\psi}}(\hh) =
\rho^2(\hh)$. The associated multivariate density $f_{\GG_{\psi}}$ was discussed earlier by \citet{krishnamoorthy1951}, and its properties have been studied since then by several authors; see \citep{Kri:rao:1961,Royen:2004}. It turns out that the analytical expressions of the multivariate density can be derived only in some special case. However, in the bivariate case, the pdf has general expression given by \citep{kibble1941two}
\begin{equation}\label{pairchi2}
f_{\GG_{\psi;ij}}(g_{i},g_j)=\frac{(g_ig_j)^{\psi/2-1}e^{-\frac{(g_i+g_j)}{1-\rho^2(\bh)}}}{\Gamma\left(\psi/2\right)\{1-\rho^2(\bh)\}^{\nu/2}}
\left[\frac{\sqrt{\rho^2(\bh)g_ig_j}}{1-\rho^2(\bh)}\right]^{1-\psi/2}I_{\psi/2-1}
\left[ \frac{ 2\sqrt{\rho^2(\bh)g_ig_j} } {1-\rho^2(\bh)}\right],
\end{equation}
where  $I_{a}(x)$  is the modified Bessel function of the first kind of order a.
Now consider the random field $\{Y_{\nu,\alpha}(\ss), \ss \in A\}$  defined as:
\begin{equation}\label{repbetgen}
Y_{\nu,\alpha}(\ss)=\dfrac{H_{\nu}(\ss)}{H_{\nu}(\ss)+N_{\alpha}(\ss)}.
\end{equation}
where $H_{\nu}$, and $N_{\alpha}$ are two independent copies of the Gamma random field $G_{\psi}$, $\psi \in \{ 1,2,\ldots \}$ sharing an underlying common correlation function $\rho(\bh)$.
By construction
$Y_{\nu,\alpha}(\ss)$ has  beta marginal distribution $\mathcal{B}(\nu/2,\alpha/2)$
with mean and variance given by $\EE\left\{Y_{\nu,\alpha}(\ss)\right\}=\nu/(\nu+\alpha)$ and $\VV\left\{Y_{\nu,\alpha}(\ss)\right\}=2\nu \alpha/\{(\nu+\alpha)^2(\nu+\alpha+2)\}$, respectively. Hereafter, we call $Y_{\nu,\alpha}$ the auxiliary beta random field with underlying correlation function $\rho(\hh)$.

A possible drawback for the gamma random fields used in (\ref{repbetgen}) is that it is a limited model due to the restrictions to the half-integers for the shape parameter. Actually, in some special  cases, it can assume any positive value greater than zero. This feature is intimately related to the infinite divisibility of the squared of the underlying Gaussian random fields $Z^2$ as shown in \cite{Kri:rao:1961}. Characterization of the infinite  divisibility of $Z^2$ has been studied in \cite{Bapat:1989,Eisenbaum:Kaspi:2006,Griffiths:1970,Vere-Jones:1997}. In particular, \citet{Bapat:1989} provides a characterization based on $\Omega_n$, the correlation matrix associated with $\rho(\hh)$. Specifically,  $\psi >0$ if and only if there exists a matrix $S_n$ such that $S_n\Omega_n^{-1}S_n$ is an $M$-matrix \citep{plee:1977}, where $S_n$ is a signature matrix, i.e., a diagonal matrix of size $n$ with entries either $1$ or $-1$. This condition is satisfied, for instance, by a stationary Gaussian random process $Z$ defined on $A=\RR$ with  an exponential correlation function. However, it is not guaranteed to be satisfied for any covariance matrix in any dimension.

These sorts of restrictions on the shape parameter of the Gamma random field $G_{\nu}$ are clearly inherited by the   auxiliary beta random field, which means that $Y_{\nu,\alpha}$ is well defined for $\nu,\alpha\in\{1,2,\ldots\}$. This certainly restricts the flexibility of this kind of model.

In what follows, we make use of  the generalized hypergeometric function defined in \citet{Gradshteyn:Ryzhik:2007} by:
\begin{equation}\label{pfq}
{}_pF_q(a_1,a_2,\ldots,a_p;b_1,b_2,\ldots,b_q;x)=\sum\limits_{k=0}^{\infty}
\frac{(a_1)_k,(a_2)_k,\ldots,(a_p)_k}{(b_1)_k,(b_2)_k,\ldots,(b_q)_k}\frac{x^k}{k!}\;\;\;\text{for}\;\;\;p,q\in\{0,1,\ldots\},
\end{equation}
where $(a)_{k}=  \Gamma(a+k)/ \Gamma(a)$, for $k\in \NN \cup \{0\} $, is the Pochhammer symbol.
In particular we focus on  the Gaussian hypergeometric function

\begin{equation}\label{2f1}
{}_2F_1(a,b,c;x)=\sum\limits_{k=0}^{\infty}\frac{(a)_k (b)_k}{(c)_k}\frac{x^k}{k!},\;\;\; |x|<1,
\end{equation}
and we  also consider  the Appell hypergeometric  function of the fourth type  \citep{Gradshteyn:Ryzhik:2007} defined as
\begin{equation}\label{ap4}
F_4(a,b;c,c';w,z)=\sum\limits_{k=0}^{\infty}\sum\limits_{m=0}^{\infty}\frac{(a)_{k+m}(b)_{k+m}w^kz^m}{k!m!(c)_k(c')_m},\;\;\;\;|\sqrt{w}|+|\sqrt{z}|<1,
\end{equation}
which is a special function  of two variables.
The  special functions (\ref{2f1}) and (\ref{ap4}) are related through the identity
\begin{equation}\label{apell4}
F_4(a,b;c,c';w,z)=\sum\limits_{k=0}^{\infty}\frac{(a)_{k}(b)_{k}z^k}{k!(c')_k}{}_2F_1(a+k,b+k;c;w),\;\;\;\;|\sqrt{w}|+|\sqrt{z}|<1;
\end{equation}
see \citep{Brychkov:Saad:2017}.

Equation  (\ref{apell4}) is useful for the computation of the $F_4$ function given that
efficient numerical computation of the ${}_2F_1$ function  can be found in different libraries
of statistical softwares including \textsf{R}, \textsf{MATLAB} and \textsf{Python}. Furthermore, we introduce a two-variable   power series generalization called  the Kampé de Feriet function  defined as
\begin{equation}\label{KF}
F_{E;G;H}^{A;B;C}\left[ \begin{array}{c}
                                  (a);(b_B);(c_C)\\
                                  (e_E);(g_G);(h_H)
                                \end{array}\middle\vert x,y\right]
=\sum\limits_{k=0}^{\infty}\sum\limits_{m=0}^{\infty}\frac{\prod\limits_{j=1}^A(a_j)_{k+m}\prod\limits_{j=1}^B(b_j)_{k}\prod\limits_{j=1}^C(c_j)_m x^ky^m}{k!m!\prod\limits_{j=1}^E(e_j)_{k+m}\prod\limits_{j=1}^G(g_j)_{k}\prod\limits_{j=1}^R(h_j)_m}.
\end{equation}

This special  function can be viewed as a generalization of the $F_4$ and other types of Appell functions.  For instance, it can be shown that $F_{2;0;0}^{0;1;1}=F_4$  \citep{Srivastava:Karlsson}.

The following theorem provide the pdf of the bivariate distribution of the auxiliary beta random field defined in (\ref{repbetgen}) in terms of the Appell $F_4$ function.
\begin{theorem}\label{theo1}
Let $Y_{\nu,\alpha}$ the auxiliary beta random field defined in  \eqref{repbetgen} with underlying correlation function $\rho(\hh)$. Then the pdf of
$\YY_{\nu,\alpha;ij}$ is given by
\begin{equation*}\label{pairb}
f_{\YY_{\nu,\alpha;ij}}(\yy_{ij})=
\frac{(y_iy_j)^{\nu/2-1}\{(1-y_i)(1-y_j)\}^{\alpha/2-1}\Gamma^2\left(c\right)}{\Gamma^2\left(\nu/2\right)\Gamma^2\left(\alpha/2\right)\{1-\rho^2(\hh)\}^{-c}}
F_4\left(c,c;\nu/2,\alpha/2;\rho^2(\hh)y_iy_j,\rho^2(\hh)(1-y_i)(1-y_j)\right),
\end{equation*}
with $c=(\nu+\alpha)/2$.
\end{theorem}

Note that $f_{\YY_{\nu,\alpha;ij}}$ is well defined for $\nu,\alpha>0$  irrespectively of the correlation function, as it is obtained from a bivariate Gamma distribution. Moreover, when $\rho(\hh)=0$, it can be easily shown that $f_{\YY_{\nu,\alpha;ij}}(\yy_{ij})=f_{Y_{\nu,\alpha}}(y_{i})f_{Y_{\nu,\alpha}}(y_{j})$. This implies that zero pairwise correlation implies pairwise independence.

The following theorem provides the  correlation function of the  auxiliary  beta random field $Y_{\nu,\alpha}$ in terms of the Kampé de Fériet function defined in (\ref{KF}).
\begin{theorem}\label{theo2}
Let $Y_{\nu,\alpha}$ the auxiliary   beta random field defined in  \eqref{repbetgen} with underlying correlation $\rho(\hh)$. Then:
\begin{equation}\label{CC}
\rho_{Y_{\nu,\alpha}}(\hh)=\frac{\nu (c+1)}{\alpha}\left[\{1-\rho^2(\hh)\}^{c}A-1\right],
\end{equation}
    where $c=(\nu+\alpha)/2$ and
\begin{equation*}\label{A1}
A=F_{2;0;1}^{2;1;2}\left[ \begin{array}{c}
                                  c;\frac{\alpha}{2};\frac{\nu}{2}+1\\
                                  c+1;-;\frac{\nu}{2}
                                \end{array}\middle\vert \rho^{2}(\hh),\rho^{2}(\hh)\right].
\end{equation*}
\end{theorem}

\section{Clayton  random fields}\label{sec:3}
The goal of this section is to introduce what we call the  Clayton  random field, which is a random field with standard uniform marginal distributions with a flexible type of dependence allowing reflection (a)symmetry.
We call the proposed random field `Clayton' because it  can be viewed as a generalization, of the classical Clayton copula.

To clarify the link between our construction and the classical Clayton copula, let us first  consider a random vector ${\mathbf{U}}=(U_1,\dots,U_n)^\top$, where

\begin{equation}\label{pppo}
\mathbf{U}=\varphi\left(\mathbf{E}/M\right).
\end{equation}
Here  ${\bf E}=(E_1,\dots,E_n)^\top$ is a vector of  i.i.d. random variables with standard exponential distribution and $M$ an independent positive  random variable. The function  $\varphi:   \R^+ \to [0,1]$ in (\ref{pppo}) is applied pointwise
 and it is the Laplace transform of the random variable $M$, i.e., a completely monotone function \citep{Miller2001}.
  Then,  $\mathbf{U}$ is a vector with uniform marginals $U_i$  \citep{MO1988, Mai2014}.
  The cdf associated to the random vector $\mathbf{U}$ has been widely studied
  and it belongs to a sub-class of the so-called Archimedean copulas.

  An  $n$-dimensional Archimedean copula is defined with  a generator function  $\varphi$ through:
\begin{equation}\label{pppo2}
F_{\mathbf{U}}(\uu)=\varphi\{\varphi^{-1}(u_1)+\dots+\varphi^{-1}(u_n)\}
\end{equation}
and it has been shown  by  \citet{McNeil:2009} that a necessary and sufficient condition
for $\varphi$ to  define a $n$-dimensional copula
  is that $\varphi$  is a $n$-monotone  function on $[0,\infty]$.
 For the precise definition of $n$-monotonicity we refer the interested reader to \cite{Malov2001,McNeil:2009,Williamson:1956}.
If $D_p$ is the set of all $p$-monotone functions then  it is known that
$D_2 \supsetneq D_3 \supsetneq \cdots  \supsetneq D_{\infty}$, where $D_{\infty}$ denotes the set of all completely monotone functions.

This implies that any  function $\varphi \in D_{\infty}$ can be used to define an Archimedean copula in arbitrary dimension $n\geq2$ and
it has been shown in  \cite{MO1988} that the random vector $\mathbf{U}$ in $(\ref{pppo})$ has cdf of type  $(\ref{pppo2})$
when $\varphi \in D_{\infty}$.
For instance, choosing  $M\sim \mathcal{G}(1/a,1)$ with Laplace transform given by  $\psi(x)=(1+x)^{-1/a}$, $a>0$
with inverse $\psi^{-1}(y)=y^{-a}-1$, then the Clayton copula
\begin{equation*}\label{cck}
F_{\mathbf{U}}(\uu)= \left( u_{1}^{-a}+\dots+u_{n}^{-a}-n+1\right)^{1/a},
\end{equation*}
is obtained as a special case.

Inspired by the stochastic representation  in (\ref{pppo})
we now propose a new random field with uniform marginals. Specifically, our idea is to
obtain a random field with uniform marginals
generalizing the stochastic representation in (\ref{pppo})  to the spatial setting. To achieve this task, we first  need to relax  the independence assumption of the exponential random variables.
This basically  prevents the generation of a random field with ``a forced'' nugget effect. In addition we need to replace the positive random variable $M$ with a positive random field  $\{M(\ss), \ss \in A\}$. This  allows to avoid identifiability problems (similar to the ones described in \cite{genton:Zhang:2012} and \cite{bevilacqua2021non}) when estimating with just one realization of the random field (the typical setting for spatial data). Note that when independent replicates are available,
our construction  in principle  could be simplified assuming  $M$ as a sequence of independent random variables of gamma type.

Following these arguments, we generalize the model in (\ref{pppo}) by defining  the Archimedean random field, i.e., a new class of random fields $\{U(\ss), \ss \in A\}$ with uniform marginals as
\begin{equation}\label{arch}
U(\ss)=\psi\left\{ E(\ss)/M(\ss) \right\},
\end{equation}
where $E$ is a random field with standard exponential marginal distribution, $M$ is a positive random field  and $\psi$ is the  Laplace transform of the random variable $M(\ss)$.

In this work we focus on the special case where  $ E \equiv G_{2}$ and $M$ is a copy of the Gamma random field  defined in Section \ref{sec:2}, that is $M \equiv G_{\nu}$ with marginal distribution $\mathcal{G}(\nu/2,1)$. Since the associated Laplace transform is given by $\psi(t)=(1+t)^{-\nu/2}$, the random field  in (\ref{arch}) assumes the form
\begin{equation*}\label{clayton}
U_{\nu}(\ss)=\left\{\frac{G_{\nu}(\ss)}{G_{\nu}(\ss)+G_{2}(\ss)}\right\}^{\nu/2}, \quad \nu\in\{1,2,\ldots\}.
\end{equation*}
It turns out that, under this specific choice, $U_{\nu}$ can be written as a power transformation of a special case of the auxiliary beta random field introduced in Section \ref{sec:2} that is
\begin{equation*}\label{clayton3}
U_{\nu}(\ss) =\{Y_{\nu,2}(\ss)\}^{\nu/2}, \quad \nu\in\{1,2,\ldots\},
\end{equation*}
where $Y_{\nu,2}$ has  marginal distribution $\mathcal{B}(\nu/2,1)$.
Since we choose the Laplace transform of a  Gamma marginal distribution in analogy with the classical Clayton copula,
we call  $U_{\nu}$ a Clayton  random field.

The multivariate pdf  of $U_{\nu}$ is not explicitly known.  However, the pdf  of the bivariate random vector  $\UU_{\nu;ij}=(U_{\nu}(\ss_i), U_{\nu}(\ss_j))^\top$ (bivariate copula density) can be easily obtained using Theorem \ref{theo1}, viz.
\begin{align}\label{pairbU}
f_{\UU_{\nu;ij}}(\uu_{ij})=
\{1-\rho^2(\hh)\}^{\nu/2+1}
F_4\left(\nu/2 +1,\nu/2+1;\nu/2,1;\rho^2(\hh)(u_iu_j)^{2/\nu},\rho^2(\hh)(1-u^{2/\nu}_i)(1-u^{2/\nu}_j)\right).
\end{align}
If the bivariate random vector has uniform marginals, reflection symmetry can be easily  checked
using  the associated bivariate density through $f_{\UU_{\nu;ij}}(u_i,u_j)=f_{\UU_{\nu;ij}}(1-u_i,1-u_j)$; see \citep{Joe:2014}.
Recalling that the $F_4$  function is symmetric, i.e., $F_4(\cdot,\cdot,\cdot,x,y)=F_4(\cdot,\cdot,\cdot,y,x)$,
it is apparent that  $f_{\UU_{\nu;ij}}(u_{j},u_{j})=f_{\bm{U}_{\nu;ij}}(1-u_{i},1-u_{j} )$  only when $\nu=2$, that is reflection asymmetry is obtained when  $\nu\neq 2$.

Note that the parameter $\nu$ only affects the bivariate distribution. That is, it can be viewed as dependence parameter. In addition, the bivariate cdf of $U_{\nu}$ (i.e.,  the associated bivariate copula function) is given in the following theorem.
\begin{theorem}\label{theo3}
Let $U_{\nu}$ the Clayton  random field in Equation \eqref{clayton3} with underlying correlation $\rho(\hh)$. Then the cdf of the bivariate random vector $\UU_{\nu;ij}=(U_{\nu}(\ss_i),U_{\nu}(\ss_j))^\top$
is given by
\begin{eqnarray}\label{copu}
F_{\UU_{\nu;ij}}(\bm{t}_{ij})&=&
4t_i t_j\{1-\rho^2(\hh)\}^{\nu/2+1}
\sum\limits_{k=0}^{\infty}\sum\limits_{m=0}^{\infty}\frac{\left(\nu/2\right)_k(t_it_j)^{2k/\nu}(\rho^{2}(\hh))^{k+m} }{k!(2k+\nu)^2 B^2\left(k+\nu/2,m+1\right)}
{}_2F_1\left(k+\nu/2,-m;k+\nu/2+1;t^{2/\nu}_i\right)\nonumber\\
&\quad&\times\quad
{}_2F_1\left(k+\nu/2,-m;k+\nu/2+1;t^{2/\nu}_j\right).
\end{eqnarray}
\end{theorem}

From the copula point of view, if the dependence parameter $\rho(\hh)\in[0,1)$ is equal to $0$, then it can be shown that independence is achieved irrespective of $\nu$, that  is $F_{\UU_{\nu;ij}}(\bm{t}_{ij})=F_{U_{i}}(t_i)F_{U_{j}}(t_j)=t_i t_j$.

We now compare the type of bivariate dependence induced by the Gaussian copula random field
$U^G(\ss)=\Phi\{Z(\ss)\}$
 and the chi-squared copula
random field    $U^{CS}(\ss)=|1-2\Phi\{Z(\ss)\}|$ \citep{QUESSY201640} with  the one induced by  the proposed  Clayton random field
$U_{\nu}(\ss) =\{Y_{\nu,2}(\ss)\}^{\nu/2}$.
 Although copula theory uses transformations to uniform marginals, it is better to consider transformation to Gaussian marginals for identifying the type of dependence (\cite{Joe:2014}, p. 9). In this case, the plot of  the bivariate  densities
associated with  the proposed model, the Gaussian copula and the chi-squared copula
can be obtained by computing
\begin{equation}\label{pairkUU}
f_{\SS_{ij}}(\ss_{ij})=
f_{A_{ij}}\{F_S(\ss_i),F_S(\ss_j)\}f_S(\ss_i)f_S(\ss_j),
\end{equation}
for $A_{ij}=\UU_{\nu;ij}, \UU^G_{ij}, \UU^C_{ij}$
setting  $F_S=\Phi$ and $f_S=\phi$.

\begin{figure}[h!]
\hspace{-0.8cm}
\centering{
\begin{tabular}{lll}
\includegraphics[width=5.0cm, height=5.3cm]{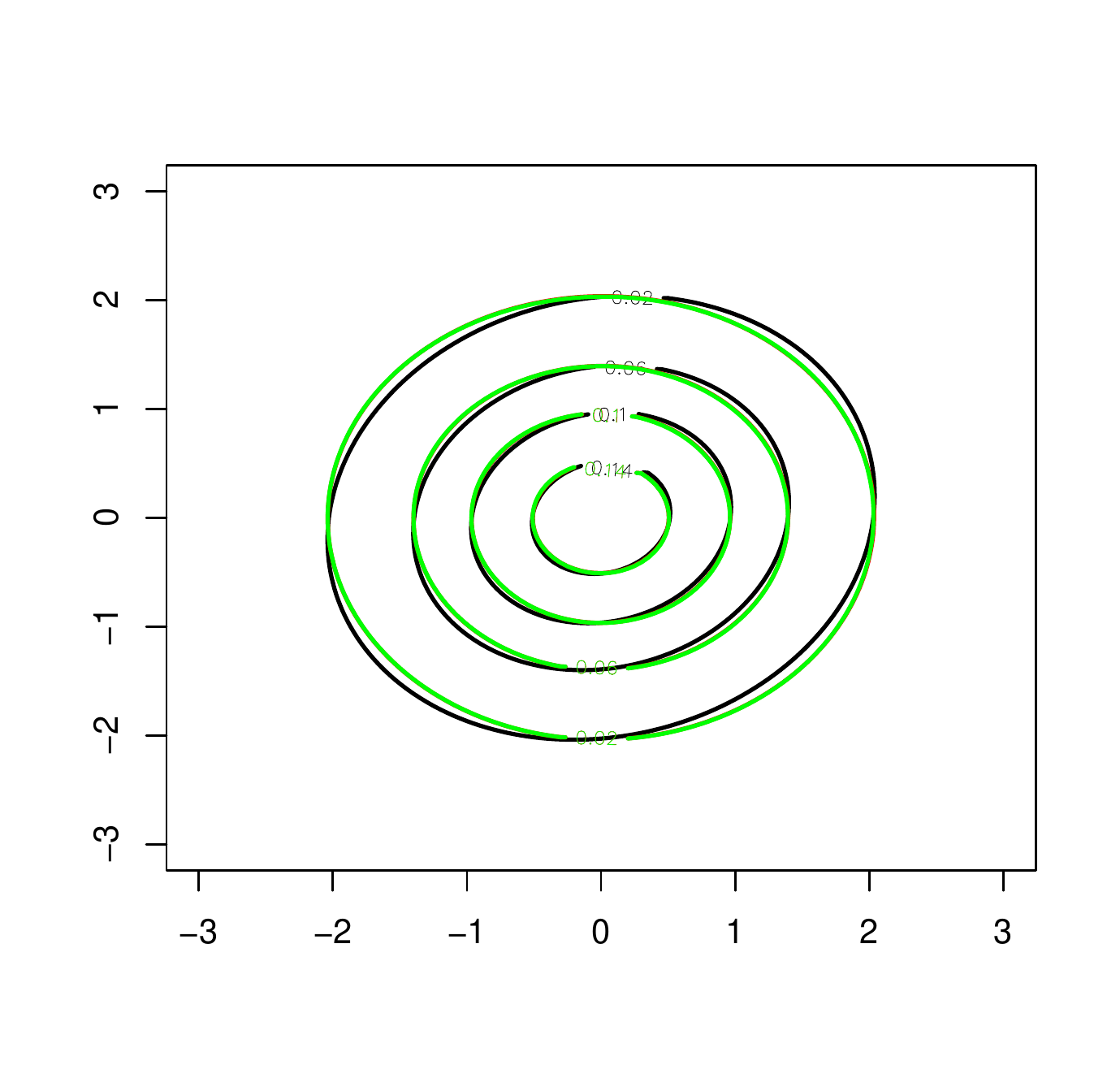} & \includegraphics[width=5.0cm, height=5.3cm]{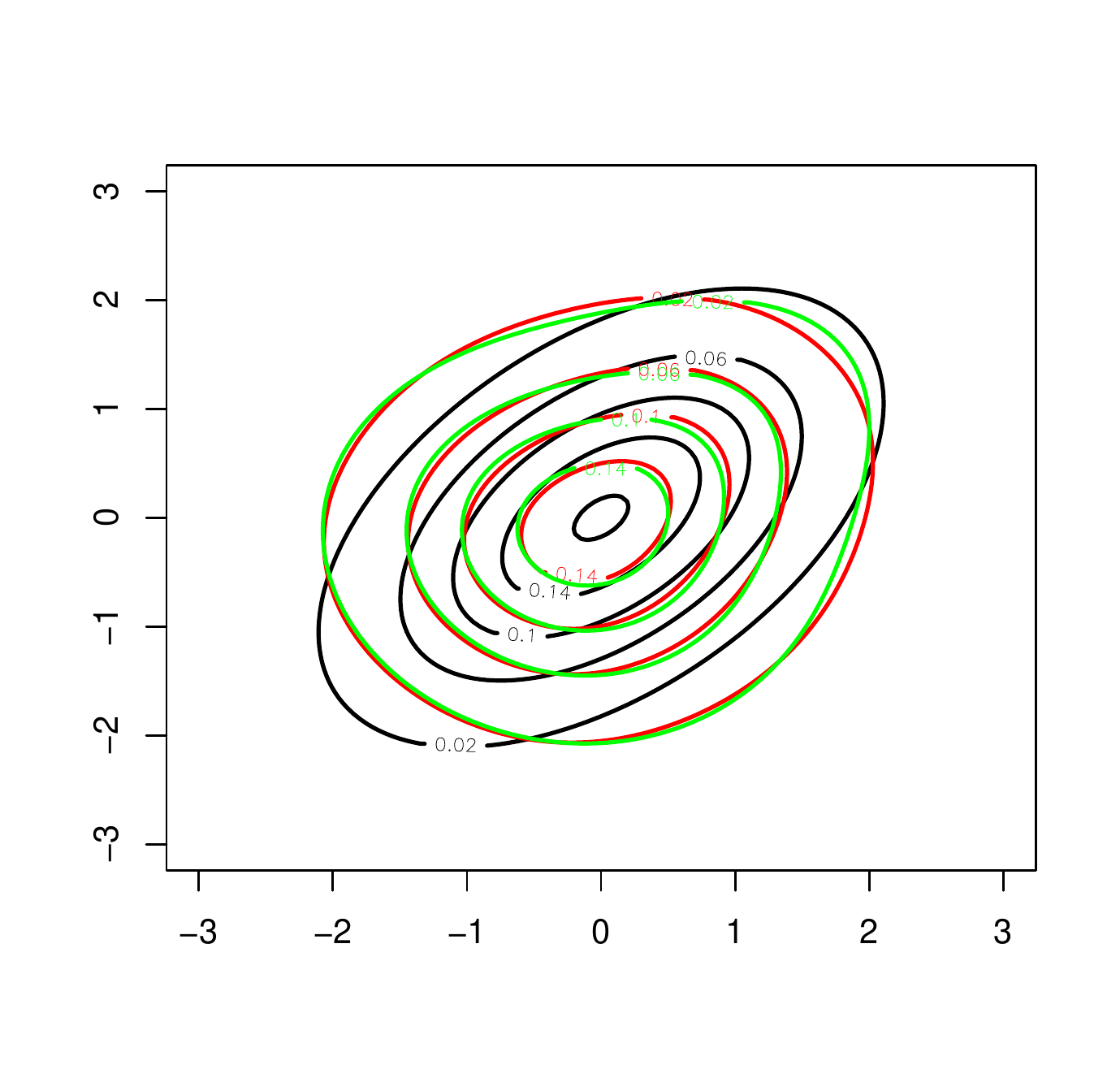} & \includegraphics[width=5.0cm, height=5.3cm]{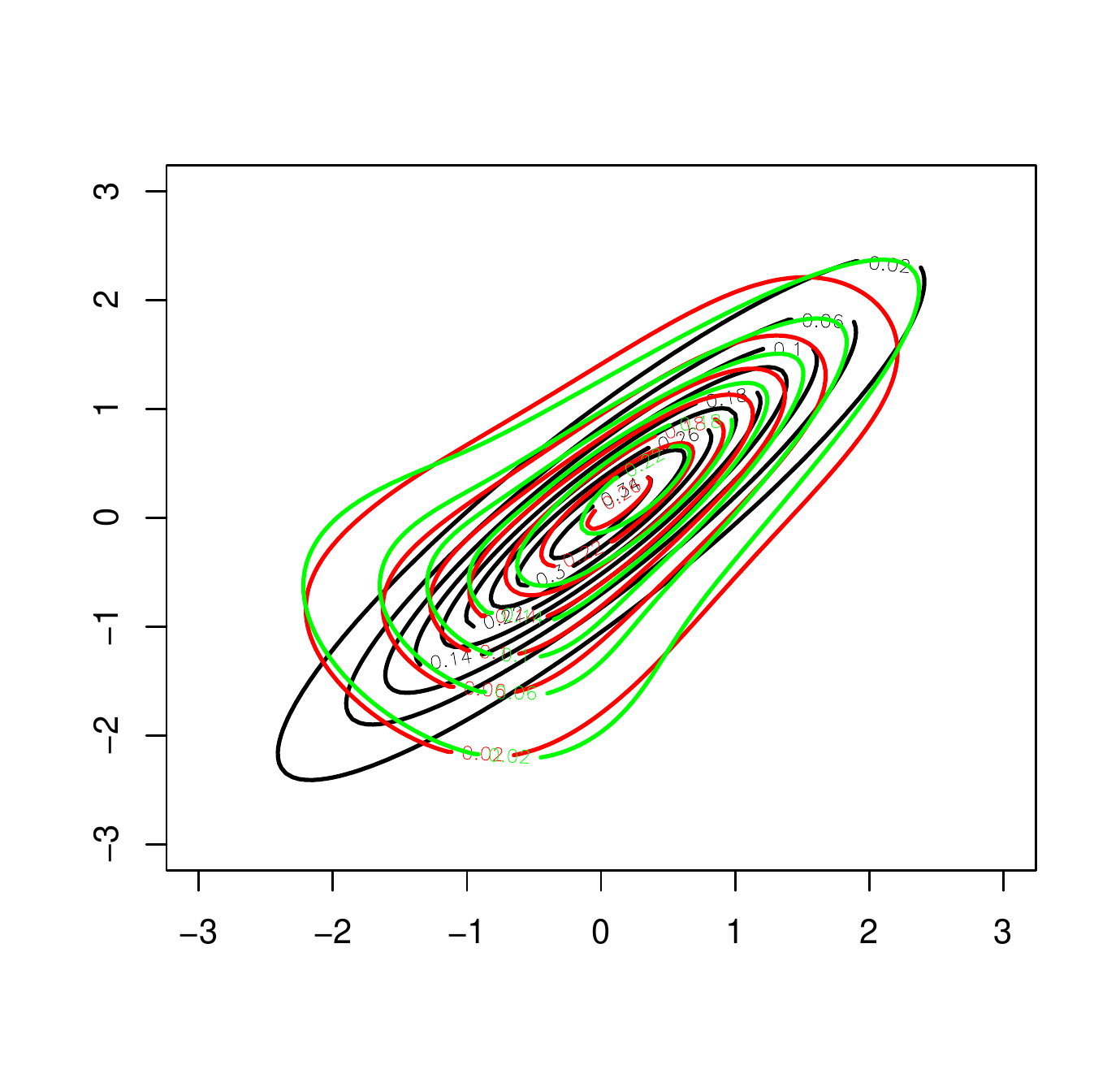}\\
\includegraphics[width=5.0cm, height=5.3cm]{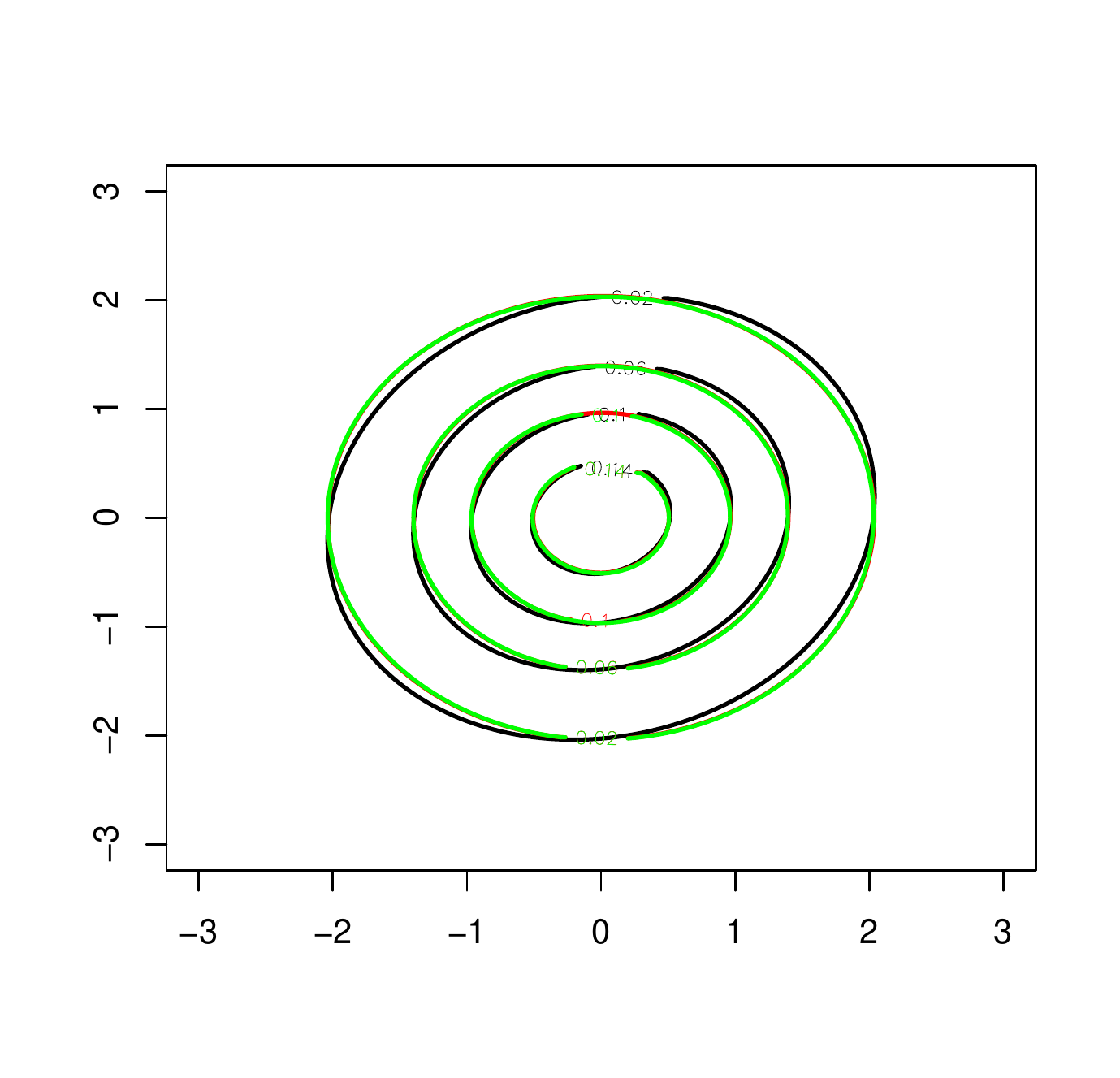} & \includegraphics[width=5.0cm, height=5.3cm]{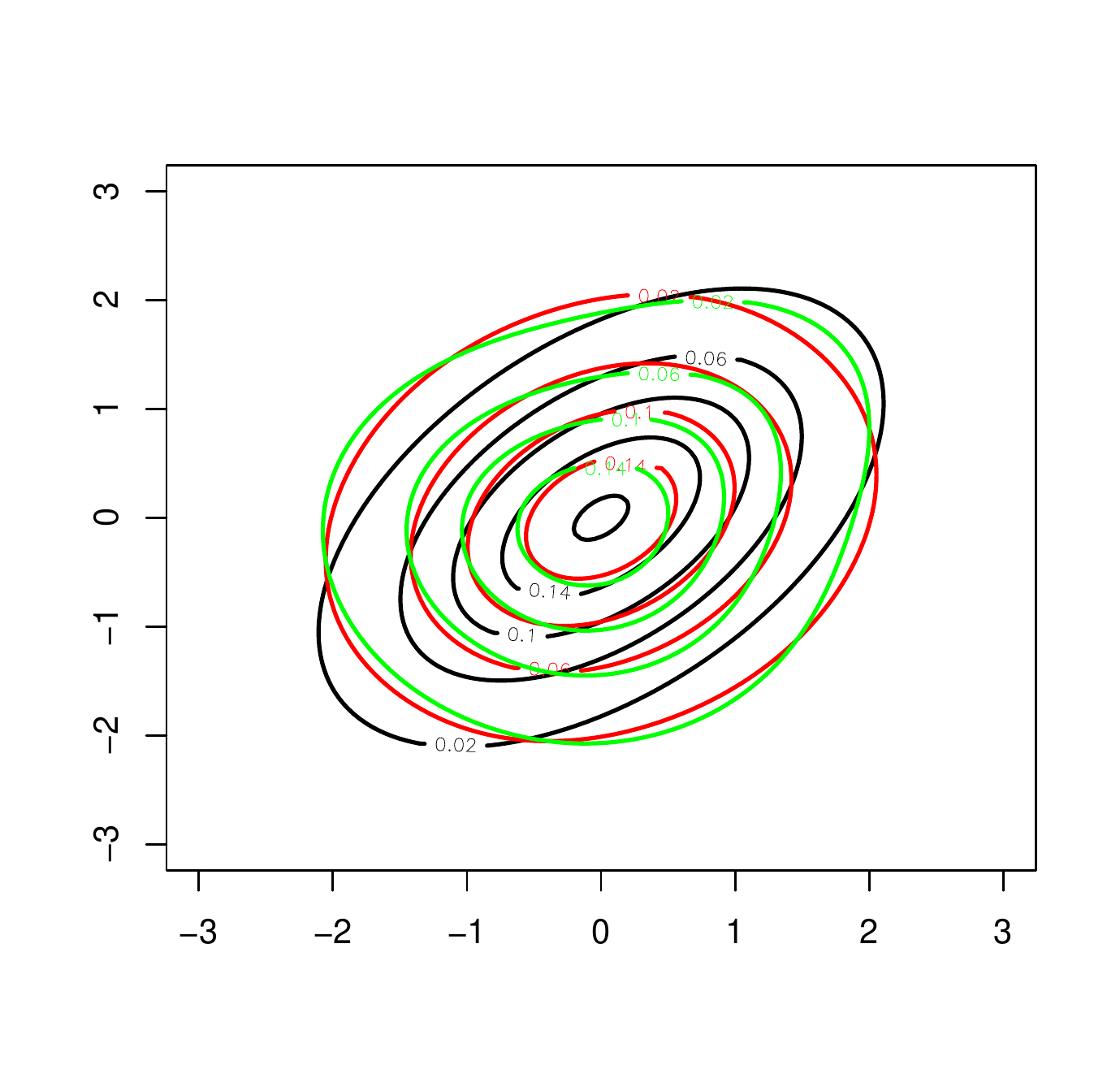} & \includegraphics[width=5.0cm, height=5.3cm]{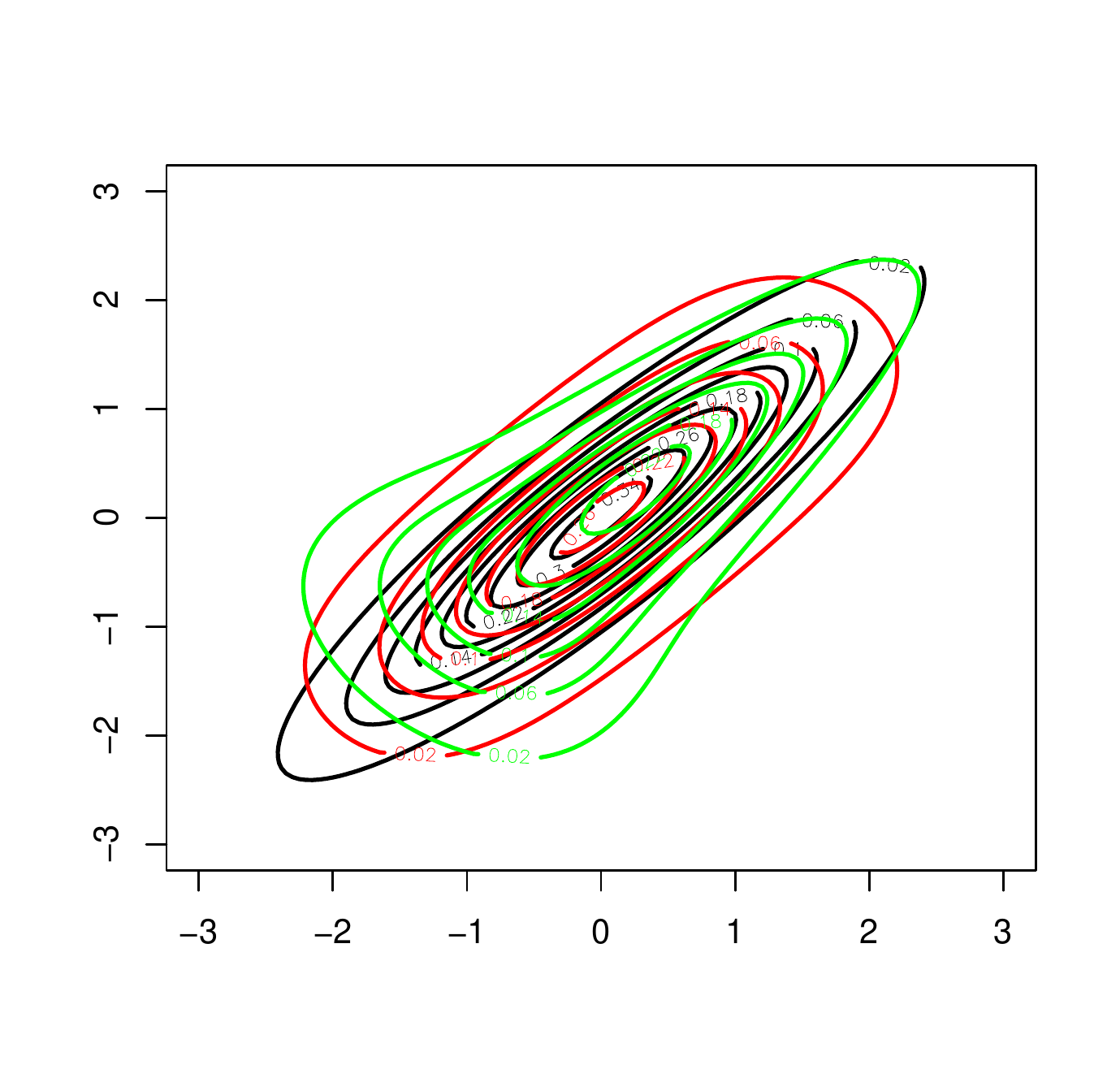}\\
\includegraphics[width=5.0cm, height=5.3cm]{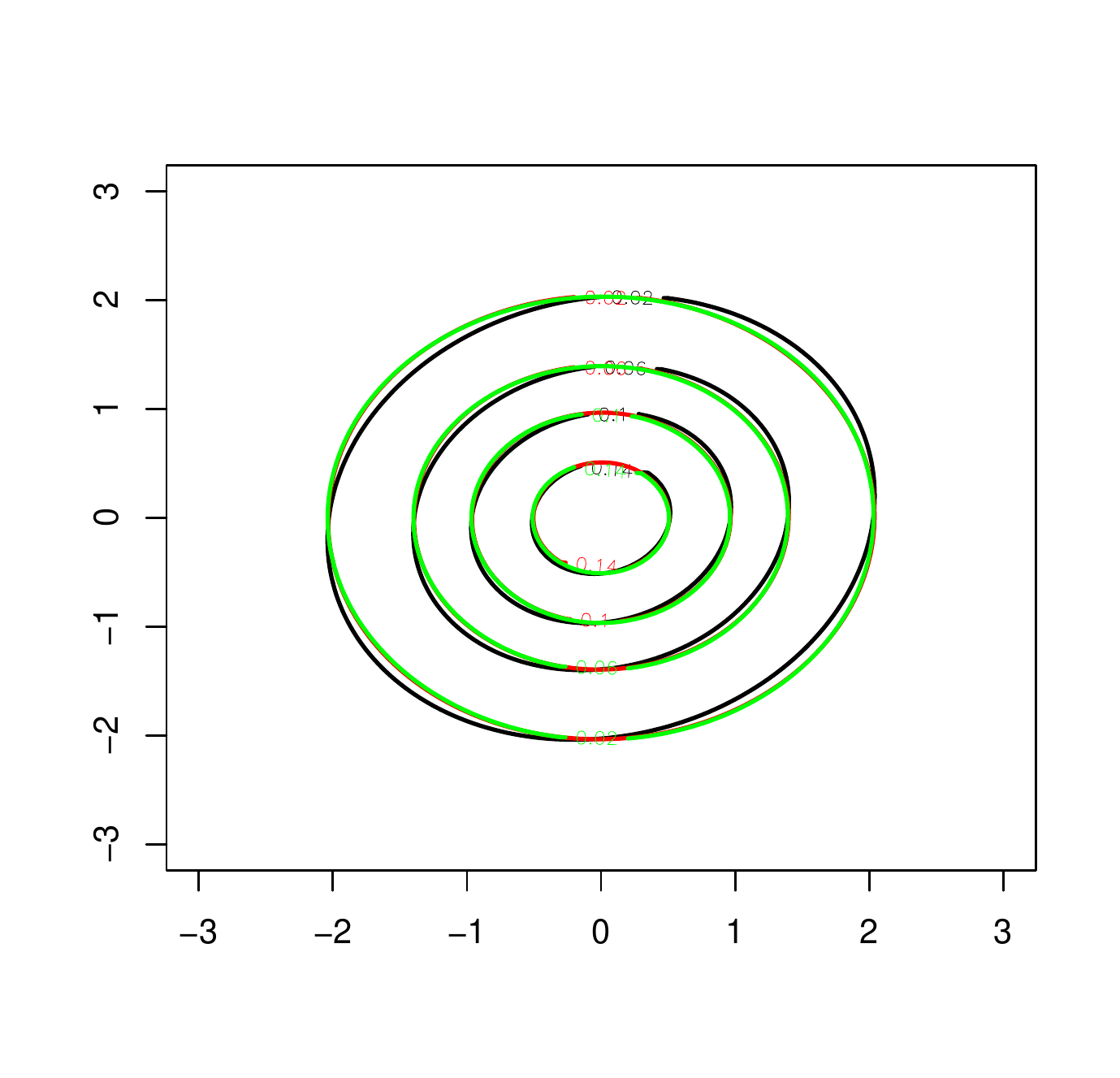} & \includegraphics[width=5.0cm, height=5.3cm]{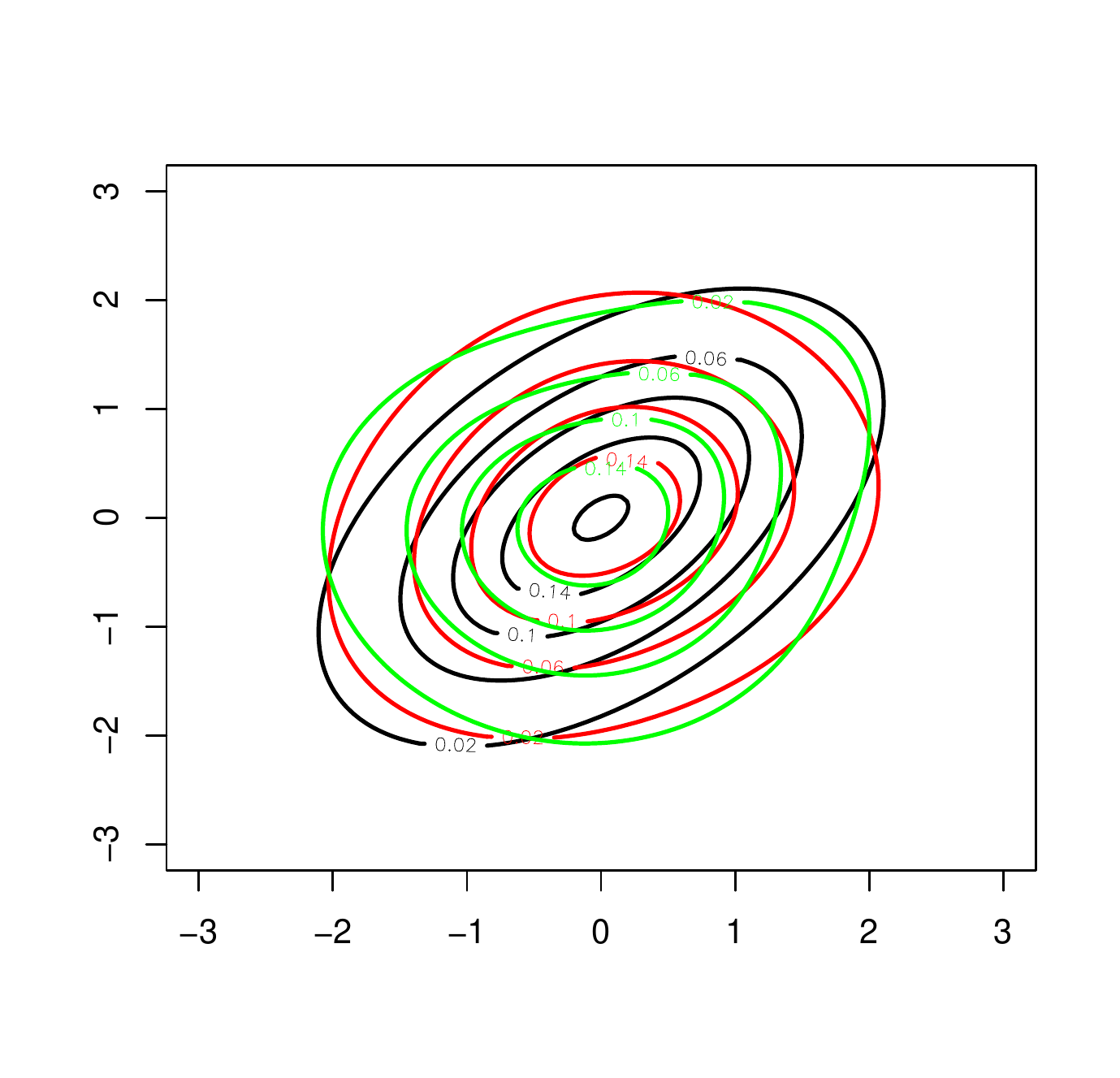} & \includegraphics[width=5.0cm, height=5.3cm]{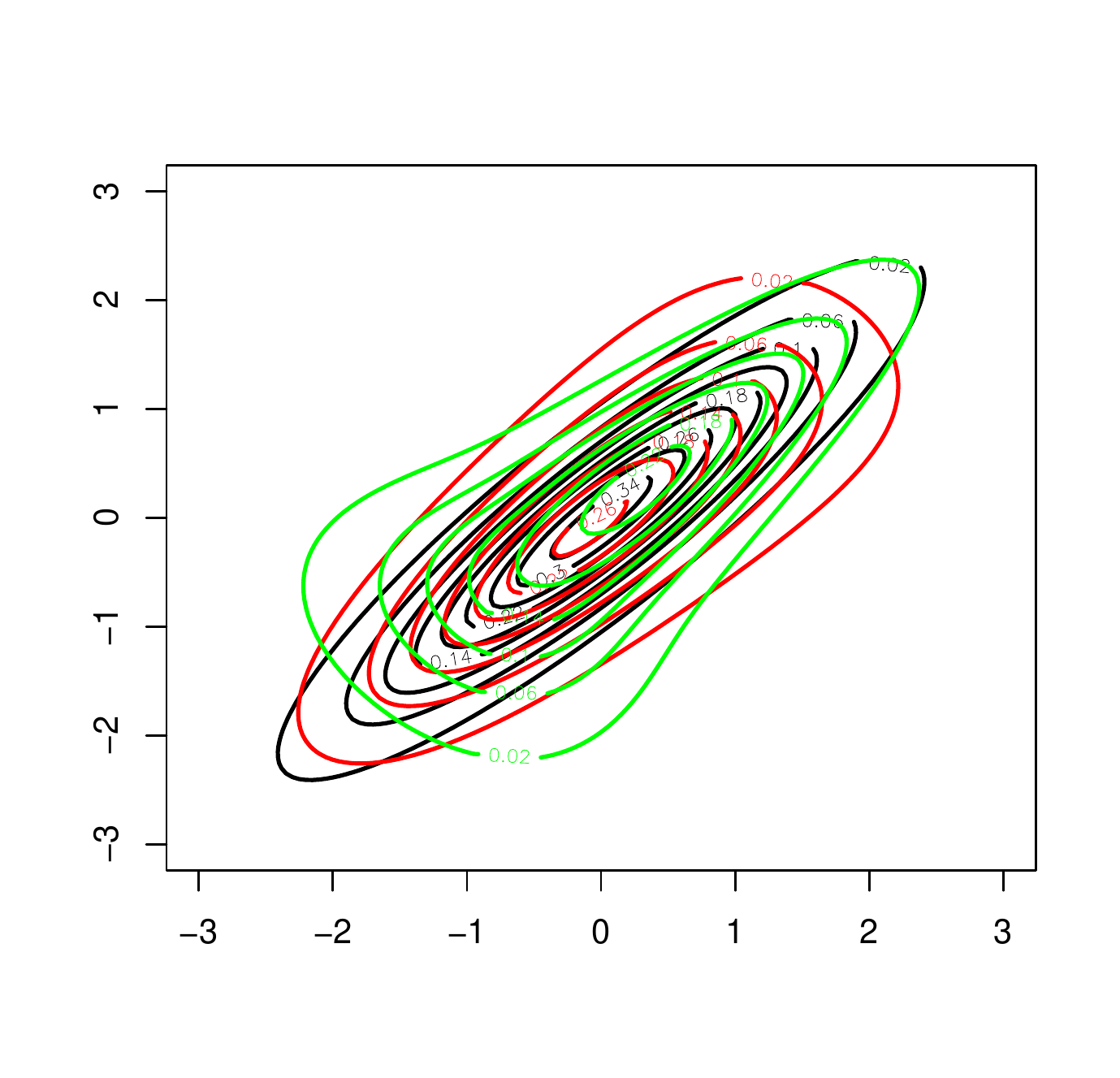}
\end{tabular}

\caption{Contour plots of the bivariate Gaussian density (black color), the bivariate chi-squared copula density (green color) and of the bivariate density (\ref{pairkUU}) with $F_S=\Phi$ and $f_S=\phi$ (red color)  with underlying correlation $\rho(\hh)\in\{0.1, 0.5, 0.9\}$ (from  left to right) when $\nu=1$ (first row), $\nu=2$ (second row), $\nu=5$ (third row).
}  \label{copufig}}
\end{figure}

 Fig. \ref{copufig} depicts the contour plots of the  three models increasing the underlying correlation $\rho(\hh)\in\{0.1, 0.5, 0.9\}$ (from left to right). The red, black  and  green colors
 are associated with  the  proposed model, the Gaussian copula and chi-squared copula respectively.
 The cases $\nu\in\{1, 2, 5\}$ (first, second, and third row, respectively) are also considered for the proposed model.

It turns out that when $\nu=2$, both the Gaussian and Clayton cases are reflection symmetric irrespective of the correlation. However, the Clayton  model shows a non-elliptical dependence. This can be clearly appreciated when the correlation is stronger while when the correlation is low the two distributions tend to be similar, as expected. In fact, when the correlation is zero we obtain independence in both cases.

If $\nu=1$ and $\nu=5$, then it can be appreciated that both the Clayton and chi-squared  models are not reflection symmetric, but our proposed methodology is more flexible because it is able to model
asymmetry to lower tail  ($\nu=1$) and asymmetry to upper tail ($\nu=5$), while the chi-squared copula is able  to model only
asymmetry to lower tail; see  \cite{Joe:2014}, Sec. 2.14. Hence, our model  is more flexible when modeling reflection (a)symmetries
and, as a consequence  can be applied  to a wider range of spatial data.
Note that  other  copula models  can capture both lower and upper tail asymmetry, such as mixture of Clayton and reflected Clayton copulas \citep{Joe:2014}.
However, when applying them in the spatial context, they are not identifiable with just one realization of the random field.

\subsection{Correlation function and geometric properties}

The following Theorem provides the correlation of the Clayton  random field $U_{\nu}$ in terms of the Kampé de Fériet function.

\begin{figure}[h!]
\centering{
\begin{tabular}{c}
\includegraphics[width=5.5cm, height=5.5cm]{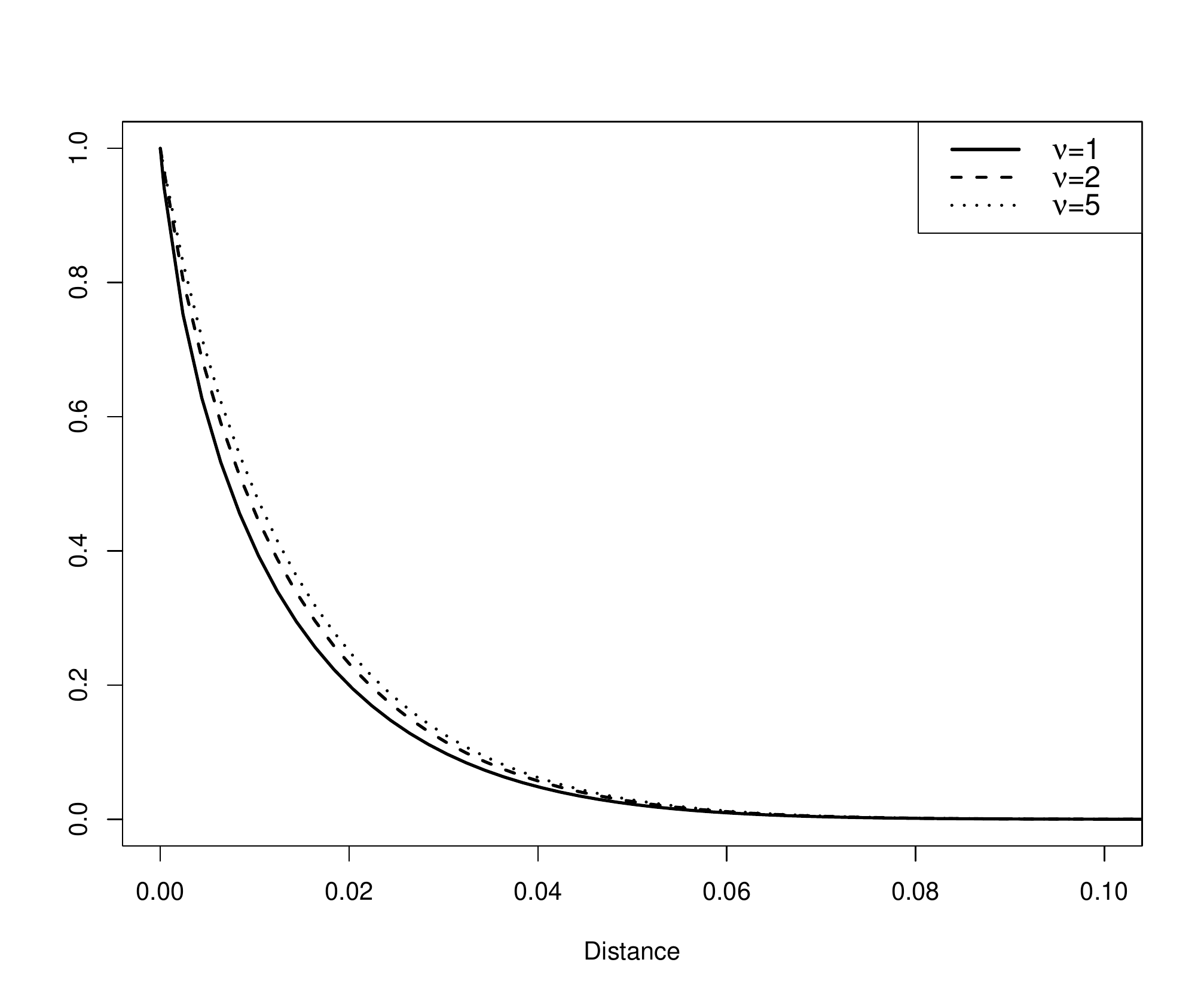}
\end{tabular}
\caption{The correlation function $\rho_{U_{\nu}}(\hh)$
in (\ref{corru})
 for $\nu\in\{1, 2, 5\}$ when the  underlying correlation model is
$ \rho(\hh)={\cal GW}_{0,4,0.15}(\hh).$}
\label{ccc_f3}}
\end{figure}

\begin{theorem}\label{theo4}
Let $U_{\nu}$ the Clayton   random field  in \eqref{clayton3} with underlying  correlation $\rho(\hh)$. Then
\begin{equation}\label{corru}
\rho_{U_{\nu}}(\hh)=3\left( \{1-\rho^2(\hh)\}^{\nu/2+1}F_{2;1;0}^{2;2;1}\left[ \begin{array}{c}
                                 \nu/2+1;\nu;1\\
                                  \nu+1;\frac{\nu}{2};-
                                \end{array}\middle\vert \rho^{2}(\hh),\rho^{2}(\hh)\right] - 1\right).
\end{equation}
\end{theorem}
Fig. \ref{ccc_f3} depicts the correlation function
$\rho_{U_{\nu}}(\hh)$ varying the asymmetry parameter $\nu\in\{1, 2, 5\}$ when the underlying correlation is the Generalized Wendland model
\citep{bb2019}
defined as
\begin{equation*} \label{WG4*}
	{\cal GW}_{\delta,\mu, b}(\hh)=
	\begin{cases}
	K \left\{ 1- \left(||\hh||/b \right)^2 \right\}^{\delta+\mu}
	    {}_2F_1\left(\mu/2,(\mu+1)/2;\delta+\mu+1;1- \left(||\hh||/b \right)^2 \right),
	& 0\leq ||\hh|| \leq b,\\
	0, & ||\hh|| > b,\end{cases}
	\end{equation*}
with $K=\Gamma(\delta)\Gamma(2\delta+\mu+1)/\{\Gamma(2\delta)\Gamma(\nu+\mu+1)2^{\mu+1}\}$, $\delta\geq0$, $\mu\geq 0.5(d+1)+\delta$
and $b>0$ is the compact support.
	This kind  of  correlation model is very flexible since  it allows to parametrize the mean squared differentiability of the  underlying spatial field and it includes the Matérn correlation as a special case model after a suitable reparametrization \citep{Bevilacqua2022}.
Specifically, we set  $ \rho(\hh)={\cal GW}_{0,4,0.15}(\hh)=\left(1- ||\hh||/0.15\right)^4_+$ and we plot $\rho_{U_{\nu}}(\hh)$  versus the distance $||\hh||$
varying the parameter $\nu\in\{1, 2, 5\}$. It can be seen that  $\rho_{U_{\nu}}(\hh)$ slightly  increases when increasing $\nu$ for each $||\hh||$.

We now focus on the special case $\nu=2$ that is the reflection symmetric case. The following corollary provide  a closed form expression for the correlation of the Clayton  random field $U_{2}$.

\begin{corollary}\label{theo5}
Let $U_{2}$ the symmetric Clayton random field with underlying correlation $\rho(\hh)$. Then
\begin{equation}\label{corru_nu2}
\rho_{U_2}(\hh)=\frac{2[\rho^2(\hh)\{3\rho^2(\hh)-1\}-\{\rho^2(\hh)-1\}^2 \ln\{1-\rho^2(\hh)\}]}{\rho^4(\hh)}-3.
\end{equation}
\end{corollary}

The simple form of the correlation function $\rho_{U_2}(\hh)$ allows to study the geometrical properties of the Clayton random field. It turns out that nice properties such as stationarity, mean-square continuity, degrees of mean-square differentiability, long-range dependence and compact support can be inherited from the underlying Gaussian random field $Z$ with correlation $\rho(\hh)$. The following theorem resumes all these properties.
\begin{theorem}\label{theo6}
Let $U_{2}$ the symmetric Clayton  random field    with underlying Gaussian random field $Z$ with correlation $\rho(\hh)$. Then,
\begin{enumerate}
\item[(i)] $U_2$ is weakly stationary;
\item[(ii)] $U_2$ is mean-square continuous if $Z$ is mean-square continuous;
\item[(iii)] $U_2$ is $0$-times mean-square differentiable if  $Z$  is $0$-times mean-square differentiable;
     for $m\geq 1$, if $\rho^{(1)}(\bm{0})=0$,  $U_{\nu}$ is $m$-times mean-square differentiable if  $Z$  is $m$-times mean-square differentiable;
\item[(iv)]  $U_2$ is  a long-range dependent process  if  $Z$ is a  long-range dependent process;
\item[(v)] $\rho(\hh)=0 \implies \rho_{U}(\hh)=0$.
\end{enumerate}
\end{theorem}

\begin{figure}[h!]
\hspace{-0.8cm}
\centering{
\begin{tabular}{ccc}
$\nu=1$&$\nu=2$&$\nu=5$\\
\includegraphics[width=5cm, height=3.5cm]{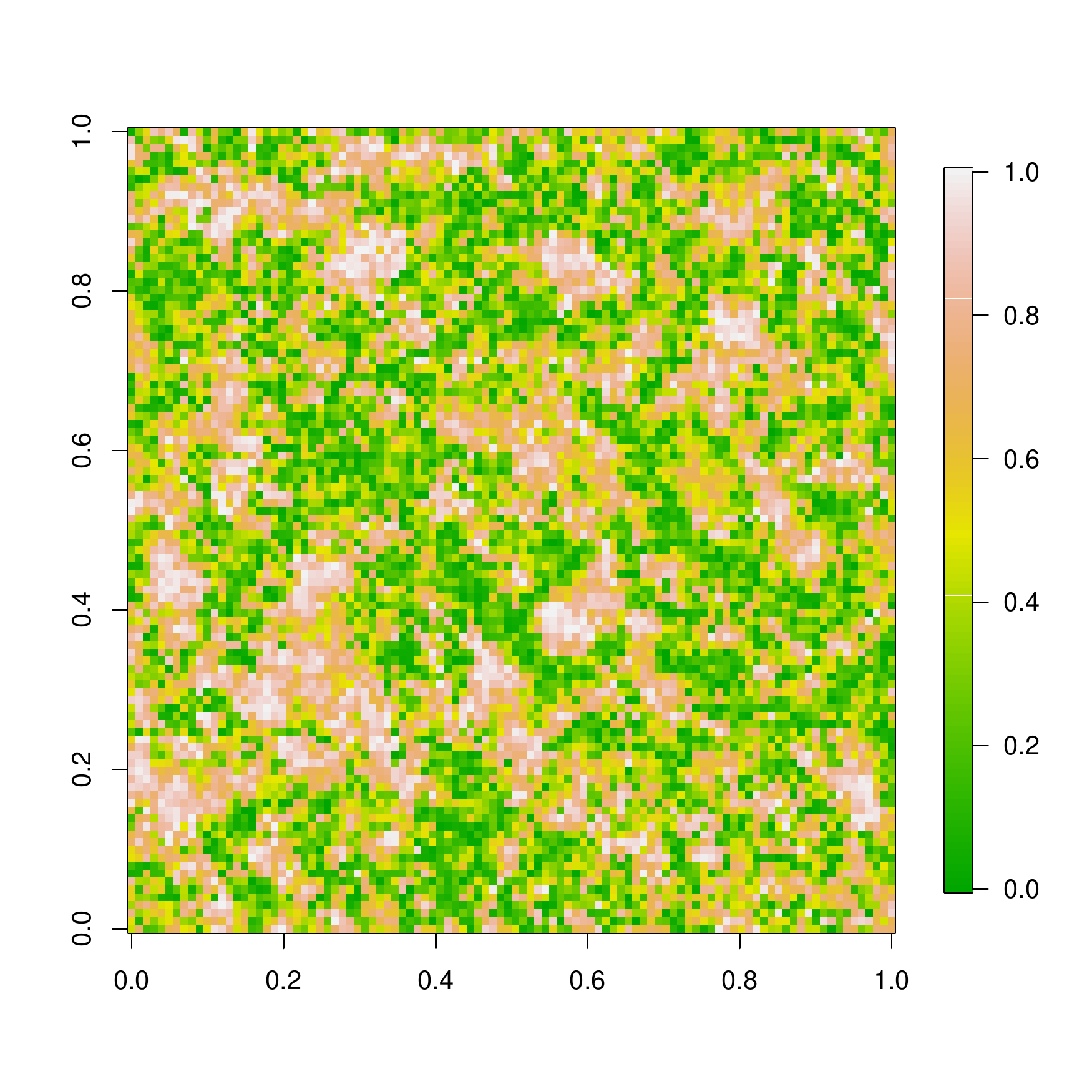} & \includegraphics[width=5cm, height=3.5cm]{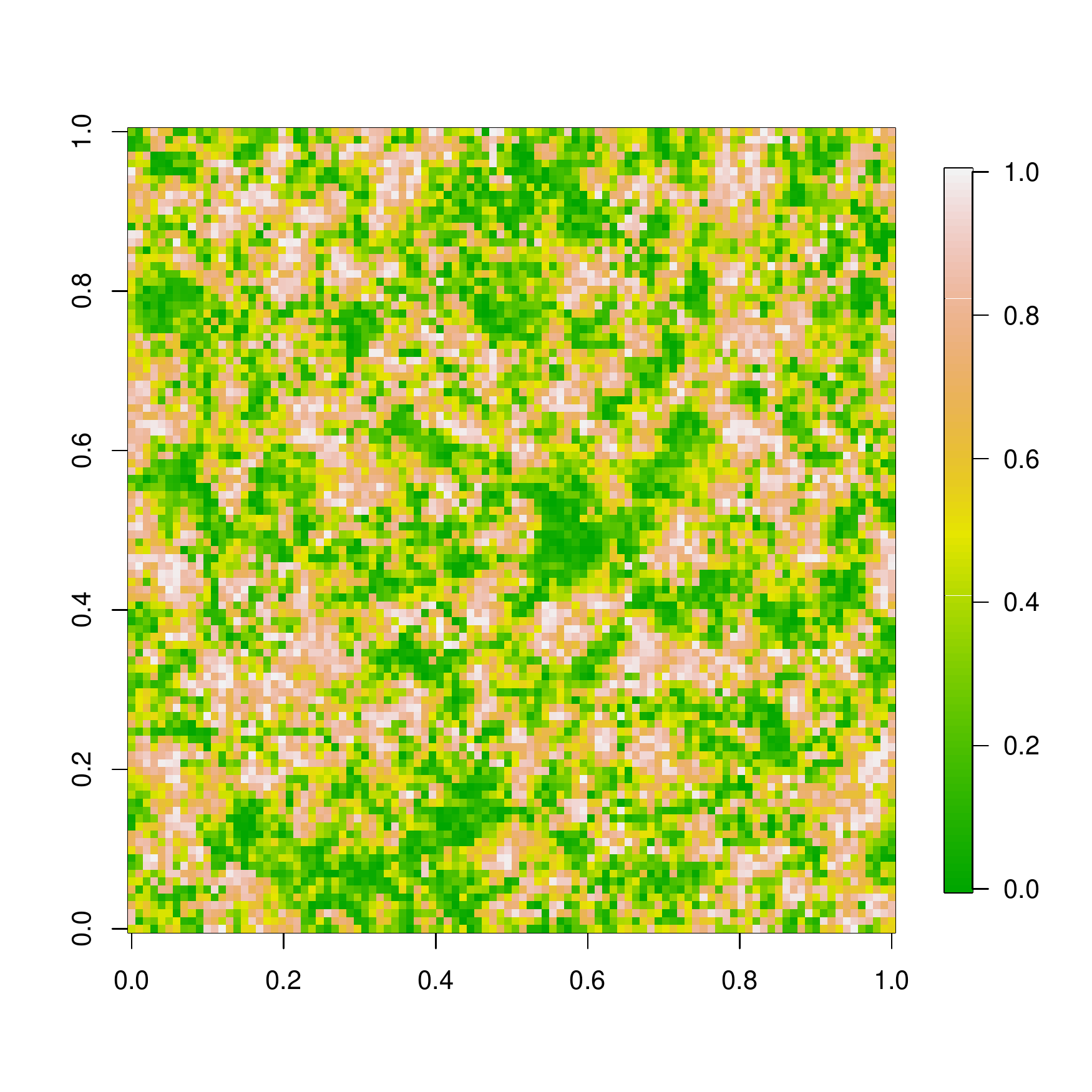}& \includegraphics[width=5cm, height=3.5cm]{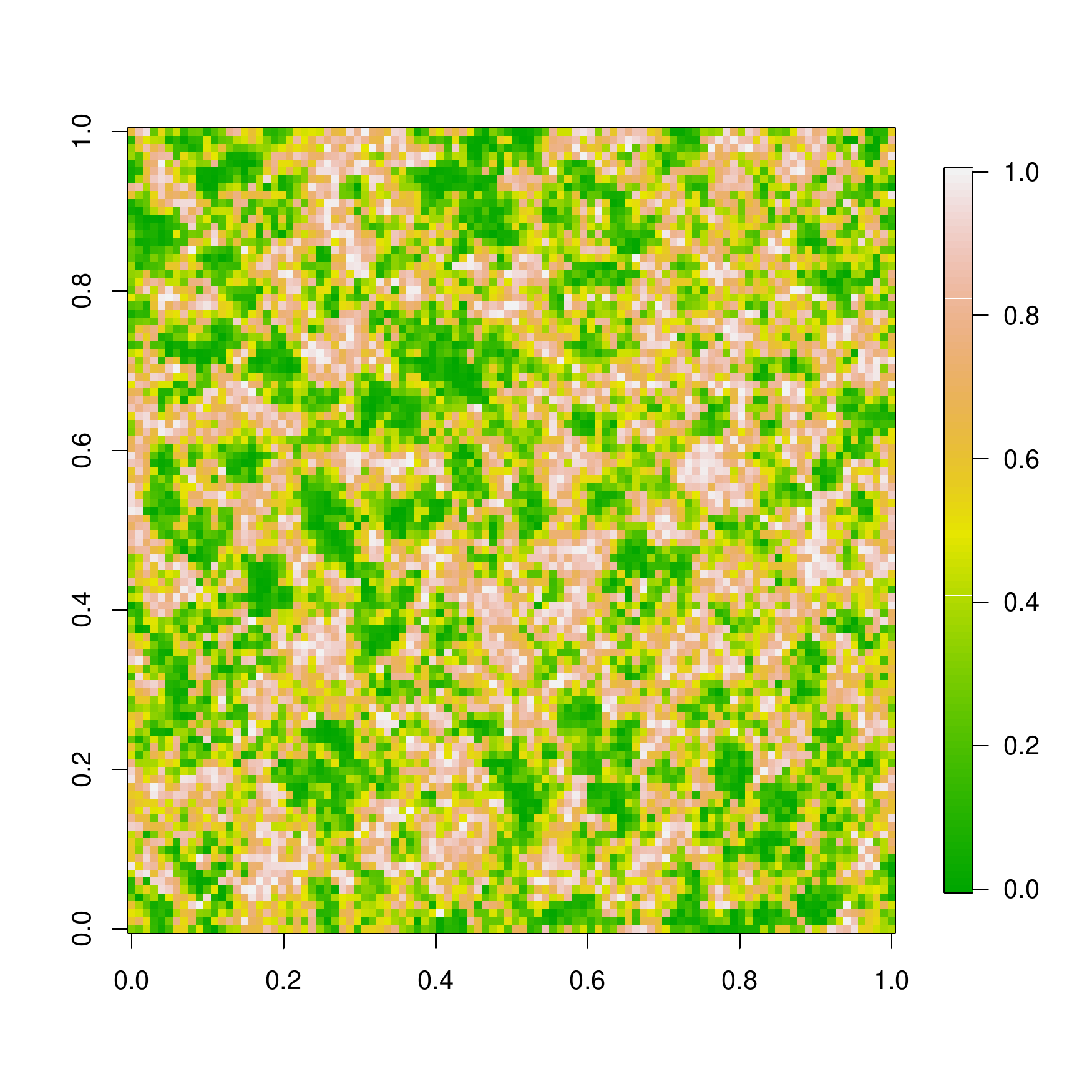}  \\
\includegraphics[width=5cm, height=3.5cm]{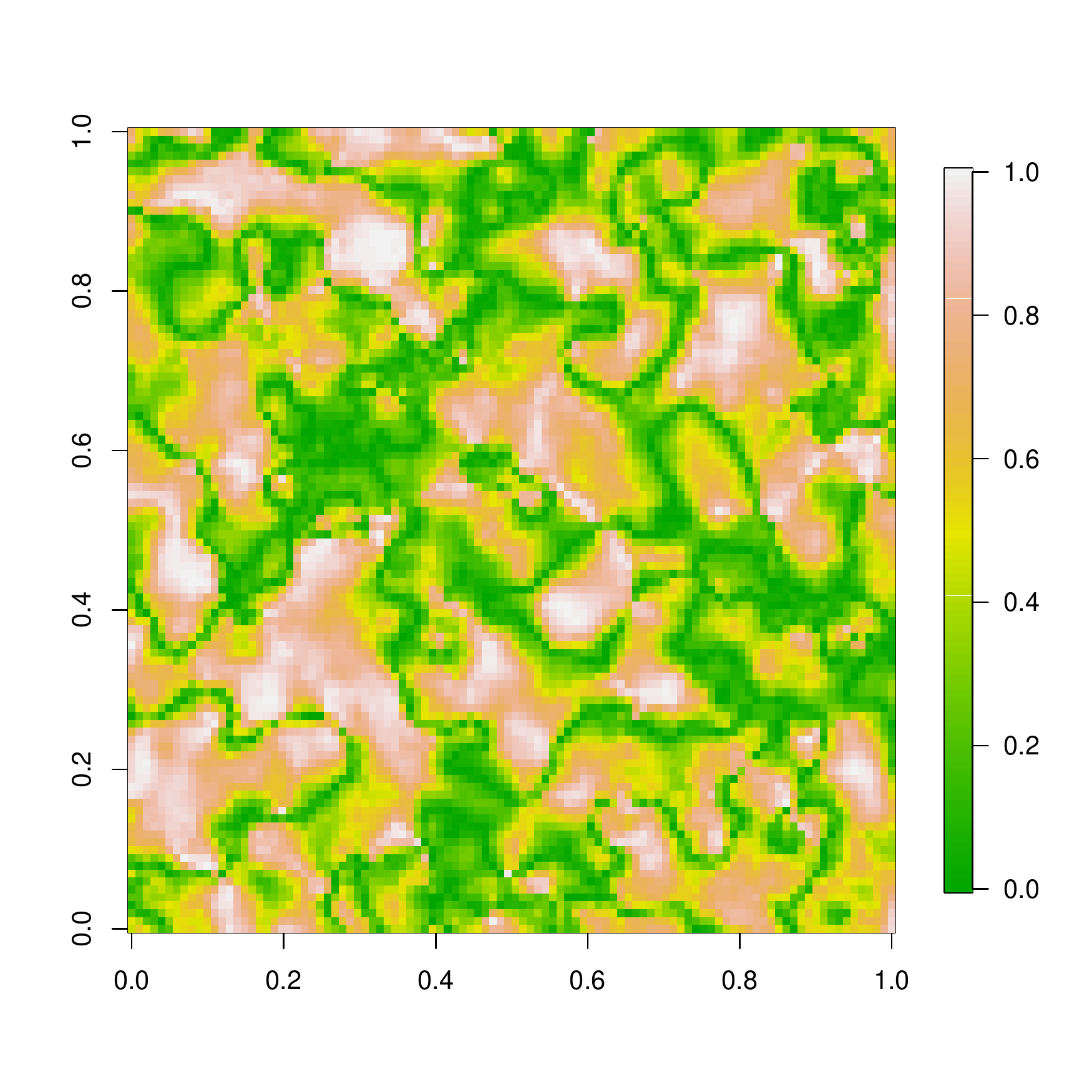} & \includegraphics[width=5cm, height=3.5cm]{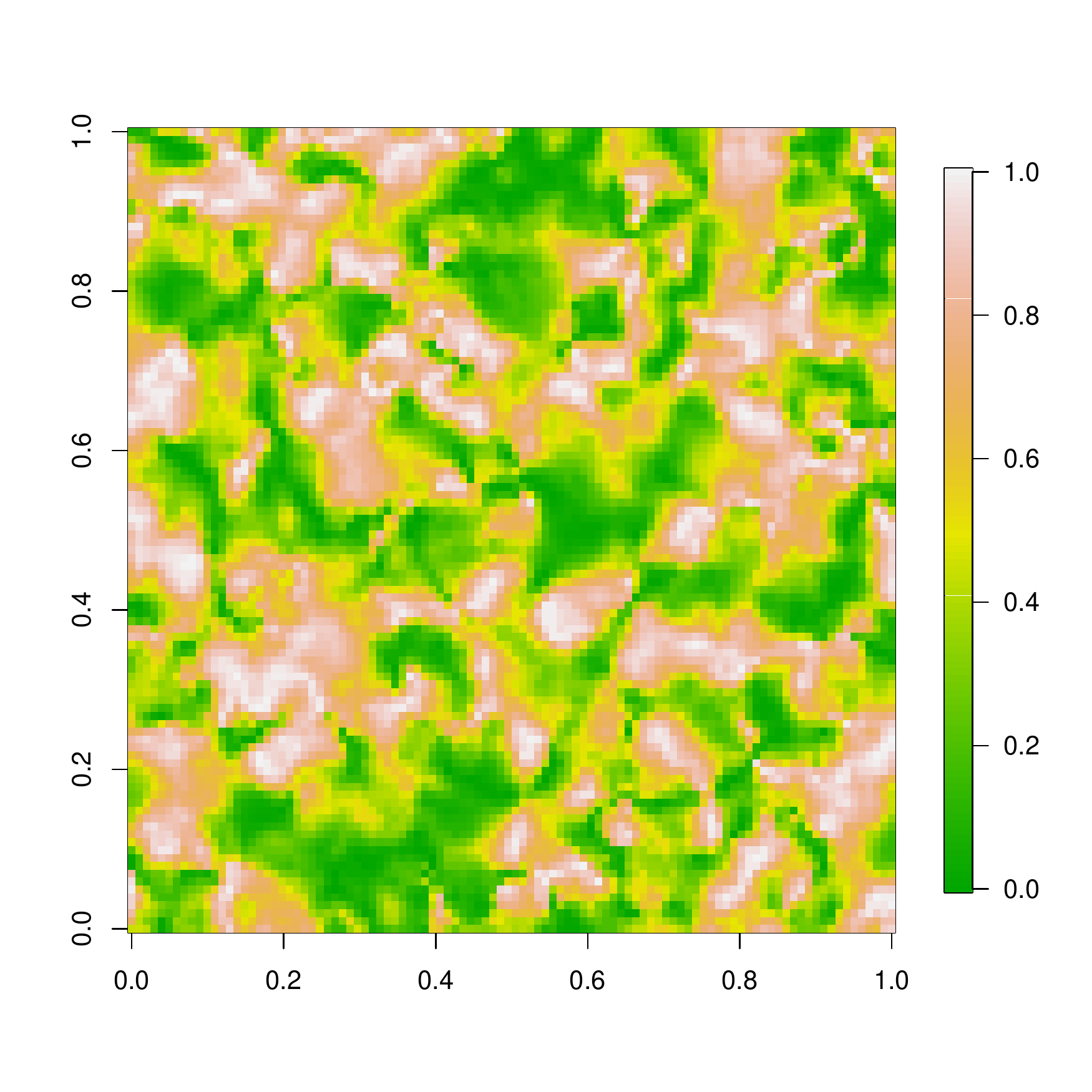} & \includegraphics[width=5cm, height=3.5cm]{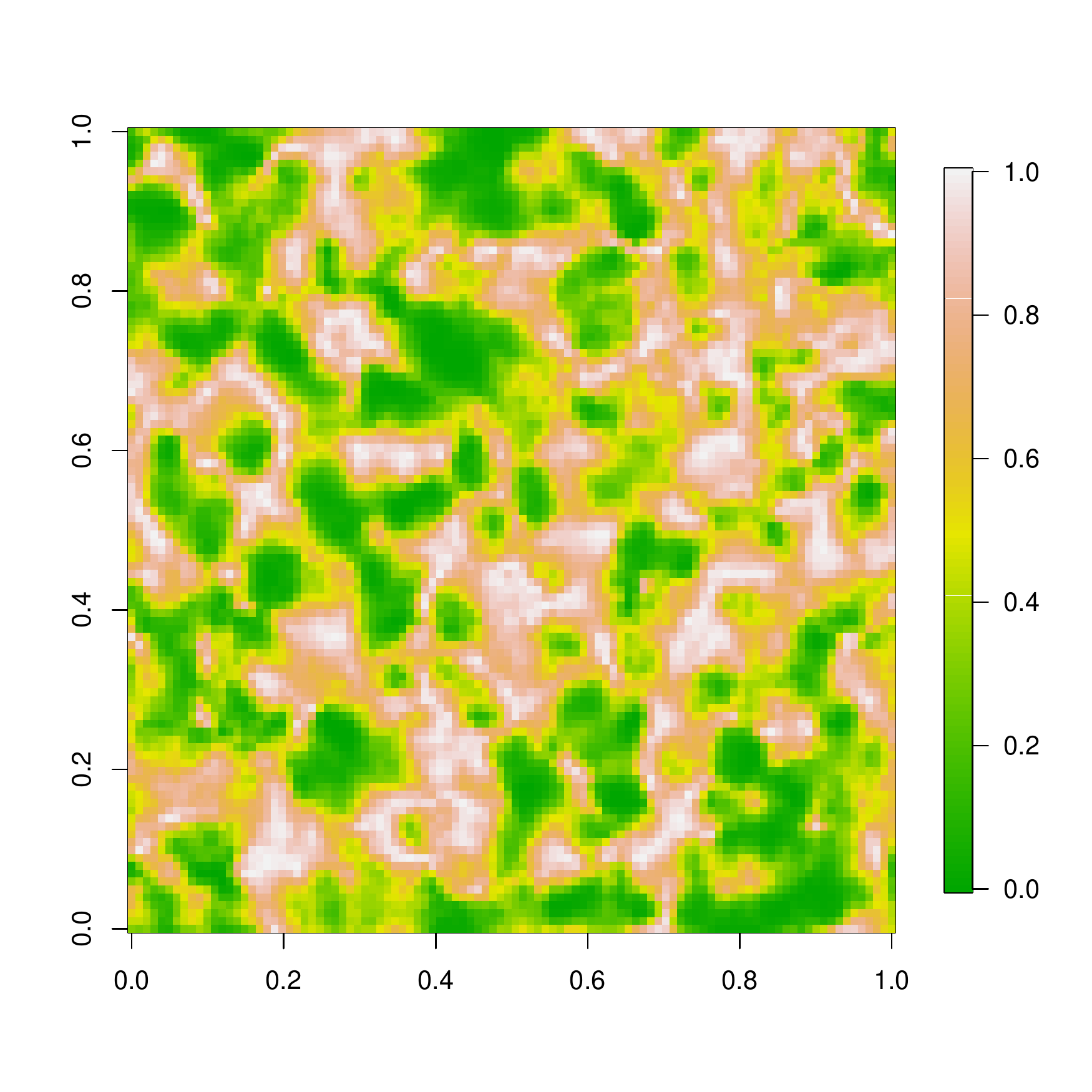}  \\
\end{tabular}
\caption{
Six realizations of the Clayton random field $U_\nu$. First row: using an underlying  zero-time differentiable Generalized Wendland  correlation model ${\cal GW}_{0,4,0.2}$ and  reflection asymmetry parameter equal to $\nu\in\{1, 2, 5\}$ (from left to right). Second row: using an underlying  one-time differentiable Generalized Wendland  ${\cal GW}_{1,4,0.2}$  correlation model   and  asymmetry parameter equal to
 $\nu\in\{1, 2, 5\}$ (from left to right).} \label{fig:f2}}
\end{figure}
The consequence of Theorem \ref{theo6} is that, as in the Gaussian case, the geometrical properties, the long range dependence and the compact support  of the Clayton random field can be modeled using a suitable correlation function. For instance long-range dependence can be achieved using correlations with specific features such as the Generalized Cauchy model \citep{GneitingS:2004,LiTe09,berg2008}.
In addition, it can be shown that  for the Generalized Wendland model ${\cal GW}_{\delta,\mu, \beta}$ the condition  $\rho^{(1)}(\bm{0})=0$ is satisfied when $\delta\geq1$; that is, $U_2$ inherits the mean square differentiability of $Z$ if the underlying correlation model is
Generalized Wendland.

Even though Theorem \ref{theo6} is restricted to the case $\nu=2$ we conjecture that the results are still valid for  $U_{\nu}$, $\nu \neq 2$. Some empirical evidence can be viewed looking at the sample paths of $U_{\nu}$. Fig. \ref{fig:f2}, first row, depicts three realizations of $U_{\nu}$ with $\nu\in\{1, 2, 5\}$ (from left to right) when the underlying correlation is $\varphi_{0,4,0.2}$ (zero time mean square differentiable). The second row depicts three realizations of $U_{\nu}$ with $\nu\in\{1, 2, 5\}$ (from left to right) when the underlying correlation is $\varphi_{1,4,0.2}$ (one time mean square differentiable).

It is important to stress that a not mean-square continuous   Clayton random field
 can be obtained by introducing a nugget effect, i.e.,
a discontinuity at the origin of $\rho(\hh)$.
 This can be achieved by replacing the underlying
correlation function $\rho(\hh)$ with $\rho^*(\hh) = \rho(\hh)(1-\tau^2)+\tau^2 \mathbb{1}_0(\hh)$, where $0 \leq \tau^2 \leq1$
 represents
the  nugget effect.

Note that, in addition to the correlation function $\rho_{U_{\nu}}(\hh)$, other measures of dependence can be considered.
For instance it can be shown that  the Spearman correlation coefficient is equal to the correlation function in this case. Similarly, using (\ref{copu}) it is possible to obtain the Kendall rank correlation coefficient and the Blomqvist medial correlation coefficient defined as \citep{Genest1986,nelsen2007introduction}. They are respectively given by
\begin{equation*}\label{kendall}
\tau_{\UU_{\nu}}(\hh)=4\left\{\int_{[0,1]^2}F_{\UU_{\nu}}(\mathbf{t}_{ij})dF_{\UU_{\nu}}(\mathbf{t}_{ij})-\dfrac{1}{4}\right\}
\end{equation*}
and
\begin{equation*}\label{blomqvist}
\beta_{\UU_{\nu}}(\hh)=4\left\{F_{\UU_{\nu}}\left(\dfrac{1}{2},\dfrac{1}{2}\right)-\dfrac{1}{4}\right\}.
\end{equation*}
Finally, two common measures
 of extremal dependence  of a bivariate distribution
 are the lower and upper tail  dependence coefficients \citep{Sibuya:1960,Joe:2014} defined respectively as
$$\lambda_L=\lim\limits_{t \to 0^+} \frac{F_{\UU_{ij}}(t,t)}{t}\;\;\hbox{and}\;\;\lambda_U=\lim\limits_{t \to 1} \frac{1-2t+F_{\UU_{ij}}(t,t)}{1-t},$$

\noindent for a given bivariate copula $F_{\UU_{ij}}(t_{i},t_j)$.
Finding closed form expressions  for
$\lambda_L$ and $\lambda_U$ is challenging from a mathematical point of view when considering the proposed copula  $F_{\UU_{\nu;ij}}$,
since  it involves  a double infinite sum  of a  squared hypergeometric function. Based on some numerical evidence, we conjecture that the proposed copula is tail-independent, that is, $\lambda_L=\lambda_U=0$.

\section{Application to  beta spatial regression model}\label{sec:4}

As an application of the proposed Clayton random field, we focus on a random field with marginal distributions of the (reparametrized) beta type that allows to perform spatial mean  regression for spatial  data defined on a bounded support.
A  random field $S$ with an  arbitrary cdf $F_S$ using the Clayton  random field, can be obtained as
\begin{equation*}\label{copula2}
S(\ss)=F_S^{-1}\{U_{\nu}(\ss)\},
\end{equation*}
with  bivariate distribution given by
\begin{equation}\label{bb44}
f_{\SS_{ij}}(\ss_{ij})=
f_{\UU_{\nu;ij}}\{F_S(\ss_i),F_S(\ss_j)\}f_S(\ss_i)f_S(\ss_j)
\end{equation}
and  correlation function given by
\begin{equation}\label{corruII}
\rho_{S_{\nu}}(\hh)=\frac{\int\limits_{0}^1\int\limits_{0}^1 F_{S}^{-1}(u_i)F_{S}^{-1}(u_j)f_{\bm{U}_{\nu;ij}}(\uu_{ij})d\uu_{ij}   -\EE\left\{S(\ss_i)\right\}\EE\left\{S(\ss_i)\right\}}{\sqrt{\VV\left\{S(\ss_i)\right\}\VV\left\{S(\ss_j)\right\}}}.
\end{equation}
Note that, if we assume continuous random variable then
the generalized inverse distribution function $F_{S}^{-1}$ is Lipschiz of order 1 \citep{qq2008} and, using results in \cite{BANERJEE200385}, this implies that mean square continuity and degrees of mean square differentiability of $S$ are inherited from the uniform random field, i.e., using Theorem \ref{theo6}, from the underlying Gaussian random field at least for the case $\nu=2$.

As a special case, we focus on  a random field $\{B_{\xi,\delta}(\ss), \ss \in A\}$ with beta marginal distribution. It can be   defined using the Clayton random field, as follows:
\begin{equation}\label{betaa}
B_{\xi,\delta}(\ss)=F^{-1}_{B_{\xi,\delta}}\{U_{\nu}(\ss)\}.
\end{equation}
It is important to stress that the parameters  of the beta random field (\ref{betaa}), unlike the beta random field defined in Section \ref{sec:2}, are  not restricted to be positive integer values.

By construction $B_{\xi,\delta}(\ss)\sim \mathcal{B}(\xi,\delta)$  and the marginal density and cdf are given respectively by
\begin{equation*}\label{ppl4}
f_{B_{\xi,\delta}}(y)=\frac{y^{\xi-1} (1-y)^{\delta-1}}{B(\xi,\delta)},\qquad F_{B_{\xi,\delta}}(y)=I(y,\xi,\delta), \qquad0<y<1,
\end{equation*}
where $\xi>0, \delta>0$ are two shape parameters and  $I(y,\xi,\delta)$ is the regularized incomplete beta function defined as
\begin{equation*}
I(y,\xi,\delta)=\int_{0}^{y}t^{\xi-1}(1-t)^{\delta-1}dt=\frac{x^{\xi}}{\xi}{}_2F_1(\xi,1-\delta;\xi+1;y).
\end{equation*}
The mean and variance are given respectively by $\EE\left\{B_{\xi,\delta}(\ss)\right\}=\xi/(\xi+\delta)$ and $\VV\left\{B_{\xi,\delta}(\ss)\right\}=2\xi \delta/\{(\xi+\delta)^2(\xi+\delta+2)\}$.

Here we consider the parametrization proposed in \citep{ferrari2004beta} that is the random field $\{B_{\mu(\ss)\delta, (1-\mu(\ss))\delta}, \ss \in A\}$  with $0<\mu(\ss)<1$. Using this parametrization, it can be shown that the (spatially varying) mean of $B_{\mu(\ss)\delta, (1-\mu(\ss))\delta}$ is given by $\mu(\ss)$ that can be specified through a regression model with a logistic link, i.e.,
\begin{equation*}\label{popo}
E\{B_{\mu(\ss)\delta, (1-\mu(\ss))\delta}\}=\mu(\ss)=1/(1+e^{-X(\ss)^\top \bm{\beta}}),
\end{equation*}
with $\bm{\beta}\subseteq \R^k$ a vector of regression parameters.
Finally the  bivariate distribution and the correlation function can be obtained using (\ref{bb44}) and (\ref{corruII}), respectively.

\subsection{Nearest neighbors weighted pairwise composite likelihood estimation} \label{sec:CL}

Let us assume a random field with arbitrary marginals $S=\{S(\ss),\ss\in A\}$ defined on $A \in \RR^d$, with $\EE\left\{S(\ss)\right\}=\mu(\ss)$ and $\VV\{S(\ss)\}=\sigma^2$, with a parametric model for the underlying correlation that is $\rho(\hh)=\rho(\hh,\bvartheta)$, where $\bvartheta$ is the vector of correlation parameters. Then the vector of dependence parameters is given by $\btheta=(\nu,\bvartheta)^\top$, where $\nu$ is the reflection symmetry parameter.

For any set of  distinct points $(\ss_1,\ldots,\ss_n)^\top$,  $\ss_i \in A$, $n\in \NN$, we denote by $\SS_{ij}=(S(\ss_i),S(\ss_j))^\top$, $i \neq j$ and  with    $f_{\SS_{ij}}$  we denote  the associated probability density functions.

Following \cite{Lindsay:1988}, the log-composite likelihood is an objective function defined as a sum of $K$ sub-log-likelihoods, viz.
\begin{equation*}\label{eq:CL}
CL(\btheta)=\sum_ {k=1}^K \ell(\btheta; B_k)w_k,
\end{equation*}
where $\btheta$ is the vector of unknown parameters, $B_k$ is a marginal or conditional set of $\SS$ and $\ell(\btheta; B_k)$ is a log-likelihood calculated by considering only the random variables in $B_k$ and $w_k$ are suitable  weights that do not depend on $\btheta$. The maximum CL estimate is given by $\hat\btheta=\textrm{argmax}_{\btheta}\, CL(\btheta)$.

The weighted pairwise composite likelihood  estimation method \citep{Bevilacqua:Gaetan:2015} is obtained by setting $B_k=\SS_{ij}$   and in this  case  we obtain the pairwise log-likelihood function $\ell_{ij}(\btheta)=\ln\{f_{\SS_{ij}}(\btheta)\}$.
The corresponding weighted pairwise composite log-likelihoods  function is given by
\begin{equation*}\label{eq:PP}
wpl(\btheta)=\sum_{i=1}^n \sum_{j \neq i}^n \ell_{ij}(\btheta) w_{ij}
\end{equation*}
and $\hat\btheta=\textrm{argmax}_{\btheta}\, wpl(\btheta)$ is the associated estimator.

 In general, a loss of statistical efficiency is expected for both cases with respect to the maximum likelihood (ML) estimation and the role of the weights $w_{ij}$ is to minimize this loss.
Using theory of optimal estimating equations \citep{Heyde1997}, it can be easily seen \citep{Bevilacqua2012} that  the optimal weights require the computation of the inverse of an  $n(n-1)\times  n(n-1)$ matrix which is computationally even harder than what is required for ML estimation.
Some approximations of the optimal weights have been proposed in literature as, for instance, in \citep{Li2018, Pace2019}. However, the computation of these sorts of weights can be computationally demanding  for large $n$.

To avoid this computational problem, in this work, we adopt  a weight function based on nearest neighbors as proposed in \cite{Caaman_et_al:2022}. Specifically, let $N_m (\ss_l)$ the set of the neighbors of order $m\in\{1,\ldots,n\}$ of the point $\ss_l \in A$.  We make use of the following asymmetric  weight function for $i,j\in\{1,\ldots,n\}$ and $i\neq j$
\begin{equation}\label{neigh}
w_{ij}(m)=\begin{cases}
1, &\ss_i\in N_m(\ss_j), \\
0, & \text{otherwise}.\end{cases}
\end{equation}

Following the same arguments as in \cite{Bevilacqua:Gaetan:2015}, it can be shown that,
under increasing domain asymptotic framework,
 $\hat\btheta=\textrm{argmax}_{\btheta}\, wpl(\btheta)$  is consistent and asymptotically normal with asymptotic covariance matrix given by the inverse of the Godambe information matrix defined as
\begin{equation*}
G(\btheta)=H(\btheta)J(\btheta)^{-1}H(\btheta)^\intercal,
\end{equation*}
where
\begin{equation*}
H(\btheta)=-\EE\left\{\nabla^{2}wpl(\btheta)\right\},\qquad
J(\btheta)=\EE\left\{\nabla wpl(\btheta)\nabla wpl(\btheta)\,^\intercal \right\}.
\end{equation*}

Standard error estimation can be obtained by considering the square root diagonal  elements of $G^{-1}(\widehat{\bm{\theta}})$.
Moreover, model selection can be performed by considering  the information criterion, defined as
\begin{equation*}\label{plic}
\mbox{PLIC}= - 2wpl(\hat{\bm{\theta}})  + 2\mathrm{tr}(
H(\hat{\bm{\theta}})G^{-1}(\hat{\bm{\theta}})),
\end{equation*}
which is the composite likelihood version of
the Akaike information
criterion (AIC) 
\citep{Varin:Vidoni:2005}.
Note that the computation of standard errors and  \mbox{PLIC} 
 require  evaluation of the matrices $H(\hat{\bm{\theta}})$ and $J(\hat{\bm{\theta}})$. However, the
evaluation  of $J(\hat{\bm{\theta}})$ is  computationally
unfeasible for large datasets and  in this case subsampling
techniques can be used  to estimate $J(\bm{\theta})$ as in \citep{Bevilacqua2012, Heagerty:Lele:1998}.
A straightforward  and  more robust alternative that we adopt in  Section \ref{sec:5} is  the parametric bootstrap estimation of $G^{-1}_a(\bm{\theta})$.

\subsection{A numerical example}\label{spatial_settings}

We simulate, using Cholesky decomposition, $1000$ realizations of $B_{\mu(\ss)\delta, (1-\mu(\ss))\delta}$
observed at $N=1000$ spatial location sites uniformly distributed in the unit square.
In particular we consider a reparametrized beta random field
$B_{\mu(\ss)\delta, (1-\mu(\ss))\delta}$ obtained using the Clayton random field $U_{\nu}$
setting $\nu\in\{1, 2, 6\}$.
As  underlying correlation model we consider
\begin{equation}\label{askey}
{\cal GW}_{0,4, b}(\hh)=	\left(1- ||\hh||/b\right)^4_+.
\end{equation}
The use of a compactly supported correlation function is a good strategy from a computational viewpoint because in this case, when using the pairwise composite likelihood method of estimation,
the bivariate distribution factorizes as the product of two marginal distributions when the underlying correlation is zero.
We consider two mean regression parameters, i.e., we assume a spatially varying mean as
$$\mu(\ss )=\dfrac{1}{1+\exp \left[ -\{\beta_0+\beta_1 u_1( \ss)\} \right]},$$
\noindent with $\beta_0=0.2$ and $\beta_1=-0.2$, where $u_1( \ss )$ is  a $\mathcal{U}(0,1)$ random variable. Finally, we set the  shape parameters as $\delta\in\{0.5,1.5,2.5\}$ and the compact support  parameter $b=0.2$.
The estimation is performed using the weighted pairwise composite likelihood method using the weight function  (\ref{neigh}) setting $m=2$.

It is important to stress that,  in principle, the parameter $\nu$ can be estimated because
the bivariate distributions in  (\ref{pairbU}) and  (\ref{bb44}) are  defined for each $\nu>0$. However, recall that the existence of the random field
$U_{\nu}$ is guaranteed only for $\nu\in\{1,2,\ldots\}$. As a consequence, in this numerical example and in the real data applications
the reflection asymmetry parameter $\nu$ is assumed fixed. An alternative strategy is a two-step estimation. In the first step,  all the parameters including
the asymmetry parameter $\nu$ are estimated. In the second step
$\nu$ is fixed equal to the rounded value of the estimate obtained at first
step and then the  marginal and  correlation dependence parameters are estimated.

\begin{figure}[h!]
\centering{
\begin{tabular}{ccc}
\includegraphics[width=10.2cm, height=10.2cm]{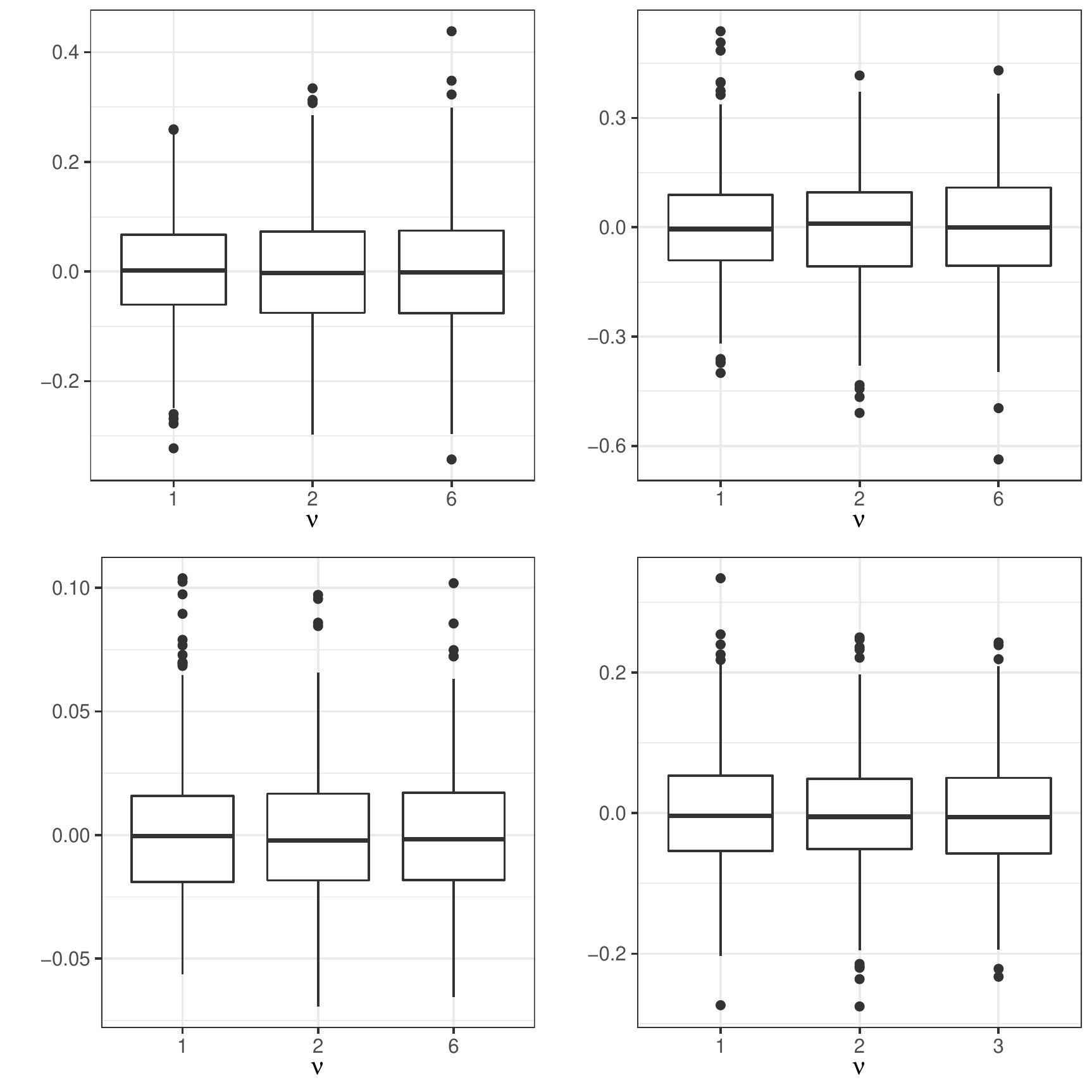}
\end{tabular}
\caption{From left to right: centered boxplots of $\hat{\beta}_0,\hat{\beta}_1$ (first row),  $\hat{b}$ and $\hat{\delta}$ (second row) for $\nu\in\{1,2,6\}$, respectively, when $\delta=1.5$
when estimating
the RF $B_{\mu(\ss)\delta, (1-\mu(\ss))\delta}$
with marginal beta distribution
 using weighted  pairwise composite likelihood. The underlying correlation function is ${\cal GW}_{0,4, b}(\hh)=	\left(1- ||\hh||/b\right)^4_+$ with $b=0.2$ and the mean is specified as $\mu(\ss )=(1+\exp \left[ -\{\beta_0+\beta_1 u_1( \ss)\} \right])^{-1}$  with  $\beta_0=0.2$ and $\beta_1=-0.2$.}  \label{boxplots1}}
\end{figure}

Table \ref{tab:tab411} depicts Bias and MSE for each parameter of the nine scenarios.
 Some general patterns can be observed. For instance the MSE of $\hat{b}$ is not affected by the choice of $\delta$ and $\nu$ and
 the MSE of $\hat{\delta}$  and of the  estimated regression parameters increases when increasing  $\delta$, irrespective of   $\nu$.
Finally, as a general comment, the
estimates are overall unbiased and symmetric distributed.
As an example, Fig.  \ref{boxplots1} shows the centered boxplots of the  weighted pairwise likelihood   estimates for the mean regression parameters $\beta_0$, $\beta_1$,  the shape parameter when $\delta=1.5$ and  the spatial compact support parameter $b$ for different  $\nu\in\{1,2,6\}$.

\begin{table}[!hbtp]
\caption{Bias and MSE of each parameter for different setting of $\nu$ and $\delta$ when estimating
the RF $B_{\mu(\ss)\delta, (1-\mu(\ss))\delta}$
with marginal beta distribution
 using weighted  pairwise composite likelihood. The underlying correlation function is ${\cal GW}_{0,4, b}(\hh)=	\left(1- ||\hh||/b\right)^4_+$ with $b=0.2$ and the mean is specified as $\mu(\ss )=(1+\exp \left[ -\{\beta_0+\beta_1 u_1( \ss)\} \right])^{-1}$  with  $\beta_0=0.2$ and $\beta_1=-0.2$.}
\label{tab:tab411}
\vskip-0.3cm\hrule

\smallskip
\centering\small
\scalebox{0.66}{
\begin{tabular}{rrrrrrlrrrrrlrrrrrl}
\multicolumn{1}{c}{$\nu$}      & \multicolumn{6}{c}{$1$}                                                                 & \multicolumn{6}{c}{$2$}                                                                  & \multicolumn{6}{c}{$6$}                                                                 \\ \hline
\multicolumn{1}{c}{$\delta$} & \multicolumn{2}{c}{0.5} & \multicolumn{2}{c}{1.5} & \multicolumn{2}{c}{2.5}              & \multicolumn{2}{c}{0.5} & \multicolumn{2}{c}{1.5} & \multicolumn{2}{c}{2.5}              & \multicolumn{2}{c}{0.5} & \multicolumn{2}{c}{1.5} & \multicolumn{2}{c}{2.5}              \\
                               & Bias        & MSE        & Bias        & MSE        & Bias       & \multicolumn{1}{r}{MSE} & Bias         & MSE       & Bias        & MSE        & Bias       & \multicolumn{1}{r}{MSE} & Bias        & MSE        & Bias        & MSE        & Bias       & \multicolumn{1}{r}{MSE} \\ \hline
$\hat{\beta}_0$ & 0.0091	&	0.0140	&	0.0091	&	0.0103	&	0.0083	&	0.0080	&	-0.0050	&	0.0138	&	-0.0031	&	0.0104	&	-0.0028	&	0.0080	&	-0.0057	&	0.0169	&	-0.0030	&	0.0129	&	-0.0030	&	0.0099  \\
$\hat{\beta}_1$ &  -0.0070	&	0.0310	&	-0.0066	&	0.0220	&	-0.0058	&	0.0167	&	0.0093	&	0.0287	&	0.0051	&	0.0223	&	0.0047	&	0.0171	&	-0.0025	&	0.0321	&	-0.0074	&	0.0235	&	-0.0067	&	0.0179 \\
$\hat{b}$ & -0.0245	&	0.0014	&	-0.0247	&	0.0014	&	-0.0247	&	0.0014	&	-0.0017	&	0.0007	&	-0.0025	&	0.0007	&	-0.0026	&	0.0007	&	-0.0155	&	0.0009	&	-0.0156	&	0.0009	&	-0.0156	&	0.0009	\\
$\hat{\delta}$ & 0.0034	&	0.0006 	&	0.0087	&	0.0068	&	0.0152	&	0.0209	&	0.0039	&	0.0006	&	0.0139	&	0.0070	&	0.0243	&	0.0217	&	0.0014	&	0.0006	&	0.0036	&	0.0069	&	0.0058	&	0.0214\\
\end{tabular}}
\hrule
\end{table}

\section{Application to Vegetation Indexes}\label{sec:5}

In the study of ecological phenomena and dynamics, the use of remote sensing measurements and associated transformations is a common practice. One particularly stands out from the rest, the Normalized Difference Vegetation Index (NDVI); as reviewed in \citep{pettorelli2005using}, the use of the NDVI gained traction in recent years. This index is derived from the red near-infrared reflectance ratio defined as
$$NDVI=\dfrac{NIR-RED}{NIR+RED},$$
\noindent where $NIR$ and $RED$ are the amounts of near-infrared and red light, respectively, reflected by the vegetation and measured by satellites. This index ranges from $-1$ to $+1$, where negatives values represent an absence of vegetation.

\begin{figure}[h!]
\centering{
\begin{tabular}{ccc}
\includegraphics[width=0.33\textwidth]{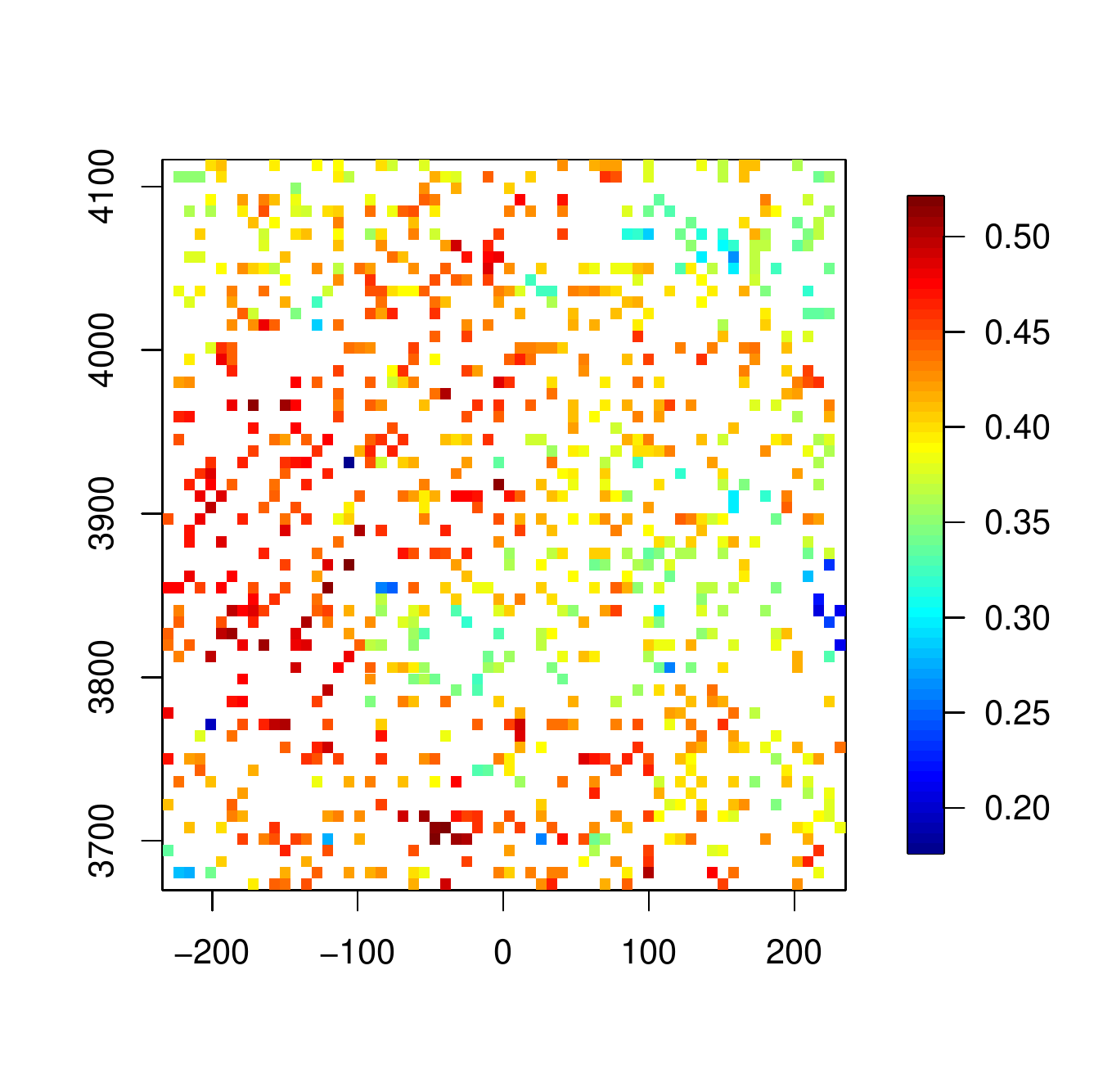} &
 \includegraphics[width=0.33\textwidth]{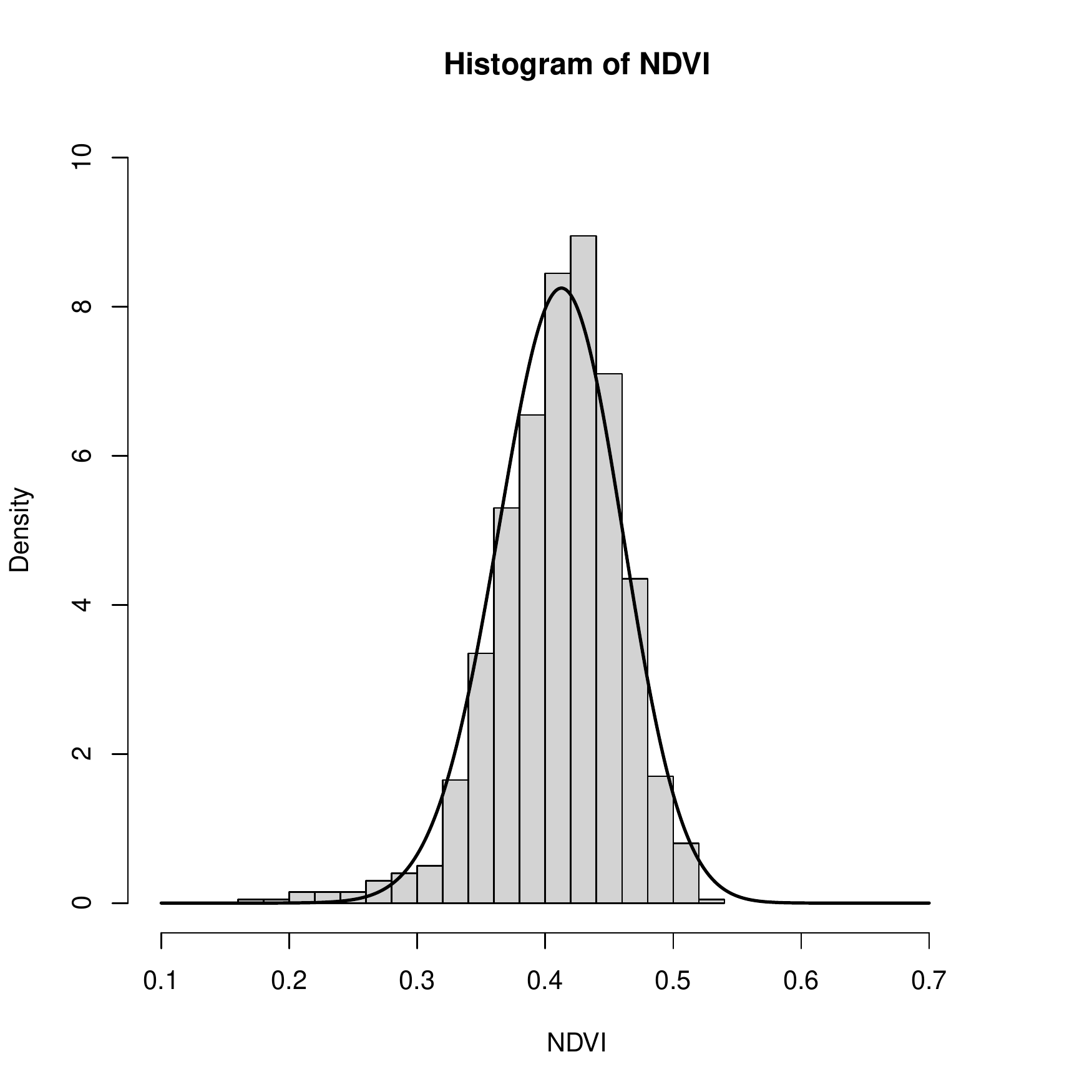} &
\raisebox{-0.12\height}{\includegraphics[width=0.33\textwidth]{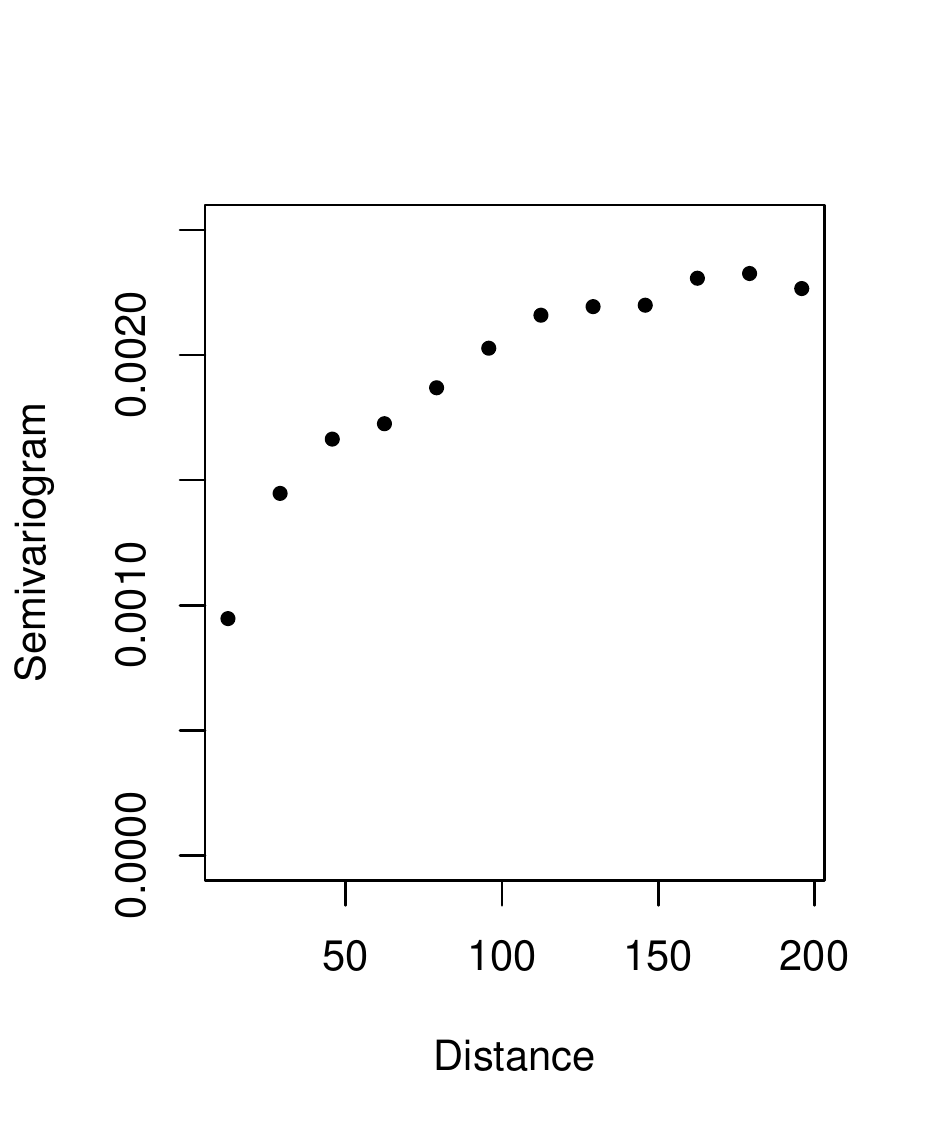}}\\
(a) & (b) & (c)
\end{tabular}
\caption{From left to right: a)  coloured map of the NDVI data; b) normalized histogram of the NDVI data  compared with the  fitted  density  of the beta  distribution; 
c) empirical semivariogram of the NDVI data.}
\label{app_f1}}
\end{figure}

To illustrate our proposed methodology, we make use of the NDVI extracted from Google Earth Engine \citep{gorelick2017google} using data provided from The National Oceanic and Atmospheric Administration (NOAA) Climate Data Record (CDR) from May 1st, 2018.
We analyze our data using \textsf{R} and the \texttt{GeoModels} package \citep{Bevilacqua:2018aa} in which our methodology was implemented. In this example, we focus our study  on 1000 randomly sampled NDVI data between longitude $-92$ and $-97$, and latitude between $33$ and $37$, corresponding roughly to the east side of Texas, US. Fig. \ref{app_f1} show a
coloured map after a suitable projection, the normalized histogram, and the empirical semi-variogram of NDVI data.

In our analysis, we consider two random fields sharing the same marginal distribution.
In particular  we consider  the (reparametrized) beta random field (assuming a constant mean) obtained from the Clayton random field and from the Gaussian copula  random field  that is we consider: $$B^C_{\mu\delta, (1-\mu)\delta}(\ss)=F^{-1}_{B_{\mu\delta, (1-\mu)\delta}}\{U_{\nu}(\ss)\}, \qquad
B^G_{\mu\delta, (1-\mu)\delta}(\ss)=F^{-1}_{B_{\mu\delta, (1-\mu)\delta}}\left[\Phi\{Z(\ss)\}\right],$$
\noindent where $\mu=1/(1+e^{-\beta})$ and $\delta>0$.
Note that  NDVI data are bounded supported   from $-1$ to $+1$. However, the beta random field which is defined  on $(0,1)$
 can be easily extended  to an arbitrary bounded support $(a_1,a_2)$ through the transformation
$(B_{\mu\delta, (1-\mu)\delta}-a_1)/(a_2-a_1)$ with $a_2>a_1$.

We first examine the suitability of the beta marginal model for the  NDVI data. In order to check the reliability of our marginal distribution, we compare the histogram with the fitted beta distribution, where the marginal
parameters $\beta$ and $\delta$ have been estimated through maximum likelihood assuming independence ($\hat{\beta}=0.8721, \hat{\delta}=355.52$).
 From  Fig. \ref{app_f1}b, it can be appreciated that our marginal model seems to be a viable hypothesis. Note that, even in the NDVI data potentially ranges from $-1$ to $1$, our data are basically  concentrated  between $0.1$ and $0.7$ approximatively.

\begin{figure}[h!]
\centering{
\begin{tabular}{cc}
\includegraphics[width=5.2cm, height=5.2cm]{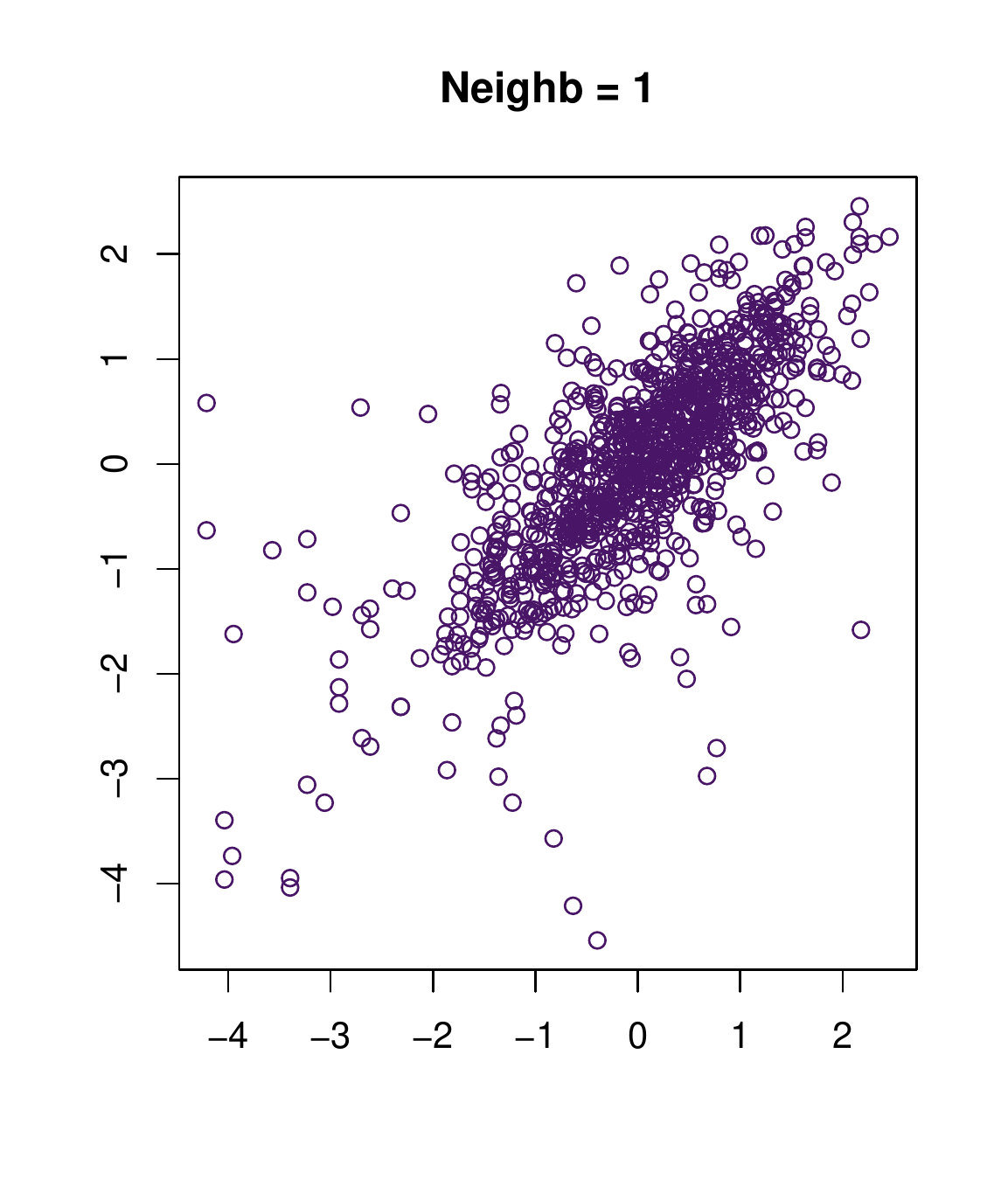} & \includegraphics[width=5.2cm, height=5.2cm]{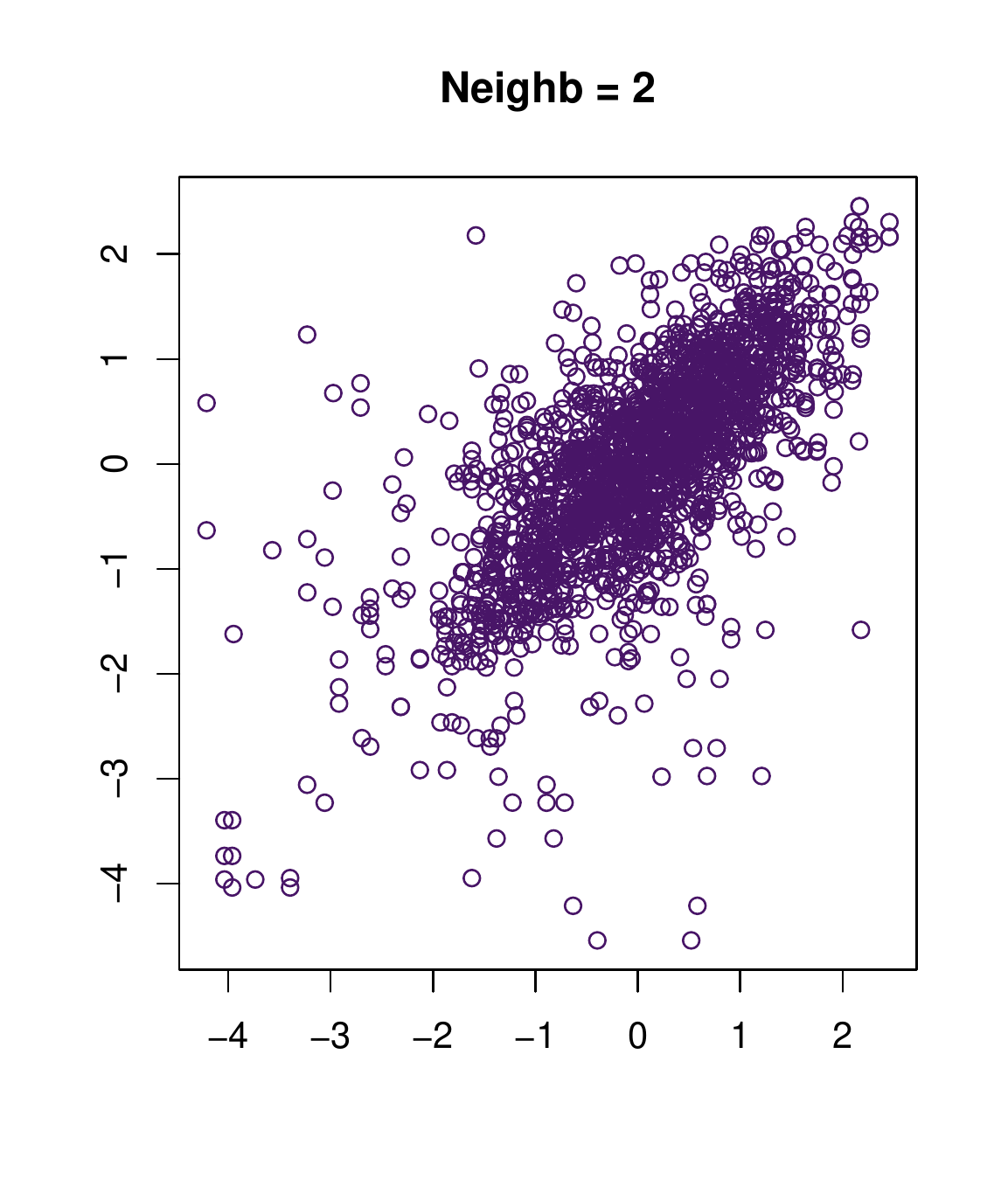} \\
\includegraphics[width=5.2cm, height=5.2cm]{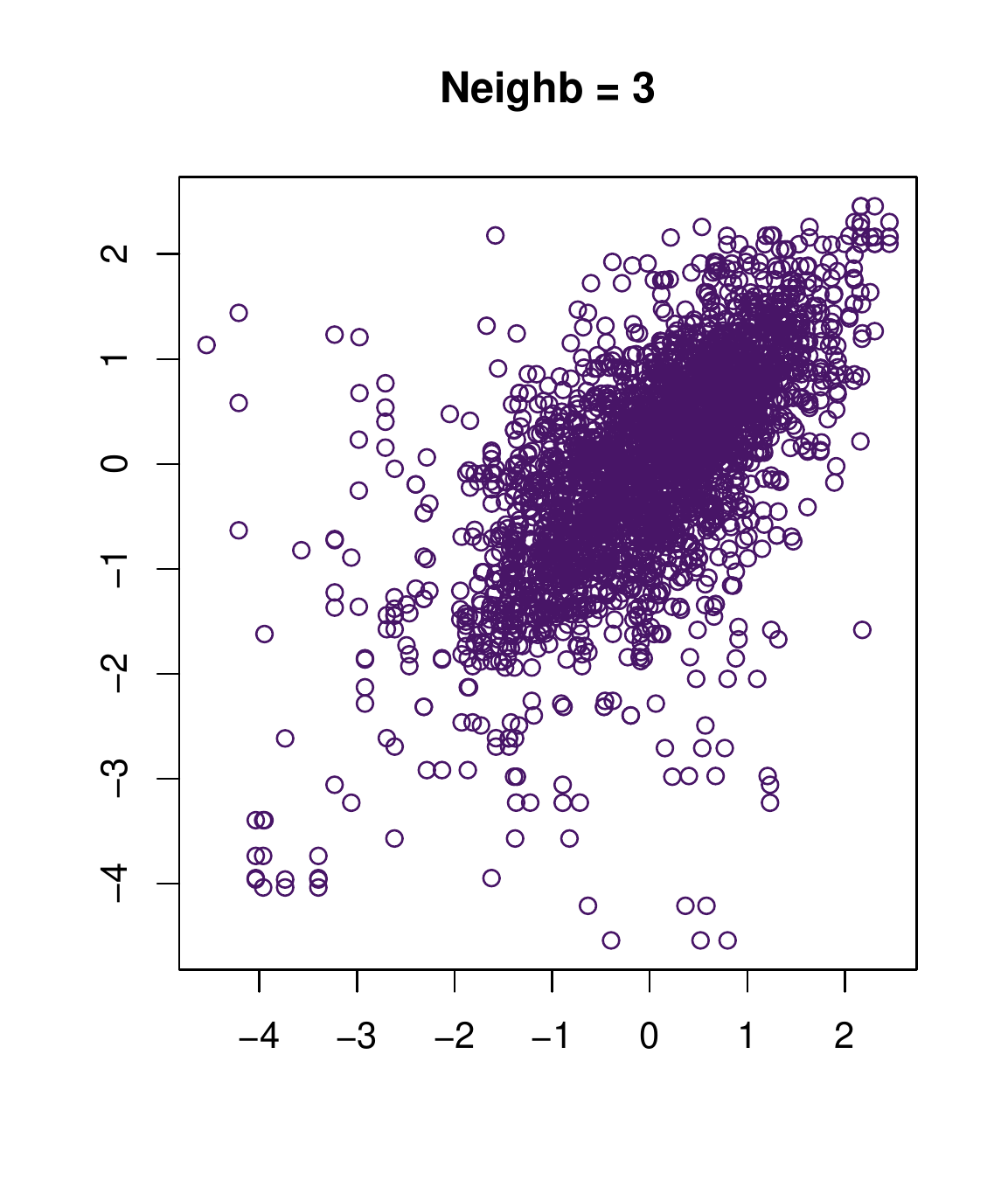} & \includegraphics[width=5.2cm, height=5.2cm]{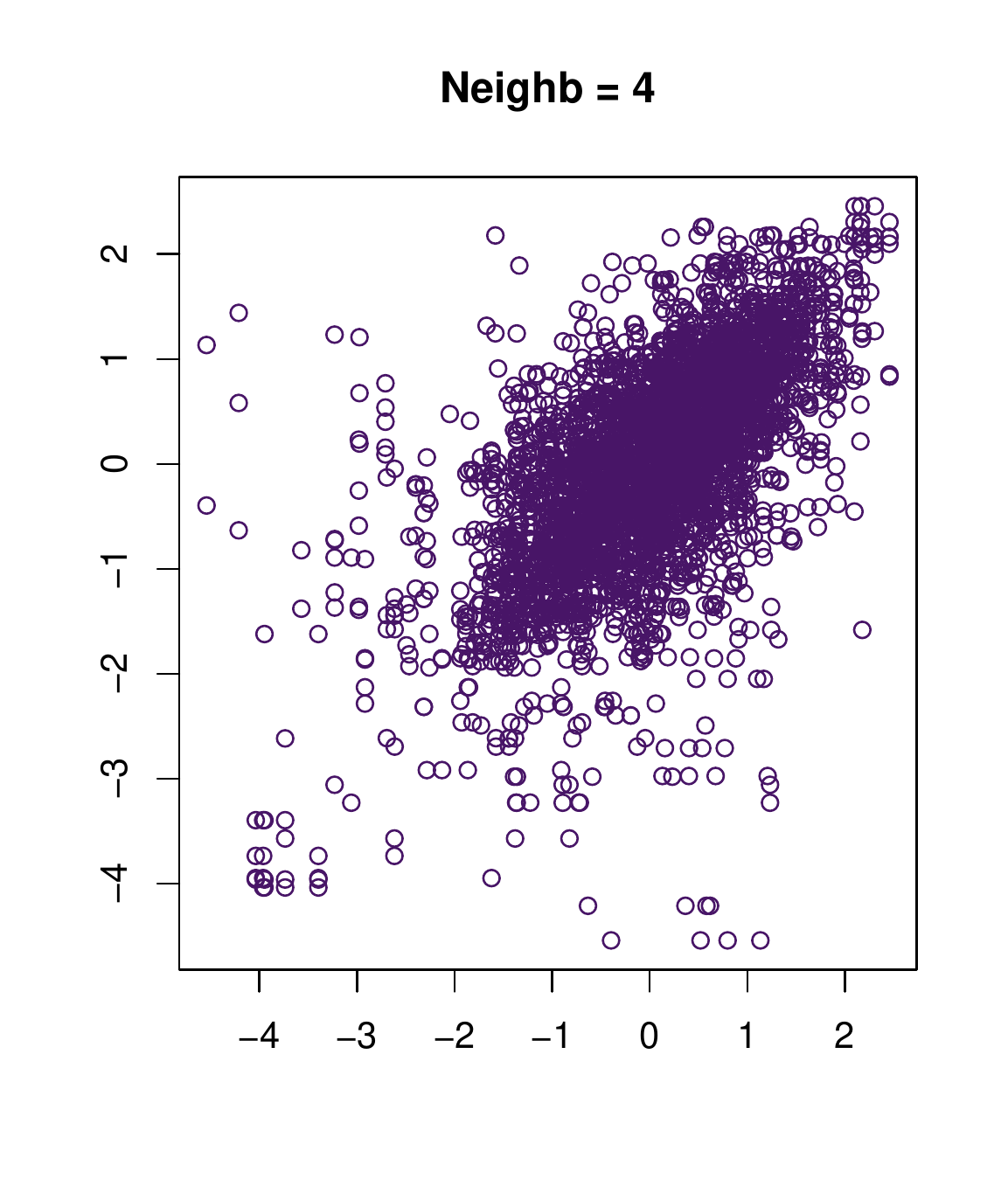}
\end{tabular}
\caption{Spatial scatterplots of NDVI data for four increasing neighborhood orders (from left to right) after normal score transformation.}
\label{app_f2}}
\end{figure}

 In addition,  the semivariogram depicted in Fig. \ref{app_f1}c shows  a clear spatial dependence and a negligible  nugget effect in the data. 
To identify the presence of reflection asymmetry, we apply the normal score transformation to the NDVI data and we compute the scatterplot.
 Fig. \ref{app_f2} depicts the spatial scatterplot for four different neighborhood orders, which clearly show the presence of
 reflection asymmetry in the data; in particular, asymmetry to upper tail   can be detected.

We estimate both beta random fields $B^C_{\mu\delta, (1-\mu)\delta}$ and $B^G_{\mu\delta, (1-\mu)\delta}$
with the weighted pairwise composite likelihood estimation method
using, as in the simulation study, the weight function  (\ref{neigh}) setting $m=2$.
We assume that the underlying isotropic correlation function is of the generalized Wendland type as defined in (\ref{askey}).

For the proposed $B^C_{\mu\delta, (1-\mu)\delta}$   model, the (a)symmetry parameter   is fixed equal to $\nu\in\{2, 4, 6\}$. We consider these specific values based on the empirical evidence of upper tail asymmetry  (see Fig. \ref{app_f2})
and taking into account the discussion on  Fig. \ref{copufig} in Section 3.

\begin{figure}[h!]
\centering{
\begin{tabular}{c}
\includegraphics[width=5.2cm, height=5.2cm]{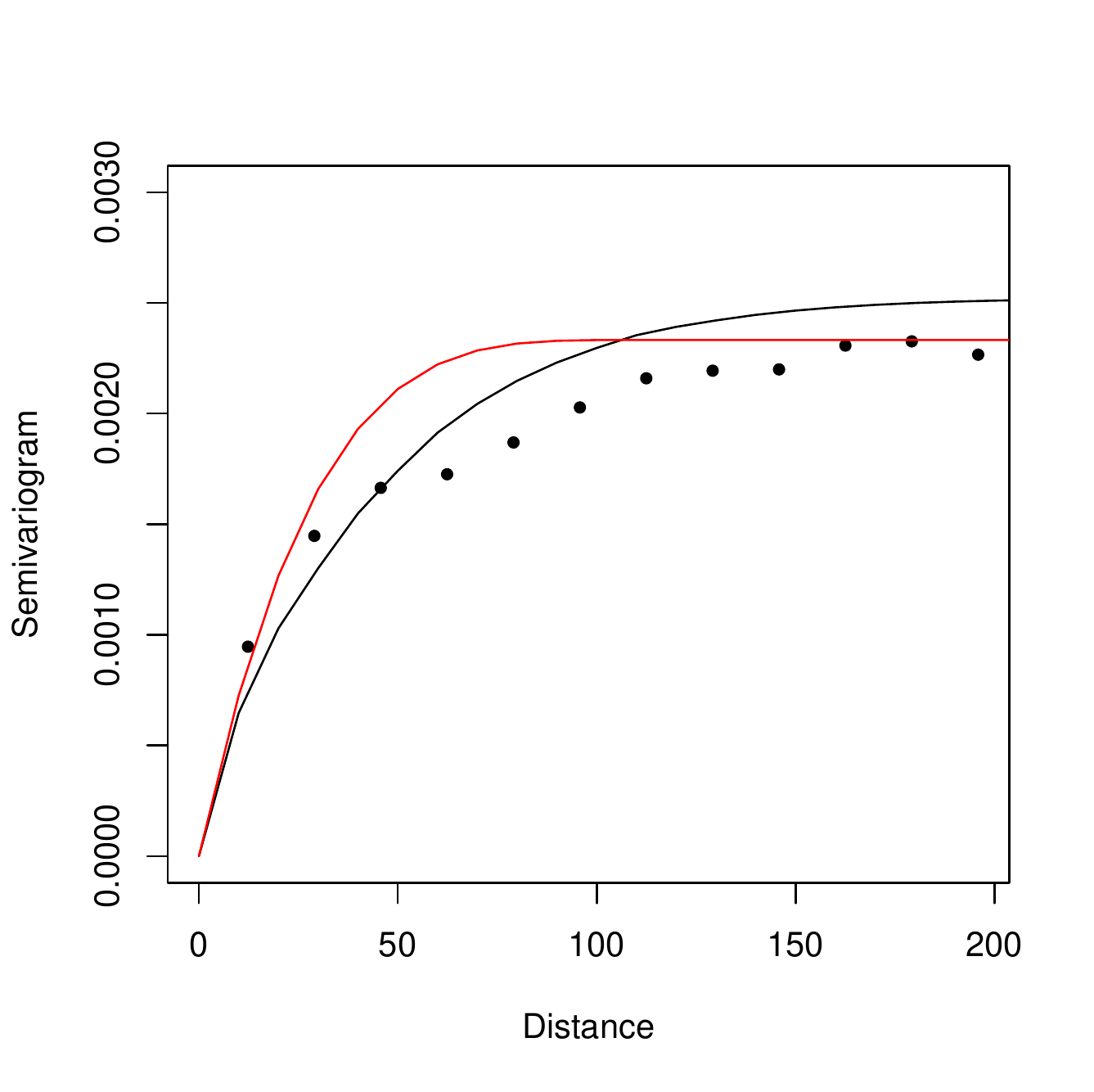}
\end{tabular}
\caption{Empirical and estimated semivariogram of NDVI of
$B^C_{\mu\delta, (1-\mu)\delta}$   (the proposed model, black line) when $\nu=4$
and $B^G_{\mu\delta, (1-\mu)\delta}$   (the Gaussian copula, red line).}
\label{app_f3}}
\end{figure}

Table \ref{performancepp} depicts the composite likelihood estimates with associated standard errors  and, in addition, the composite likelihood AIC (PLIC) as defined in (\ref{plic}) for both types of models.
They are computed using parametric bootstrap as explained in Section 4.2.
 It turns out that the estimation of the marginal  parameters  $\beta$ and $\delta$ are quite similar, as expected.
However, the compact support parameter estimate, i.e., the spatial dependence, is clearly greater for  the proposed model; see also Fig. \ref{app_f3}.
More importantly, the PLIC criterion selects the $B^C_{\mu\delta, (1-\mu)\delta}$ model with $\nu=4$. Note that,
also in the reflection symmetric case ($\nu=2$) the PLIC criterion selects the $B^C_{\mu\delta, (1-\mu)\delta}$ model
with respect to the  $B^G_{\mu\delta, (1-\mu)\delta}$  model.

\begin{table}[h!]
\caption{Estimates
of the beta random fields $B^C_{\mu\delta, (1-\mu)\delta}$   (the proposed model)
for $\nu\in\{2,4,6\}$ and $B^G_{\mu\delta, (1-\mu)\delta}$ (the Gaussian copula model) with
$\mu=1/(1+e^{-\beta})$
when the underlying correlation model is ${\cal GW}_{0,4, b}(\hh)=	\left(1- ||\hh||/b\right)^4_+$.
 In parenthesis,  for each parameter, standard error estimation is reported. Last line shows the composite likelihood AIC information criterion (PLIC).}\label{performancepp}

\vskip-0.3cm\hrule

\smallskip
\centering\small

\begin{tabular}{ccccc}
\multirow{2}{*}{Parameter} & \multirow{2}{*}{$B^G_{\mu\delta, (1-\mu)\delta}$ }                                    & \multicolumn{3}{c}{$B^C_{\mu\delta, (1-\mu)\delta}$}                                                                                                                                                                                                         \\ \cline{3-5}
                           &                                                              & \multicolumn{1}{c}{$\nu=2$}                                                      & \multicolumn{1}{c}{$\nu=4$}                                                      & $\nu=6$                                                      \\ 
$\hat{\beta}$                & \begin{tabular}[c]{@{}c@{}}0.8733\\ (0.0127)\end{tabular}    & \multicolumn{1}{c}{\begin{tabular}[c]{@{}c@{}}0.8727\\ (0.0238)\end{tabular}}    & \multicolumn{1}{c}{\begin{tabular}[c]{@{}c@{}}0.8715\\ (0.0245)\end{tabular}}    & \begin{tabular}[c]{@{}c@{}}0.8713\\ (0.0245)\end{tabular}    \\ 
$\hat{b}$             & \begin{tabular}[c]{@{}c@{}}112.4101\\ (12.1615)\end{tabular} & \multicolumn{1}{c}{\begin{tabular}[c]{@{}c@{}}473.6463\\ (95.6832)\end{tabular}} & \multicolumn{1}{c}{\begin{tabular}[c]{@{}c@{}}433.3901\\ (81.6638)\end{tabular}} & \begin{tabular}[c]{@{}c@{}}415.3676\\ (77.1154)\end{tabular} \\ 
$\hat{\delta}$             & \begin{tabular}[c]{@{}c@{}}355.9033\\ (35.7105)\end{tabular} & \multicolumn{1}{c}{\begin{tabular}[c]{@{}c@{}}331.6079\\ (41.5803)\end{tabular}} & \multicolumn{1}{c}{\begin{tabular}[c]{@{}c@{}}330.5262\\ (46.4087)\end{tabular}} & \begin{tabular}[c]{@{}c@{}}329.7499\\ (48.4774)\end{tabular} \\ 
PLIC                    & -14349                                                       & \multicolumn{1}{c}{-14372}                                                       & \multicolumn{1}{c}{-14380}                                                       & -14368                                                       \\ 
\end{tabular}
\hrule
\end{table}
Finally, in Fig. \ref{app_f3} we provide a graphical comparison between the empirical and fitted semivariogram
of the $B^G_{\mu\delta, (1-\mu)\delta}$ model  (red line) and the $B^C_{\mu\delta, (1-\mu)\delta}$ model (black line)  when $\nu=4$.
In can be seen  that the proposed model is able to better capture the spatial dependence when compared with the Gaussian copula case.

\section{Concluding remarks}\label{sec:6}
We have introduced a new kind of spatial  random field with uniform marginal distribution   as a viable alternative to the  spatial Gaussian copula.
The main feature of our model is that, unlike the Gaussian copula case, the type of dependence can be reflection symmetric or not.
In particular  the dependence structure of the proposed random field is indexed by a correlation function with an additional parameter that characterizes the type of reflection (a)symmetry and, as a consequence, it can be useful for modeling Gaussian and non-Gaussian spatial data exhibiting reflection asymmetries.
To the best of our knowledge, this is the first copula that can be used for spatial data that allows both left and right bivariate asymmetries

As an application of the proposed model we considered a random field with beta  marginals
and we applied it to the analysis of  Normalized Difference Vegetation Indexes data.
In  particular we showed that the
beta random field obtained using the proposed
Clayton random field performs better than our benchmark (the Gaussian copula) in terms of Composite Likelihood Akaike Information Criterion, providing a practical example of the feasibility of our results.

For simplicity of presentation, we restrict the treatment to the spatial Euclidean setting.
However, the results presented in this paper can be applied to more general spaces. The key is the specification of a well-defined correlation function in a specific space for the underlying Gaussian random field such as correlation models defined on the space-time setting or on the sphere \citep{gneiting2013,porcubev}.

A possible limitation of the proposed model is the lack of amenable expressions of the associated multivariate distributions, particularly due to the natural occurrence of  involved special functions, such as the Kampé de Fériet function. This prevents an inference approach based on the full likelihood and the computation of the optimal predictor. In the first case, our simulation study shows that an inferential approach based on weighted pairwise composite likelihood using the bivariate density given in (\ref{pairbU}) could be an effective solution for estimating the unknown parameters.
In the second case,
even if the prediction has not been investigated in this paper,
it  can be performed using the results presented in this work coupled with the methods proposed for instance in \cite{Kazianka2010}.

Finally, the general class proposed in  (\ref{arch}) potentially opens the door to
new copula models for spatial data with complex dependencies. In this paper we have focused in a specific instance  of the class. However other types of complex copula models  for spatial data can be developed starting from   (\ref{arch})
and this topic will be subject to future research.

\section*{Acknowledgments}
Moreno Bevilacqua acknowledges financial support from grant FONDECYT 1200068 and ANID/PIA/ANILLOS ACT210096
and  ANID project Data Observatory Foundation DO210001 from the Chilean government
and project MATH-AMSUD 22-MATH-06 (AMSUD220041). The work of Christian Caama\~no-Carrillo was partially supported by grant FONDECYT 11220066 from the Chilean government and DIUBB 2120538 IF/R from the University of B\'io-B\'io.  The work of Eloy Alvarado was partially supported by CONICYT-PFCHA/DOCTORADO-BECAS-CHILE/2018-21180953.

\section*{Appendix}

\subsection*{Proof of Theorem \ref{theo1}}\label{proof_theo1}
\noindent For simplicity of notation we use $\bm{W}=H_{\nu}(\ss)$ and  $\bm{R}=N_{\alpha}(\ss)$. Under the transformation $g_i=y_ir_i(1-y_i)^{-1}$ and
$g_j=y_jr_j(1-y_j)^{-1}$ in (\ref{pairchi2}) with Jacobian
$J((g_i,g_j)\to (y_i,y_j))=r_ir_j\{(1-y_i)(1-y_j)\}^{-2}$ and using
series expansion of the hypergeometric function ${}_0F_1$, we have
\begin{small}
\begin{align}\label{degam}
f_{\bm{Y}_{\nu,\alpha;ij}}(\bm{y}_{ij})&=\int\limits_{\mathbb{R}_+^2}f_{\bm{W}_{ij}|\bm{R}_{ij}}(\bm{w}_{ij}|\bm{r}_{ij})f_{\bm{R}_{ij}}(\bm{r}_{ij})Jd\bm{r}_{ij} 
=\frac{(y_iy_j)^{\nu/2-1}\{(1-y_i)(1-y_j)\}^{-\nu/2-1}}{\Gamma^2\left(\nu/2\right)\Gamma^2\left(\alpha/2\right)(1-\rho^2(\hh))^{(\nu+\alpha)/2}}\int\limits_{\mathbb{R}_+^2}(r_ir_j)^{(\nu+\alpha)/2-1}e^{-\frac{1}{1-\rho^2(\hh)}\left(\frac{r_iy_i}{1-y_i}+\frac{r_jy_j}{1-y_j}\right)}e^{-\frac{r_i+r_j}{1-\rho^2(\hh)}} \nonumber \\
&\quad\times\quad  {}_0F_1\left(\nu/2;\frac{\rho^2(\hh)y_iy_jr_ir_j}{(1-y_i)(1-y_j)\{1-\rho^2(\hh)\}^2}\right){}_0F_1\left(\alpha/2;\frac{\rho^2(\hh)r_ir_j}{\{1-\rho^2(\hh)\}^2}\right)d\bm{r}_{ij} \nonumber \\
&=\frac{(y_iy_j)^{\nu/2-1}\{(1-y_i)(1-y_j)\}^{-\nu/2-1}}{\Gamma^2\left(\nu/2\right)\Gamma^2\left(\alpha/2\right)\{1-\rho^2(\hh)\}^{(\nu+\alpha)/2}}
\sum\limits_{k=0}^{\infty}\sum\limits_{m=0}^{\infty}\frac{I(k,m)}{k!m!\left(\nu/2\right)_k \left(\alpha/2\right)_m}\left\{\frac{\rho^2(\hh)y_iy_j}{(1-y_i)(1-y_j)\{1-\rho^2(\hh)\}^2}\right\}^k 
 \left\{\frac{\rho^2(\hh)}{\{1-\rho^2(\hh)\}^2}\right\}^m,
\end{align}
\end{small}
where, using Fubini's Theorem and (3.381.4) of \cite{Gradshteyn:Ryzhik:2007}, we obtain
\begin{align}\label{degam1}
I(k,m)&=\int\limits_{\mathbb{R}_+}r_i^{(\nu+\alpha)/2+k+m-1}e^{-\left\{\frac{1}{(1-\rho^2(\hh))(1-y_i)}\right\}r_i}dr_i\int\limits_{\mathbb{R}_+}r_j^{(\nu+\alpha)/2+k+m-1}e^{-\left\{\frac{1}{(1-\rho^2(\hh))(1-y_j)}\right\}r_j}dr_j\nonumber\\
&=\Gamma^2\left((\nu+\alpha)/2+k+m\right)\left\{(1-\rho^2(\hh))(1-y_i)\right\}^{(\nu+\alpha)/2+k+m}\left\{(1-\rho^2(\hh))(1-y_j)\right\}^{(\nu+\alpha)/2+k+m}
\end{align}
and combining (\ref{degam1}) and (\ref{degam}), we obtain
\begin{footnotesize}
\begin{align*}
f_{\bm{Y}_{\nu,\alpha;ij}}(\yy_{ij})=&\frac{(y_iy_j)^{\nu/2-1}\{(1-y_i)(1-y_j)\}^{\alpha/2-1}\Gamma^2\left((\nu+\alpha)/2\right)}{\Gamma^2\left(\nu/2\right)\Gamma^2\left(\alpha/2\right)\{1-\rho^2(\hh)\}^{(\nu+\alpha)/2}}
\sum\limits_{k=0}^{\infty}\sum\limits_{m=0}^{\infty}\frac{\left((\nu+\alpha)/2\right)_{k+m}^2}{k!m!\left(\nu/2\right)_k\left(\alpha/2\right)_m}
\left\{\rho^2(\hh)y_iy_j\right\}^k\left\{\rho^2(\hh)(1-y_i)(1-y_j)\right\}^m\nonumber\\
=&\frac{(y_iy_j)^{\nu/2-1}\{(1-y_i)(1-y_j)\}^{\alpha/2-1}\Gamma^2\left((\nu+\alpha)/2\right)}{\Gamma^2\left(\nu/2\right)\Gamma^2\left(\alpha/2\right)\{1-\rho^2(\hh)\}^{-(\nu+\alpha)/2}}
F_4\left((\nu+\alpha)/2,(\nu+\alpha)/2;\nu/2,\alpha/2;\rho^2(\hh)y_iy_j,\rho^2(\hh)(1-y_i)(1-y_j)\right),
\end{align*}
\end{footnotesize}
where $F_4(a,b;c,c';w,z)$ is the fourth Appell hypergeometric function of two variables defined in (\ref{ap4}).

\subsection*{Proof of Theorem \ref{theo2}}\label{proof_theo2}
\noindent We first cosider the $(a,a)$-th product moment $\EE\left\{Y_{\nu,\alpha}^a(\ss_i)Y_{\nu,\alpha}^a(\ss_j)\right\}$ of the beta random field $Y_{\nu,\alpha}$ with underlying correlation $\rho(\hh)$. Using $c=(\nu+\alpha)/2$, then by definition
\begin{small}
\begin{align}\label{cal}
\EE\left\{Y_{\nu,\alpha}^a(\ss_i)Y_{\nu,\alpha}^a(\ss_j)\right\}
&=\frac{\Gamma^2\left(c\right)}{\Gamma^2\left(\nu/2\right)\Gamma^2\left(\alpha/2\right)\{1-\rho^2(\hh)\}^{-(\nu+\alpha)/2}}
\int\limits_{[0,1]^2}y_i^{\nu/2+a-1}y_j^{\nu/2+a-1}\{(1-y_i)(1-y_j)\}^{\alpha/2-1}\nonumber\\
&\quad\times \sum\limits_{k=0}^{\infty}\sum\limits_{m=0}^{\infty}\frac{\left(c\right)_{k+m}^2}{k!m!\left(\nu/2\right)_k\left(\alpha/2\right)_m}
\left\{\rho^2(\hh)y_iy_j\right\}^k\left\{\rho^2(\hh)(1-y_i)(1-y_j)\right\}^m d\yy_{ij}\nonumber\\
&=\frac{\Gamma^2\left(c\right)}{\Gamma^2\left(\nu/2\right)\Gamma^2\left(\alpha/2\right)\{1-\rho^2(\hh)\}^{-c}}
\sum\limits_{k=0}^{\infty}\sum\limits_{m=0}^{\infty}\frac{\left(c\right)_{k+m}^2}{k!m!\left(\nu/2\right)_k\left(\alpha/2\right)_m}I(k,m)\rho^{2}(\hh)^{k+m},
\end{align}
\end{small}
where, using Fubini's Theorem,
\begin{equation*}
I(k,m)=\int\limits_0^1y_i^{\nu/2+a+k-1}(1-y_i)^{\alpha/2+m-1}dy_i
\int\limits_0^1y_j^{\nu/2+a+k-1}(1-y_j)^{\alpha/2+m-1}dy_j.
\end{equation*}
Then using (3.251) of \cite{Gradshteyn:Ryzhik:2007}, we obtain
\begin{equation}\label{res1}
I(k,m)=\dfrac{\Gamma\left(\nu/2+a+k\right)\Gamma^2\left(\alpha/2+m\right)\Gamma\left(\nu/2+a+k\right)}
{\Gamma\left(c+a+m+k\right)\Gamma\left(c+a+m+k\right)}
\end{equation}
and combining (\ref{res1}) with (\ref{cal}), we obtain
\begin{equation*}
\EE\left\{Y_{\nu,\alpha}^a(\ss_i)Y_{\nu,\alpha}^a(\ss_j)\right\}
=\frac{\Gamma^2\left(c\right)\Gamma\left(\nu/2+a\right)^2 \{1-\rho^2(\hh)\}^{c}}
{\Gamma^2\left(\nu/2\right)\Gamma\left(c+a\right)^2}
\sum\limits_{k=0}^{\infty}\sum\limits_{m=0}^{\infty}\frac{\left(c\right)_{k+m}^2\left(\nu/2+a\right)^2_k \left(\alpha/2\right)_m^2\{\rho^{2}(\hh)\}^{k+m}}{k!m!\left(\nu/2\right)_k\left(\alpha/2\right)_m\left(c+a\right)^2_{k+m}}.\nonumber
\end{equation*}
The latter equation can be reduced to
\begin{equation}\label{et2}
\EE\left\{Y_{\nu,\alpha}^a(\ss_i)Y_{\nu,\alpha}^a(\ss_j)\right\}= \frac{\Gamma^2\left(c\right)\Gamma\left(\nu/2+a\right)^2 \{1-\rho^2(\hh)\}^{c}}
{\Gamma^2\left(\nu/2\right)\Gamma\left(c+a\right)^2}
F_{2;1;0}^{2;2;1}\left[ \begin{array}{c}
                                  c;\frac{\nu}{2}+a;\frac{\alpha}{2}\\
                                  c+a;\frac{\nu}{2};-
                                \end{array}\middle\vert \rho^{2}(\hh),\rho^{2}(\hh)\right],
\end{equation}
where $F_{E;G;H}^{A;B;C}$ is the Kampé de Fériet function defined in (\ref{KF}).

Using (\ref{et2}) with $a=1$ and the fact that $\EE\left\{Y_{\nu,\alpha}(\ss)\right\}=\nu/(\nu+\alpha)$ and $\VV\left\{Y_{\nu,\alpha}(\ss)\right\}=2\nu \alpha/\{(\nu+\alpha)^2(\nu+\alpha+2)\}$, the correlation can be written as
\begin{equation*}
\rho_{Y_{\nu,\alpha}}(\hh)=\frac{\nu (c+1)}{\alpha}\left[\{1-\rho^2(\hh)\}^{c}A-1\right],
\end{equation*}
where
\begin{equation*}
A=F_{2;1;0}^{2;2;1}\left[ \begin{array}{c}
                                  c;\frac{\nu}{2}+1;\frac{\alpha}{2}\\
                                  c+1;\frac{\nu}{2};-
                                \end{array}\middle\vert \rho^{2}(\hh),\rho^{2}(\hh)\right]=
F_{2;0;1}^{2;1;2}\left[ \begin{array}{c}
                                  c;\frac{\alpha}{2};\frac{\nu}{2}+1\\
                                  c+1;-;\frac{\nu}{2}
                                \end{array}\middle\vert \rho^{2}(\hh),\rho^{2}(\hh)\right].
\end{equation*}

\subsection*{Proof of Theorem \ref{theo3}}\label{proof_theo3}
\noindent If $U_{\nu}$ is  a uniform random field with underlying correlation function $\rho(\hh)$, then the bivariate pdf  is given by

\begin{equation*}\label{pairu}
f_{\bm{U}_{\nu;ij}}(\uu_{ij})=
\{1-\rho^2(\hh)\}^{\nu/2+1}
F_4\left(\nu/2+1,\nu/2+1;\nu/2,1;\rho^2(\hh)(u_iu_j)^{2/\nu},\rho^2(\hh)(1-u_i^{2/\nu})(1-u_j^{2/\nu})\right).
\end{equation*}
This implies that  the bivariate cdf is given by
\begin{small}
\begin{align}\label{cal4}
F_{\bm{U}_{\nu;ij}}(\bm{t}_{ij})
&=\{1-\rho^2(\hh)\}^{\nu/2+1}
\int\limits_{0}^{t_j}\int\limits_{0}^{t_i}\sum\limits_{k=0}^{\infty}\sum\limits_{m=0}^{\infty}\frac{\left(\nu/2+1\right)_{k+m}^2}{k!m!\left(\nu/2\right)_k\left(1\right)_m}
\left\{\rho^2(\hh)(u_iu_j)^{2/\nu}\right\}^k\left\{\rho^2(\hh)(1-u_i^{2/\nu})(1-u_j^{2/\nu})\right\}^m d\bm{t}_{ij}\nonumber\\
&=\{1-\rho^2(\hh)\}^{\nu/2+1}
\sum\limits_{k=0}^{\infty}\sum\limits_{m=0}^{\infty}\frac{\left(\nu/2+1\right)_{k+m}^2}{k!m!\left(\nu/2\right)_k\left(1\right)_m}I(k,m)\{\rho^{2}(\hh)\}^{k+m},
\end{align}
\end{small}
where using Fubini's Theorem with the change of variable $x=u^{2/\nu}$ and the definition $(8.391)$ of the incomplete beta function in \cite{Gradshteyn:Ryzhik:2007}, we obtain
\begin{align}\label{res14}
I(k,m)&=\int\limits_0^{t_i}u_i^{2k/\nu}(1-u_i^{2/\nu})^{m}du_i
\int\limits_0^{t_j}u_j^{^{2k/\nu}}(1-u_j^{2/\nu})^{m}du_j\nonumber\\
&=\frac{\nu^2(t_it_j)^{2k/\nu+1}}{(2k+\nu)^2}{}_2F_1\left(k+\nu/2,-m;k+\nu/2+1;t^{2/\nu}_i\right){}_2F_1\left(k+\nu/2,-m;k+\nu/2+1;t^{2/\nu}_j\right)
\end{align}
and combining (\ref{cal4}) and (\ref{res14}), we obtain
\begin{align*}
F_{\bm{U}_{\nu;ij}}(\bm{t}_{ij})&=
\nu^2t_i t_j\{1-\rho^2(\hh)\}^{\nu/2+1}
\sum\limits_{k=0}^{\infty}\sum\limits_{m=0}^{\infty}\frac{\left(\nu/2+1\right)_{k+m}^2(t_it_j)^{2k/\nu}\{\rho^{2}(\hh)\}^{k+m} }{k!(2k+\nu)^2 \left(\nu/2\right)_k(1)^2_m}\nonumber\\
&\quad\times\quad {}_2F_1\left(k+\nu/2,-m;k+\nu/2+1;t^{2/\nu}_i\right){}_2F_1\left(k+\nu/2,-m;k+\nu/2+1;t^{2/\nu}_j\right).
\end{align*}
Finally, simplifying the Pochhammer values we obtain (\ref{copu}).

\subsection*{Proof of Theorem \ref{theo4}}\label{proof_theo4}

\noindent Since $\EE\left\{U(\ss_i)U(\ss_j)\right\}=\EE\left\{Y_{\nu,\alpha}^a(\ss_i)Y_{\nu,\alpha}^a(\ss_j)\right\}$ and     setting $\alpha=2$ and $a=\nu/2$ in (\ref{et2}) we obtain:

\begin{equation*}
\EE\left\{U(\ss_i)U(\ss_j)\right\}=\frac{ \{1-\rho^2(\hh)\}^{\nu/2+1}}{4}
F_{2;1;0}^{2;2;1}\left[ \begin{array}{c}
                                  \nu/2+1;\nu;1\\
                                  \nu+1;\frac{\nu}{2};-
                                \end{array}\middle\vert \rho^{2}(\hh),\rho^{2}(\hh)\right].
\end{equation*}
Then since  $\EE\left\{U(\ss)\right\}=0.5$ and $\VV\left\{U(\ss)\right\}=1/12$, we obtain ({\ref{corru}}):


\subsection*{Proof of corollary \ref{theo5}}\label{proof_theo5}

\noindent Replacing $\nu=2$ in  equation (\ref{corru})  we get

\begin{align*}
\rho_{U}(\hh)&=3\left[ \{1-\rho^2(\hh)\}^{2}F_{2;1;0}^{2;2;1}\left[ \begin{array}{c}
                                  2;2;1\\
                                  3;1;-
                                \end{array}\middle\vert \rho^{2}(\hh),\rho^{2}(\hh)\right] - 1\right]
=3\left[ \{1-\rho^2(\hh)\}^{2} \sum_{k=0}^{\infty}\sum_{m=0}^{\infty} \dfrac{(2)_{k+m}^{2}(2)_{k}^{2}(1)_{m}}{k!m!(3)_{k+m}^{2}(1)_{k}}\{\rho^2(\hh)\}^{k+m}-1\right]\\
&=3\left[\{1-\rho^2(\hh)\}^{2} \left\{\sum_{k=0}^{\infty}\dfrac{4\rho^{2k}(\hh)}{(2+k)^2}\sum_{m=1}^{k+1} m^2\right\}-1 \right]
=3\left[\{1-\rho^2(\hh)\}^{2} \left\{\dfrac{2}{3}\left(\sum_{k=0}^{\infty}2k\rho^{2k}(\hh)+\sum_{k=0}^{\infty}\dfrac{\rho^{2k}(\hh)}{k+2}+\sum_{k=0}^{\infty}\rho^{2k}(\hh)\right)\right\}-1 \right]\\
&=3\left[ \{1-\rho^2(\hh)\}^{2} \left\{\dfrac{-2\{\rho^2(\hh)-3\rho^4(\hh)+(\rho^2(\hh)-1)^2\ln(1-\rho^2(\hh))\}}{3\rho^4(\hh)(\rho^2(\hh)-1)^2}\right\} -1\right]\\
&=\dfrac{2[\rho^2(\hh)\{3\rho^2(\hh)-1\}-\{\rho^2(\hh)-1\}^2 \ln\{1-\rho^2(\hh)\}]}{\rho^4(\hh)}-3.
\end{align*}

\noindent For the double summation we make use of (1.4) in \cite{choi2003notes}, the Lerch's trascendent defined as
$
\Phi(z,s,a)=\sum_{n=0}^{\infty} (a+n)^{-s} z^n
$
and its link to the polylogarithm function given by
$
\Phi(z,s,2)=z^{-2}(Li_{s}(z)-z).
$ Since
$
Li_1(\rho^2(\hh))=-\ln(1-\rho^2(\hh))
$,
in our case, the term $$\sum\limits_{k=0}^{\infty}\dfrac{\rho^{2k}(\hh)}{k+2}=\Phi(\rho^2(\hh),1,2)$$ simplifies to
$$\dfrac{-\rho^2(\hh)-\ln\{1-\rho^2(\hh)\}}{\rho^4(\hh)}.$$

\subsection*{Proof of Theorem \ref{theo6}}\label{proof_theo6}

\noindent If the underlying Gaussian random field $Z$ is a weakly stationary process with correlation $\rho(\hh)$ then from (\ref{CC}) it is straightforward to see that $U$ is also weakly stationary.Following \cite{Stein:1999}, the mean-square continuity and $m$-times mean-square differentiability of $U$ are equivalent to the continuity and $2m$-times differentiability of $\rho_{U}(\hh)$ at $\hh=0$. From (\ref{corru_nu2}), using simple limits properties it can be easily seen that $\lim_{\hh\rightarrow 0} \rho_{U}(\hh)=1$ if $\rho(\hh)=1$. Hence, $U$ is mean-square continuous if  $Z$ is mean-square continuous.\\
For the mean square differentiability, let the underlying Gaussian random field $Z$ $m$-times differentiable. Then it can be shown that $\rho_{U}^{(2m)}(\hh)\big\vert_{\hh\bm{=0}}<\infty$, hence, $U$ is $m$-times times mean square differentiable. For instance, assume that the random field $Z$ is $1$-times differentiable that is  $\rho^{(i)}(\hh)\big\vert_{\hh\bm{=0}}<\infty$ for $i\in\{1,2\}$. The first derivative is given by
\begin{eqnarray*}
\rho_{U}^{(1)}(\hh)&=&\frac{-4\rho^{(1)}(\hh)}{\rho^5(\hh)}\left[-2\ln\{1-\rho^2(\hh)\}+2\rho^2(\hh)(\ln\{1-\rho^2(\hh)\}-1)+\rho^4(\hh) \right].
\end{eqnarray*}
Computing the limit at $\hh\rightarrow 0$ and using L'H\^{o}spital's rule conveniently, we obtain $\rho_{U}^{(1)}(\hh)\big\vert_{\hh\bm{=0}}=4 \rho^{(1)}(\hh)$. The second derivative is given by
\begin{eqnarray*}
\rho_{U}^{(2)}(\hh)&=&\frac{4}{\rho^6(\hh)}\Bigg[ (\rho^{(1)}(\hh))^2 \left[ -10\ln\{1-\rho^2(\hh)\}+2\rho^2(\hh)(3\ln\{1-\rho^2(\hh)\}-5)+\rho^{4}(\hh)\right]\\
&\quad&- \rho(\hh)\rho^{(2)}(\hh)\left[ -2\ln\{1-\rho^2(\hh)\}+2\rho^2(\hh)\left[\ln\{1-\rho^2(\hh)\}-1\right]+\rho^{4}(\hh)\right]\Bigg].
\end{eqnarray*}
Taking the limit at $\hh\rightarrow 0$ with  L'H\^{o}spital's rule   and assuming $\rho'(\bm{0})=0$, it can be shown that $\rho_{U}^{(2)}(\hh)\big\vert_{\hh\bm{=0}}<\infty$ and hence the uniform random field is $1-$times differentiable. Similarly for  $m> 1$,  the $(2m)$-derivatives of $\rho_{U}(\hh)$  evaluated at $\hh=\bm{0}$
 takes finite values if $\rho'(\bm{0})=0$.
\\
Following \cite{LiTe09}, the process $U$ is long-range dependent if the correlation of $U$ is such that $\int_{\mathbb{R}_{+}^{n}}|\rho_{U}(\hh)|d^{n}\hh=\infty$.
Using the expression of $\rho_{U}(\hh)$ inside integral and integrating term by term it is sufficient to show that one of the integral diverges.
Applying the inequality  $$\int_{\mathbb{R}_{+}^{n}}\rho(\hh)d^{n}\hh \leq  \int_{\mathbb{R}_{+}^{n}}\frac{1}{\rho^2(\hh)}d^{n}\hh$$ for $\rho(\hh)\in [0,1]$ to one of the term shows that $\int_{\mathbb{R}_{+}^{n}}|\rho_{U}(\hh)|d^{n}\hh=\infty$ if and only if $\int_{\mathbb{R}_{+}^{n}}|\rho(\hh)|d^{n}\hh=\infty$. Hence, the random field $U$ has long-range dependence  if $Z$ has long-range dependence.
\\

Finally, to prove that $\rho(\hh)=0 \Rightarrow \rho_{U}(\hh)=0$, it is sufficient to apply L'H\^{o}spital's rule two times and show that:
\begin{eqnarray*}
\lim_{\rho(\hh)\rightarrow 0} \rho_{U}(\hh)=0.
\end{eqnarray*}

\bibliographystyle{myjmva}
\bibliography{clayton_jmva_bib}

\begin{thebibliography}{73}
\expandafter\ifx\csname natexlab\endcsname\relax\def\natexlab#1{#1}\fi
\providecommand{\bibinfo}[2]{#2}
\ifx\xfnm\relax \def\xfnm[#1]{\unskip,\space#1}\fi
\bibitem[{Anderes et~al.(2020)Anderes, M{\o}ller and Rasmussen}]{andmo}
\bibinfo{author}{E.~Anderes}, \bibinfo{author}{J.~M{\o}ller},
  \bibinfo{author}{J.~G. Rasmussen}, \bibinfo{title}{{Isotropic covariance
  functions on graphs and their edges}}, \bibinfo{journal}{The Annals of
  Statistics} \bibinfo{volume}{48} (\bibinfo{year}{2020}) \bibinfo{pages}{2478
  -- 2503}.
\bibitem[{Banerjee et~al.(2004)Banerjee, Carlin and
  Gelfand}]{Banerjee-Carlin-Gelfand:2004}
\bibinfo{author}{S.~Banerjee}, \bibinfo{author}{B.~P. Carlin},
  \bibinfo{author}{A.~E. Gelfand}, \bibinfo{title}{Hierarchical Modeling and
  Analysis for Spatial Data}, \bibinfo{publisher}{Chapman \& Hall/CRC Press},
  \bibinfo{address}{Boca Raton: FL}, \bibinfo{year}{2004}.
\bibitem[{Banerjee and Gelfand(2003)}]{BANERJEE200385}
\bibinfo{author}{S.~Banerjee}, \bibinfo{author}{A.~Gelfand}, \bibinfo{title}{On
  smoothness properties of spatial processes}, \bibinfo{journal}{Journal of
  Multivariate Analysis} \bibinfo{volume}{84} (\bibinfo{year}{2003})
  \bibinfo{pages}{85--100}.
\bibitem[{Bapat(1989)}]{Bapat:1989}
\bibinfo{author}{R.~B. Bapat}, \bibinfo{title}{Infinite divisibility of
  multivariate gamma distributions and m-matrices},
  \bibinfo{journal}{Sankhy$\bar{a}$ A} \bibinfo{volume}{51}
  (\bibinfo{year}{1989}) \bibinfo{pages}{73--78}.
\bibitem[{B{\'a}rdossy(2006)}]{bardossy2006copula}
\bibinfo{author}{A.~B{\'a}rdossy}, \bibinfo{title}{Copula-based geostatistical
  models for groundwater quality parameters}, \bibinfo{journal}{Water Resources
  Research} \bibinfo{volume}{42} (\bibinfo{year}{2006}).
\bibitem[{Berg et~al.(2008)Berg, Mateu and Porcu}]{berg2008}
\bibinfo{author}{C.~Berg}, \bibinfo{author}{J.~Mateu},
  \bibinfo{author}{E.~Porcu}, \bibinfo{title}{The dagum family of isotropic
  correlation functions}, \bibinfo{journal}{Bernoulli} \bibinfo{volume}{14}
  (\bibinfo{year}{2008}) \bibinfo{pages}{1134--1149}.
\bibitem[{Bevilacqua et~al.(2021)Bevilacqua, Caama{\~n}o-Carrillo,
  Arellano-Valle and Morales-O{\~n}ate}]{bevilacqua2021non}
\bibinfo{author}{M.~Bevilacqua}, \bibinfo{author}{C.~Caama{\~n}o-Carrillo},
  \bibinfo{author}{R.~B. Arellano-Valle},
  \bibinfo{author}{V.~Morales-O{\~n}ate}, \bibinfo{title}{Non-gaussian
  geostatistical modeling using (skew) t processes},
  \bibinfo{journal}{Scandinavian Journal of Statistics} \bibinfo{volume}{48}
  (\bibinfo{year}{2021}) \bibinfo{pages}{212--245}.
\bibitem[{Bevilacqua et~al.(2020)Bevilacqua, Caama{\~n}o-Carrillo and
  Gaetan}]{bevilacqua2020modeling}
\bibinfo{author}{M.~Bevilacqua}, \bibinfo{author}{C.~Caama{\~n}o-Carrillo},
  \bibinfo{author}{C.~Gaetan}, \bibinfo{title}{On modeling positive continuous
  data with spatiotemporal dependence}, \bibinfo{journal}{Environmetrics}
  \bibinfo{volume}{31} (\bibinfo{year}{2020}) \bibinfo{pages}{e2632}.
\bibitem[{Bevilacqua et~al.(2022)Bevilacqua, Caama{\~n}o-Carrillo and
  Porcu}]{Bevilacqua2022}
\bibinfo{author}{M.~Bevilacqua}, \bibinfo{author}{C.~Caama{\~n}o-Carrillo},
  \bibinfo{author}{E.~Porcu}, \bibinfo{title}{Unifying compactly supported and
  mat{\'e}rn covariance functions in spatial statistics},
  \bibinfo{journal}{Journal of Multivariate Analysis} \bibinfo{volume}{189}
  (\bibinfo{year}{2022}) \bibinfo{pages}{104949}.
\bibitem[{Bevilacqua et~al.(2019)Bevilacqua, Faouzi, Furrer and Porcu}]{bb2019}
\bibinfo{author}{M.~Bevilacqua}, \bibinfo{author}{T.~Faouzi},
  \bibinfo{author}{R.~Furrer}, \bibinfo{author}{E.~Porcu},
  \bibinfo{title}{{Estimation and prediction using generalized Wendland
  covariance functions under fixed domain asymptotics}}, \bibinfo{journal}{The
  Annals of Statistics} \bibinfo{volume}{47} (\bibinfo{year}{2019})
  \bibinfo{pages}{828 -- 856}.
\bibitem[{Bevilacqua and Gaetan(2015)}]{Bevilacqua:Gaetan:2015}
\bibinfo{author}{M.~Bevilacqua}, \bibinfo{author}{C.~Gaetan},
  \bibinfo{title}{Comparing composite likelihood methods based on pairs for
  spatial gaussian random fields}, \bibinfo{journal}{Statistics and Computing}
  \bibinfo{volume}{25} (\bibinfo{year}{2015}) \bibinfo{pages}{877--892}.
\bibitem[{Bevilacqua et~al.(2012)Bevilacqua, Gaetan, Mateu and
  Porcu}]{Bevilacqua2012}
\bibinfo{author}{M.~Bevilacqua}, \bibinfo{author}{C.~Gaetan},
  \bibinfo{author}{J.~Mateu}, \bibinfo{author}{E.~Porcu},
  \bibinfo{title}{Estimating space and space-time covariance functions for
  large data sets: A weighted composite likelihood approach},
  \bibinfo{journal}{Journal of the American Statistical Association}
  \bibinfo{volume}{107} (\bibinfo{year}{2012}) \bibinfo{pages}{268--280}.
\bibitem[{Bevilacqua et~al.(2023)Bevilacqua, Morales-Oñate and
  Caamaño-Carrillo}]{Bevilacqua:2018aa}
\bibinfo{author}{M.~Bevilacqua}, \bibinfo{author}{V.~Morales-Oñate},
  \bibinfo{author}{C.~Caamaño-Carrillo}, \bibinfo{title}{GeoModels: Procedures
  for Gaussian and Non Gaussian Geostatistical (Large) Data Analysis},
  \bibinfo{year}{2023}. \bibinfo{note}{R package version 1.0.7}.
\bibitem[{Blasi et~al.(2022)Blasi, Caamaño-Carrillo, Bevilacqua and
  Furrer}]{Blasi2022}
\bibinfo{author}{F.~Blasi}, \bibinfo{author}{C.~Caamaño-Carrillo},
  \bibinfo{author}{M.~Bevilacqua}, \bibinfo{author}{R.~Furrer},
  \bibinfo{title}{A selective view of climatological data and likelihood
  estimation}, \bibinfo{journal}{Spatial Statistics} \bibinfo{volume}{50}
  (\bibinfo{year}{2022}) \bibinfo{pages}{100596}. \bibinfo{note}{Special Issue:
  The Impact of Spatial Statistics}.
\bibitem[{Brychkov and Saad(2017)}]{Brychkov:Saad:2017}
\bibinfo{author}{Y.~A. Brychkov}, \bibinfo{author}{N.~Saad}, \bibinfo{title}{On
  some formulas for the appell function $f_2(a,b,b';c,c';w;z)$},
  \bibinfo{journal}{Journal Integral Transforms and Special Functions}
  \bibinfo{volume}{25} (\bibinfo{year}{2017}) \bibinfo{pages}{1465 --1483}.
\bibitem[{Caamaño-Carrillo et~al.(2023)Caamaño-Carrillo, Bevilacqua, López
  and Morales-Oñate}]{Caaman_et_al:2022}
\bibinfo{author}{C.~Caamaño-Carrillo}, \bibinfo{author}{M.~Bevilacqua},
  \bibinfo{author}{C.~López}, \bibinfo{author}{V.~Morales-Oñate},
  \bibinfo{title}{Nearest neighbours weighted composite likelihood based on
  pairs for (non-)Gaussian massive spatial data with an application to Tukey-hh
  random fields estimation}, \bibinfo{year}{2023}. \bibinfo{note}{Research
  Square}.
\bibitem[{Choi(2003)}]{choi2003notes}
\bibinfo{author}{J.-S. Choi}, \bibinfo{title}{Notes on formal manipulations of
  double series}, \bibinfo{journal}{Communications of the Korean Mathematical
  Society} \bibinfo{volume}{18} (\bibinfo{year}{2003})
  \bibinfo{pages}{781--789}.
\bibitem[{Cressie and Wikle(2011)}]{Cressie:Wikle:2011}
\bibinfo{author}{N.~Cressie}, \bibinfo{author}{C.~Wikle},
  \bibinfo{title}{Statistics for Spatio-Temporal Data.},
  \bibinfo{publisher}{Wiley Series in Probability and Statistics. Wiley},
  \bibinfo{year}{2011}.
\bibitem[{{De Oliveira}(2006)}]{DeOliveira:2006}
\bibinfo{author}{V.~{De Oliveira}}, \bibinfo{title}{On optimal point and block
  prediction in log-gaussian random fields}, \bibinfo{journal}{Scandinavian
  Journal of Statistics} \bibinfo{volume}{33} (\bibinfo{year}{2006})
  \bibinfo{pages}{523--540}.
\bibitem[{Diggle et~al.(1998)Diggle, Tawn and Moyeed}]{Diggle:Tawn:Moyeed:1998}
\bibinfo{author}{P.~Diggle}, \bibinfo{author}{J.~Tawn},
  \bibinfo{author}{R.~Moyeed}, \bibinfo{title}{Model-based geostatistics},
  \bibinfo{journal}{Journal of the Royal Statistical Society: Series C
  {(Applied} Statistics)} \bibinfo{volume}{47} (\bibinfo{year}{1998})
  \bibinfo{pages}{299--350}.
\bibitem[{Eisenbaum and Kaspi(2006)}]{Eisenbaum:Kaspi:2006}
\bibinfo{author}{N.~Eisenbaum}, \bibinfo{author}{H.~Kaspi}, \bibinfo{title}{A
  characterization of the infinitely divisible squared gaussian processes},
  \bibinfo{journal}{The Annals of Probability}  (\bibinfo{year}{2006})
  \bibinfo{pages}{728--742}.
\bibitem[{Erhardt et~al.(2015)Erhardt, Czado and Schepsmeier}]{ERHARDT201574}
\bibinfo{author}{T.~M. Erhardt}, \bibinfo{author}{C.~Czado},
  \bibinfo{author}{U.~Schepsmeier}, \bibinfo{title}{Spatial composite
  likelihood inference using local c-vines}, \bibinfo{journal}{Journal of
  Multivariate Analysis} \bibinfo{volume}{138} (\bibinfo{year}{2015})
  \bibinfo{pages}{74--88}. \bibinfo{note}{High-Dimensional Dependence and
  Copulas}.
\bibitem[{Ferrari and Cribari-Neto(2004)}]{ferrari2004beta}
\bibinfo{author}{S.~Ferrari}, \bibinfo{author}{F.~Cribari-Neto},
  \bibinfo{title}{Beta regression for modelling rates and proportions},
  \bibinfo{journal}{Journal of Applied Statistics} \bibinfo{volume}{31}
  (\bibinfo{year}{2004}) \bibinfo{pages}{799--815}.
\bibitem[{Gelfand and Schliep(2016)}]{GELFAND201686}
\bibinfo{author}{A.~E. Gelfand}, \bibinfo{author}{E.~M. Schliep},
  \bibinfo{title}{Spatial statistics and gaussian processes: A beautiful
  marriage}, \bibinfo{journal}{Spatial Statistics} \bibinfo{volume}{18}
  (\bibinfo{year}{2016}) \bibinfo{pages}{86--104}. \bibinfo{note}{Spatial
  Statistics Avignon: Emerging Patterns}.
\bibitem[{Genest and MacKay(1986)}]{Genest1986}
\bibinfo{author}{C.~Genest}, \bibinfo{author}{R.~J. MacKay},
  \bibinfo{title}{Copules archim{\'e}diennes et familles de lois
  bidimensionnelles dont les marges sont donn{\'e}es},
  \bibinfo{journal}{Canadian Journal of Statistics} \bibinfo{volume}{14}
  (\bibinfo{year}{1986}) \bibinfo{pages}{145--159}.
\bibitem[{Genton and Zhang(2012)}]{genton:Zhang:2012}
\bibinfo{author}{M.~G. Genton}, \bibinfo{author}{H.~Zhang},
  \bibinfo{title}{Identifiability problems in some non-{G}aussian spatial
  random fields}, \bibinfo{journal}{Chilean Journal of Statistics}
  \bibinfo{volume}{3} (\bibinfo{year}{2012}).
\bibitem[{Gneiting(2002)}]{Gneiting:2002}
\bibinfo{author}{T.~Gneiting}, \bibinfo{title}{Stationary covariance functions
  for space-time data}, \bibinfo{journal}{Journal of the American Statistical
  Association} \bibinfo{volume}{97} (\bibinfo{year}{2002})
  \bibinfo{pages}{590--600}.
\bibitem[{Gneiting(2013)}]{gneiting2013}
\bibinfo{author}{T.~Gneiting}, \bibinfo{title}{Strictly and non-strictly
  positive definite functions on spheres}, \bibinfo{journal}{Bernoulli}
  \bibinfo{volume}{19} (\bibinfo{year}{2013}) \bibinfo{pages}{1327--1349}.
\bibitem[{Gneiting and Schlather(2004)}]{GneitingS:2004}
\bibinfo{author}{T.~Gneiting}, \bibinfo{author}{M.~Schlather},
  \bibinfo{title}{Stochastic models that separate fractal dimension and the
  hurst effect}, \bibinfo{journal}{SIAM Rev.} \bibinfo{volume}{46}
  (\bibinfo{year}{2004}) \bibinfo{pages}{269--282}.
\bibitem[{Gorelick et~al.(2017)Gorelick, Hancher, Dixon, Ilyushchenko, Thau and
  Moore}]{gorelick2017google}
\bibinfo{author}{N.~Gorelick}, \bibinfo{author}{M.~Hancher},
  \bibinfo{author}{M.~Dixon}, \bibinfo{author}{S.~Ilyushchenko},
  \bibinfo{author}{D.~Thau}, \bibinfo{author}{R.~Moore}, \bibinfo{title}{Google
  earth engine: Planetary-scale geospatial analysis for everyone},
  \bibinfo{journal}{Remote Sensing of Environment}  (\bibinfo{year}{2017}).
\bibitem[{Gradshteyn and Ryzhik(2007)}]{Gradshteyn:Ryzhik:2007}
\bibinfo{author}{I.~Gradshteyn}, \bibinfo{author}{I.~Ryzhik},
  \bibinfo{title}{Table of Integrals, Series, and Products},
  \bibinfo{publisher}{Academic Press}, \bibinfo{address}{New York},
  \bibinfo{edition}{7} edition, \bibinfo{year}{2007}.
\bibitem[{Gr{\"a}ler(2014)}]{graler2014modelling}
\bibinfo{author}{B.~Gr{\"a}ler}, \bibinfo{title}{Modelling skewed spatial
  random fields through the spatial vine copula}, \bibinfo{journal}{Spatial
  Statistics} \bibinfo{volume}{10} (\bibinfo{year}{2014})
  \bibinfo{pages}{87--102}.
\bibitem[{Griffiths(1970)}]{Griffiths:1970}
\bibinfo{author}{R.~C. Griffiths}, \bibinfo{title}{Infinitely divisible
  multivariate gamma distributions}, \bibinfo{journal}{Sankhy$\bar{a}$ Ser. A}
  \bibinfo{volume}{32} (\bibinfo{year}{1970}) \bibinfo{pages}{393--404}.
\bibitem[{Guolo and Varin(2014)}]{guolo2014beta}
\bibinfo{author}{A.~Guolo}, \bibinfo{author}{C.~Varin}, \bibinfo{title}{Beta
  regression for time series analysis of bounded data, with application to
  {C}anada {G}oogle{\textregistered} flu trends}, \bibinfo{journal}{The Annals
  of Applied Statistics} \bibinfo{volume}{8} (\bibinfo{year}{2014})
  \bibinfo{pages}{74--88}.
\bibitem[{Heagerty and Lele(1998)}]{Heagerty:Lele:1998}
\bibinfo{author}{P.~Heagerty}, \bibinfo{author}{S.~Lele}, \bibinfo{title}{A
  composite likelihood approach to binary spatial data},
  \bibinfo{journal}{Journal of the American Statistical Association}
  \bibinfo{volume}{93} (\bibinfo{year}{1998}) \bibinfo{pages}{1099 --1111}.
\bibitem[{Heyde(1997)}]{Heyde1997}
\bibinfo{author}{C.~Heyde}, \bibinfo{title}{Quasi-Likelihood and Its
  Application: A General Approach to Optimal Parameter Estimation},
  \bibinfo{publisher}{Springer}, \bibinfo{address}{New York},
  \bibinfo{year}{1997}.
\bibitem[{Joe(2014)}]{Joe:2014}
\bibinfo{author}{H.~Joe}, \bibinfo{title}{Dependence modeling with copulas},
  \bibinfo{publisher}{Chapman and Hall/CRC}, \bibinfo{address}{Boca Raton, FL},
  \bibinfo{year}{2014}.
\bibitem[{Kazianka and Pilz(2010{\natexlab{a}})}]{Kazianka:Pilz:2010}
\bibinfo{author}{H.~Kazianka}, \bibinfo{author}{J.~Pilz},
  \bibinfo{title}{Copula-based geostatistical modeling of continuous and
  discrete data including covariates}, \bibinfo{journal}{Stochastic
  Environmental Research and Risk Assessment} \bibinfo{volume}{24}
  (\bibinfo{year}{2010}{\natexlab{a}}) \bibinfo{pages}{661--673}.
\bibitem[{Kazianka and Pilz(2010{\natexlab{b}})}]{Kazianka2010}
\bibinfo{author}{H.~Kazianka}, \bibinfo{author}{J.~Pilz},
  \bibinfo{title}{Spatial interpolation using copula-based geostatistical
  models}, \bibinfo{publisher}{Springer Netherlands},
  \bibinfo{address}{Dordrecht}, \bibinfo{year}{2010}{\natexlab{b}}, pp.
  \bibinfo{pages}{307--319}.
\bibitem[{Kibble(1941)}]{kibble1941two}
\bibinfo{author}{W.~Kibble}, \bibinfo{title}{A two-variate gamma type
  distribution}, \bibinfo{journal}{Sankhy{\=a}: The Indian Journal of
  Statistics}  (\bibinfo{year}{1941}) \bibinfo{pages}{137--150}.
\bibitem[{Krishnaiah and Rao(1961)}]{Kri:rao:1961}
\bibinfo{author}{P.~R. Krishnaiah}, \bibinfo{author}{M.~M. Rao},
  \bibinfo{title}{Remarks on a multivariate gamma distribution},
  \bibinfo{journal}{The American Mathematical Monthly} \bibinfo{volume}{68(4)}
  (\bibinfo{year}{1961}) \bibinfo{pages}{342--346}.
\bibitem[{Krishnamoorthy and Parthasarathy(1951)}]{krishnamoorthy1951}
\bibinfo{author}{A.~S. Krishnamoorthy}, \bibinfo{author}{M.~Parthasarathy},
  \bibinfo{title}{A multivariate gamma-type distribution},
  \bibinfo{journal}{Ann. Math. Statist.} \bibinfo{volume}{22}
  (\bibinfo{year}{1951}) \bibinfo{pages}{549--557}.
\bibitem[{Krupskii et~al.(2018)Krupskii, Huser and Genton}]{Gentcop}
\bibinfo{author}{P.~Krupskii}, \bibinfo{author}{R.~Huser},
  \bibinfo{author}{M.~G. Genton}, \bibinfo{title}{Factor copula models for
  replicated spatial data}, \bibinfo{journal}{Journal of the American
  Statistical Association} \bibinfo{volume}{113} (\bibinfo{year}{2018})
  \bibinfo{pages}{467--479}.
\bibitem[{Li and Sang(2018)}]{Li2018}
\bibinfo{author}{F.~Li}, \bibinfo{author}{H.~Sang}, \bibinfo{title}{On
  approximating optimal weighted composite likelihood method for spatial
  models}, \bibinfo{journal}{Stat} \bibinfo{volume}{7} (\bibinfo{year}{2018})
  \bibinfo{pages}{e194}.
\bibitem[{Lim and Teo(2009)}]{LiTe09}
\bibinfo{author}{S.~Lim}, \bibinfo{author}{L.~Teo}, \bibinfo{title}{Gaussian
  fields and {G}aussian sheets with {G}eneralized {C}auchy covariance
  structure}, \bibinfo{journal}{Stochastic Processes and Their Applications}
  \bibinfo{volume}{119} (\bibinfo{year}{2009}) \bibinfo{pages}{1325--1356}.
\bibitem[{Lindsay(1988)}]{Lindsay:1988}
\bibinfo{author}{B.~Lindsay}, \bibinfo{title}{Composite likelihood methods},
  \bibinfo{journal}{Contemporary Mathematics} \bibinfo{volume}{80}
  (\bibinfo{year}{1988}) \bibinfo{pages}{221--239}.
\bibitem[{Mai and Scherer(2014)}]{Mai2014}
\bibinfo{author}{J.-F. Mai}, \bibinfo{author}{M.~Scherer},
  \bibinfo{title}{Financial engineering with copulas explained},
  \bibinfo{publisher}{Springer}, \bibinfo{year}{2014}.
\bibitem[{Malov(2001)}]{Malov2001}
\bibinfo{author}{S.~V. Malov}, \bibinfo{title}{On Finite-Dimensional
  {A}rchimedean}, \bibinfo{publisher}{Birkh{\"a}user Boston},
  \bibinfo{address}{Boston, MA}, \bibinfo{year}{2001}.
\bibitem[{Marshall and Olkin(1988)}]{MO1988}
\bibinfo{author}{A.~W. Marshall}, \bibinfo{author}{I.~Olkin},
  \bibinfo{title}{Families of multivariate distributions},
  \bibinfo{journal}{Journal of the American Statistical Association}
  \bibinfo{volume}{83} (\bibinfo{year}{1988}) \bibinfo{pages}{834--841}.
\bibitem[{Masarotto and Varin(2012)}]{Masarotto:Varin:2012}
\bibinfo{author}{G.~Masarotto}, \bibinfo{author}{C.~Varin},
  \bibinfo{title}{Gaussian copula marginal regression},
  \bibinfo{journal}{Electronic Journal of Statistics} \bibinfo{volume}{6}
  (\bibinfo{year}{2012}) \bibinfo{pages}{1517--1549}.
\bibitem[{McNeil and Nešlehov{\'a}(2009)}]{McNeil:2009}
\bibinfo{author}{A.~J. McNeil}, \bibinfo{author}{J.~Nešlehov{\'a}},
  \bibinfo{title}{Multivariate {A}rchimedean copulas, $d$-monotone functions
  and $\ell_1$-norm symmetric distributions}, \bibinfo{journal}{Annals of
  Statistics} \bibinfo{volume}{37} (\bibinfo{year}{2009}) \bibinfo{pages}{3059
  -- 3097}.
\bibitem[{Miller and Samko(2001)}]{Miller2001}
\bibinfo{author}{K.~S. Miller}, \bibinfo{author}{S.~G. Samko},
  \bibinfo{title}{Completely monotonic functions}, \bibinfo{journal}{Integral
  Transforms and Special Functions} \bibinfo{volume}{12} (\bibinfo{year}{2001})
  \bibinfo{pages}{389--402}.
\bibitem[{Morales-Navarrete et~al.(2022)Morales-Navarrete, Bevilacqua,
  Caamaño-Carrillo and Castro}]{MoralesNavarrete2021}
\bibinfo{author}{D.~Morales-Navarrete}, \bibinfo{author}{M.~Bevilacqua},
  \bibinfo{author}{C.~Caamaño-Carrillo}, \bibinfo{author}{L.~M. Castro},
  \bibinfo{title}{Modeling point referenced spatial count data: A poisson
  process approach}, \bibinfo{journal}{Journal of the American Statistical
  Association} \bibinfo{volume}{0} (\bibinfo{year}{2022})
  \bibinfo{pages}{1--14}.
\bibitem[{Nelsen(2006)}]{nelsen2007introduction}
\bibinfo{author}{R.~B. Nelsen}, \bibinfo{title}{An introduction to copulas},
  \bibinfo{publisher}{Springer Science \& Business Media},
  \bibinfo{year}{2006}.
\bibitem[{Pace et~al.(2019)Pace, Salvan and Sartori}]{Pace2019}
\bibinfo{author}{L.~Pace}, \bibinfo{author}{A.~Salvan},
  \bibinfo{author}{N.~Sartori}, \bibinfo{title}{Efficient composite likelihood
  for a scalar parameter of interest}, \bibinfo{journal}{Stat}
  \bibinfo{volume}{8} (\bibinfo{year}{2019}) \bibinfo{pages}{e222}.
\bibitem[{Palacios and Steel(2006)}]{Palacios:Steel:2006}
\bibinfo{author}{M.~B. Palacios}, \bibinfo{author}{M.~F.~J. Steel},
  \bibinfo{title}{Non-{G}aussian {B}ayesian geostatistical modeling},
  \bibinfo{journal}{Journal of the American Statistical Association}
  \bibinfo{volume}{101} (\bibinfo{year}{2006}) \bibinfo{pages}{604--618}.
\bibitem[{Pettorelli et~al.(2005)Pettorelli, Vik, Mysterud, Gaillard, Tucker
  and Stenseth}]{pettorelli2005using}
\bibinfo{author}{N.~Pettorelli}, \bibinfo{author}{J.~O. Vik},
  \bibinfo{author}{A.~Mysterud}, \bibinfo{author}{J.-M. Gaillard},
  \bibinfo{author}{C.~J. Tucker}, \bibinfo{author}{N.~C. Stenseth},
  \bibinfo{title}{Using the satellite-derived {NDVI} to assess ecological
  responses to environmental change}, \bibinfo{journal}{Trends in ecology \&
  evolution} \bibinfo{volume}{20} (\bibinfo{year}{2005})
  \bibinfo{pages}{503--510}.
\bibitem[{Plemmons(1977)}]{plee:1977}
\bibinfo{author}{R.~Plemmons}, \bibinfo{title}{M-matrix characterizations. i --
  nonsingular m-matrices}, \bibinfo{journal}{Linear Algebra and its
  Applications} \bibinfo{volume}{18} (\bibinfo{year}{1977}) \bibinfo{pages}{175
  --188}.
\bibitem[{Porcu et~al.(2016)Porcu, Bevilacqua and Genton}]{porcubev}
\bibinfo{author}{E.~Porcu}, \bibinfo{author}{M.~Bevilacqua},
  \bibinfo{author}{M.~G. Genton}, \bibinfo{title}{Spatio-temporal covariance
  and cross-covariance functions of the great circle distance on a sphere},
  \bibinfo{journal}{Journal of the American Statistical Association}
  \bibinfo{volume}{111} (\bibinfo{year}{2016}) \bibinfo{pages}{888--898}.
\bibitem[{Quessy and Durocher(2019)}]{Quessy2019}
\bibinfo{author}{J.-F. Quessy}, \bibinfo{author}{M.~Durocher},
  \bibinfo{title}{The class of copulas arising from squared distributions:
  Properties and inference}, \bibinfo{journal}{Econometrics and Statistics}
  \bibinfo{volume}{12} (\bibinfo{year}{2019}) \bibinfo{pages}{148--166}.
\bibitem[{Quessy et~al.(2016)Quessy, Rivest and Toupin}]{QUESSY201640}
\bibinfo{author}{J.-F. Quessy}, \bibinfo{author}{L.-P. Rivest},
  \bibinfo{author}{M.-H. Toupin}, \bibinfo{title}{On the family of multivariate
  chi-square copulas}, \bibinfo{journal}{Journal of Multivariate Analysis}
  \bibinfo{volume}{152} (\bibinfo{year}{2016}) \bibinfo{pages}{40--60}.
\bibitem[{Royen(2004)}]{Royen:2004}
\bibinfo{author}{T.~Royen}, \bibinfo{title}{Multivariate {G}amma distributions
  {II}}, in: \bibinfo{booktitle}{Encyclopedia of Statistical Sciences},
  \bibinfo{publisher}{New York: John Wiley \& Sons}, \bibinfo{year}{2004}, pp.
  \bibinfo{pages}{419--425}.
\bibitem[{Sibuya(1960)}]{Sibuya:1960}
\bibinfo{author}{M.~Sibuya}, \bibinfo{title}{Bivariate extreme statistics},
  \bibinfo{journal}{Annals of the Institute of Statistical Mathematics}
  \bibinfo{volume}{11} (\bibinfo{year}{1960}) \bibinfo{pages}{195 -- 210}.
\bibitem[{Srivastava and Karlsson(1985)}]{Srivastava:Karlsson}
\bibinfo{author}{H.~M. Srivastava}, \bibinfo{author}{P.~W. Karlsson},
  \bibinfo{title}{Multiple Gaussian Hypergeometric Series},
  \bibinfo{publisher}{Ellis Horwood Ltd}, \bibinfo{year}{1985}.
\bibitem[{Stein(1999)}]{Stein:1999}
\bibinfo{author}{M.~Stein}, \bibinfo{title}{Interpolation of Spatial Data. Some
  Theory of Kriging}, \bibinfo{publisher}{Springer-Verlag},
  \bibinfo{address}{New York}, \bibinfo{year}{1999}.
\bibitem[{Sudakov(2008)}]{qq2008}
\bibinfo{author}{V.~Sudakov}, \bibinfo{title}{{Lipschitz continuity of quantile
  functions on spaces of random variables}}, \bibinfo{journal}{Journal of
  Mathematical Sciences} \bibinfo{volume}{152} (\bibinfo{year}{2008})
  \bibinfo{pages}{941 -- 943}.
\bibitem[{Suroso and Bárdossy(2018)}]{SUROSO2018685}
\bibinfo{author}{S.~Suroso}, \bibinfo{author}{A.~Bárdossy},
  \bibinfo{title}{Investigation of asymmetric spatial dependence of
  precipitation using empirical bivariate copulas}, \bibinfo{journal}{Journal
  of Hydrology} \bibinfo{volume}{565} (\bibinfo{year}{2018})
  \bibinfo{pages}{685--697}.
\bibitem[{Varin(2008)}]{Varin:2008}
\bibinfo{author}{C.~Varin}, \bibinfo{title}{On composite marginal likelihoods},
  \bibinfo{journal}{Advances in Statistical Analysis} \bibinfo{volume}{92}
  (\bibinfo{year}{2008}) \bibinfo{pages}{1--28}.
\bibitem[{Varin and Vidoni(2005)}]{Varin:Vidoni:2005}
\bibinfo{author}{C.~Varin}, \bibinfo{author}{P.~Vidoni}, \bibinfo{title}{A note
  on composite likelihood inference and model selection},
  \bibinfo{journal}{Biometrika} \bibinfo{volume}{52} (\bibinfo{year}{2005})
  \bibinfo{pages}{519--528}.
\bibitem[{Vere-Jones(1997)}]{Vere-Jones:1997}
\bibinfo{author}{D.~Vere-Jones}, \bibinfo{title}{Alpha-permanents and their
  applications to multivariate gamma, negative binomial and ordinary binomial
  distributions}, \bibinfo{journal}{New Zealand J. Math} \bibinfo{volume}{26}
  (\bibinfo{year}{1997}) \bibinfo{pages}{125--149}.
\bibitem[{Williamson(1956)}]{Williamson:1956}
\bibinfo{author}{R.~E. Williamson}, \bibinfo{title}{Multiply monotone functions
  and their laplace transforms}, \bibinfo{journal}{Duke Math. J.}
  \bibinfo{volume}{23} (\bibinfo{year}{1956}) \bibinfo{pages}{189 -- 207}.
\bibitem[{Xua and Genton(2017)}]{Xua:Genton:2017}
\bibinfo{author}{G.~Xua}, \bibinfo{author}{M.~G. Genton}, \bibinfo{title}{Tukey
  g-and-h random fields}, \bibinfo{journal}{Journal of the American Statistical
  Association} \bibinfo{volume}{112} (\bibinfo{year}{2017})
  \bibinfo{pages}{1236 --1249}.
\bibitem[{Zhang and El-Shaarawi(2010)}]{Zhang:El-Shaarawi:2010}
\bibinfo{author}{H.~Zhang}, \bibinfo{author}{A.~El-Shaarawi},
  \bibinfo{title}{On spatial skew-{G}aussian processes and applications},
  \bibinfo{journal}{Environmetrics} \bibinfo{volume}{21(1)}
  (\bibinfo{year}{2010}) \bibinfo{pages}{33--47}.

\end{thebibliography}


\end{document}